\documentclass[preprint,amsmath,amssymb,aps,prd,nofootinbib]{revtex4}
\usepackage{epsfig,graphicx,color,appendix,fge,upgreek}
\usepackage{siunitx}
\usepackage{booktabs} 
\begin{document}
\def\CP{{\it CP}~}
\def\cp{{\it CP}}
\title{\mbox{}\\[15pt]
Flavor from Consistency: Qualified-Axion, Anomaly Cancellation, and Emergent Unification}

\author{Y. H. Ahn\footnote{Email: axionahn@naver.com}
}
\affiliation{Institute of Particle and Nuclear Physics, Henan Normal University, Xinxiang, Henan 453007, China}


\begin{abstract}
We present a framework for flavored grand unification theory (flavored-GUT) in string-derived supergravity based on $G_{\rm SM}\times SL(2,\mathbb{Z})\times U(1)_X\times U(1)_{B-L}$, where gravity is intrinsically incorporated. We show that anomaly cancellation and Standard Model gauge coupling unification act as fundamental consistency conditions that determine the flavor structure, rather than treating flavor as an independent input. Mixed $SL(2,\mathbb{Z})$, $U(1)_{X}$, $U(1)_{B-L}$, and gravitational anomalies are shown to vanish, with the anomalies induced by K{\"a}hler transformations matched by those from chiral rotations of gauginos. For nontrivial $SL(2,\mathbb{Z})$ transformations of SM fermions, the anomaly-free conditions impose strong constraints on the quark and lepton flavor structures while leaving the strong CP phase unchanged. Quark and lepton mass hierarchies, mixing patterns, and the flavored Peccei-Quinn sector emerge from the same underlying structure. The consistency conditions fix the $U(1)_X$ breaking scale, identified with the Froggatt-Nielsen cutoff scale, thereby determining the QCD axion decay constant and predicting the axion mass $m_a=3.35\times10^{-8}$ eV, while simultaneously constraining the supersymmetry-breaking scale of ${\cal O}(10)$TeV and the seesaw scale.
We show that the moduli are stabilized and that gauge-invariance-induced quantized nonperturbative indices are constrained by vacuum stability. Furthermore, we demonstrate that the fundamental consistency conditions dynamically realizes a qualified QCD axion. Our results establish a predictive framework connecting flavor physics, anomaly cancellation, SM gauge coupling unification, neutrino mass generation, gravitational dynamics, and axion physics, without invoking a conventional simple unified gauge group.
\end{abstract}

\maketitle 
\section{Introduction}
The Standard Model (SM) gauge symmetry successfully describes fundamental interactions, but fails to explain the observed fermion mass hierarchies, mixing patterns, and strong CP invariance. Understanding the origin of flavor therefore remains one of the central open problems in particle physics, strongly motivating new symmetries beyond the SM. Conventional approaches typically introduce flavor symmetries\,\cite{Altarelli:2010gt,King:2013eh,Branco:2011iw,Murayama:2000dw,Branco,Ahn:2014gva}.
Although such constructions can reproduce observed fermion patterns, they often involve many free parameters and lack a fundamental organizing principle. 
Grand unified theories based on groups such as $SU(5)$, $SO(10)$, $E_6$, etc. provide partial insights into charge assignments and anomaly structures, but generally do not uniquely determine the flavor sector and require additional assumptions\,\cite{gru0,gru1,gru2}. 
Recent developments\,\cite{Binetruy:1995nt,Feruglio:2017spp,Feruglio:2023uof,Ahn:2023iqa} have shown that modular $SL(2,\mathbb{Z})$ invariance offers a minimal flavor framework in which Yukawa couplings arise as modular forms, thereby constraining quark and lepton structures without excessive scalar fields (see also Ref.\cite{Kobayashi:2020oji}). Moreover, incorporating the flavored Peccei-Quinn (PQ) symmetry\,\cite{Ahn:2014gva} into the flavor sector can simultaneously resolve the strong CP problem and provide a dark matter candidate\,\cite{Peccei-Quinn,axion, KSVZ,DFSZ}, while further constraining flavor structures and reducing the arbitrariness of Yukawa couplings\,\cite{Ahn:2016typ,Ahn:2023iqa}. 
This situation raises a fundamental question: Can the flavor structure emerge from deeper consistency conditions of the theory, rather than being imposed independently? In particular, it is natural to ask whether gauge coupling unification and anomaly cancellation can play a fundamental role in shaping flavor structures.

In this work, we propose a framework for flavored grand unification within string-derived supergravity based on $G_{\rm SM}\times SL(2,\mathbb{Z})\times U(1)_X\times U(1)_{B-L}$, where the additional $U(1)$s are gauged and gravity is intrinsically incorporated.
We show that anomaly cancellation and gauge coupling unification act as fundamental consistency conditions that determine the flavor structure rather than treating it as an independent input. The resulting modular- and gauge-invariant theory, which we refer to as a flavored grand unification theory (flavored-GUT), contains a minimal set of chiral superfields transforming under the above symmetry\footnote{The gauge symmetry $G_{\rm SM}\times U(1)_X\times U(1)_{B-L}$ may arise from three stacks of D-branes with gauge symmetry $U(3)\times U(2)\times U(1)$, see Ref.\cite{string_book}.}. In this framework, $U(1)_X$ is an anomalous flavored PQ symmetry with flavor-dependent PQ charges\,\cite{Ahn:2014gva}, whereas $U(1)_{B-L}$ is non-anomalous with respect to the SM and gravity. The modular symmetry\,\footnote{In the global supersymmetry limit ($M_P\rightarrow\infty$), modular symmetry is generally not required to be preserved at the quantum level if its origin as an exact discrete gauge symmetry in the underlying string theory is ignored (see Eqs.(\ref{kine}) and (\ref{cr000a})). Nevertheless, we argue that, although the SM fermions transform nontrivially under $SL(2,\mathbb{Z})$, this does not necessarily affect the SM strong CP phase.} $SL(2,\mathbb{Z})$, interpreted as a manifestation of string-theoretic $T$-duality, enforces invariance of the superpotential, K{\"a}hler potential, and gauge kinetic function under modular transformations. Quantum effects can nevertheless violate these symmetries through mixed modular,  gauge, and gravitational anomalies, which must cancel for the fundamental consistency of the theory\,\cite{Bilal:2008qx}. Throughout this work we assume: (i) three distinct scalar sectors are responsible for electroweak, $U(1)_X$, and $U(1)_{B-L}$ breaking. (ii) Right-handed neutrinos are introduced to cancel the $U(1)_{B-L}$ anomalies and simultaneously generate light neutrino masses through the seesaw mechanism\,\cite{Minkowski:1977sc}. (iii) In addition, the symmetry-breaking scalar fields are taken to zero modular weight; otherwise their vacuum expectation values (VEVs) would vanish\,\footnote{For vanishing modular weight, the scalar VEVs remain invariant under modular transformations, ensuring that the Yukawa structures are generated consistently without introducing flavor-dangerous modular-dependent terms.}\,\cite{Feruglio:2023uof,Ahn:2023iqa}. 

A central result of this work is that flavor is not introduced as an independent input but instead emerges from the consistency conditions (anomaly cancellation and SM gauge coupling unification). 
This differs qualitatively from conventional GUT frameworks\,\footnote{In non-supersymmetric realizations, gauge coupling unification is incompatible with precision data. Supersymmetric extensions lead to a remarkable near-unification through renormalization-group (RG) evolution\,\cite{cGUT1,Giunti:1991ta}; however, even at two-loop order the unification is not exact and typically requires threshold corrections at both the supersymmetry and GUT scales to reproduce the measured low-energy couplings\,\cite{cGUT2}.} such as minimal SU(5) or SO(10)\,\cite{cGUT}. 
In the flavored-PQ framework\,\cite{Ahn:2014gva,Ahn:2018cau}, the spontaneous breaking of $U(1)_X$ generates a pseudo Nambu-Goldstone (NG) mode, while explicit breaking of the associated global symmetry is suppressed by higher-dimensional operators characterized by a flavor scale, so-called Froggatt-Nielsen (FN) cutoff scale\,\cite{Froggatt:1978nt}. A natural question is therefore the origin and magnitude of this scale. Within a flavored-GUT framework, it is appealing to relate the flavor dynamics scale to the scale determined by the consistency conditions. RG evolution of the gauge couplings, together with the successful description of quark and lepton masses and mixings, provides indirect evidence for new interactions at a very high energy scale where gauge and flavor dynamics become correlated. 

We show that the anomalies induced by K{\"a}hler transformations align with those generated by the chiral rotation of gauginos.
When SM fermions transform nontrivially under $SL(2,\mathbb{Z})$, the cancellation of modular anomalies -- with gaugino and the gravitino contributions vanishing -- imposes stringent constraints.
These anomaly cancellation conditions, along with the guaranteed cancellation of mixed $SL(2,\mathbb{Z}) \times \{[SU(3)_C]^2, [U(1)_{\rm EM}]^2\}$ anomalies, can constrain the flavor structure of both quarks and leptons, and ensure the strong CP phase remains unmodified. 
We show that the cancellation of mixed modular, gauge, and gravitational anomalies imposes highly nontrivial constraints on modular weights and flavor-dependent charge assignments that generate fermion mass hierarchies and mixing patterns.  The anomaly coefficients for $U(1)_X \times [G_{\rm SM}]^2$ and $U(1)_X \times [gravity]^2$—determined by the $U(1)_X$ charges of SM fermions—can either vanish\,\footnote{We note that a particular anomaly-free limit reproduces the conventional $U(1)_{B-L}$ symmetry as a special case of the flavored $U(1)_X$ construction.} or remain finite, depending on the specific charge assignments across SM fermions (see Eqs.(\ref{SAo}) and (\ref{grA})).

We further demonstrate that exact SM gauge coupling unification can be realized naturally within this flavored-GUT framework. Unlike conventional GUTs, where unification is associated with embedding the SM gauge group into a simple unified gauge group\,\cite{gru0,gru1,gru2}, here it is governed by anomaly coefficients, Green-Schwarz contributions, Abelian kinetic mixing effects\,\cite{bob,gcuk0}, and flavored $U(1)$ gauge sectors associated with the underlying flavor structure. As a result, the three SM gauge couplings unify at a common scale while remaining consistent with the measured values of $\sin^2\theta_W(M_Z)$ and $\alpha_3(M_Z)$ (see the current status in Refs.\cite{PDG, BelleII}). Remarkably, the same consistency conditions that determine the flavor structure also fix the $U(1)_X$ gauge boson mass $M_X$, identified with the FN cutoff scale. Since the ratio $F_a/M_X$ is determined by the flavor structure, the QCD axion decay constant $F_a$ is no longer an independent parameter. The framework simultaneously contains the $U(1)_{B-L}$ breaking (seesaw) scale and the supersymmetry-breaking scale. Consequently, unlike the conventional axion models\,\cite{Peccei-Quinn,axion, KSVZ,DFSZ,Ahn:2014gva}, where the axion decay constant is a free parameter, the QCD axion mass becomes a genuine prediction of the theory rather than an independent parameter. 

To complete the construction, we develop simple moduli and matter-field superpotentials that generate a scalar potential, which simultaneously determines the Yukawa couplings (especially through the modulus VEV $\langle\tau\rangle$), gauge couplings, supersymmetry-breaking scale, and cosmological constant, while also addressing the axion quality problem. Previous studies on modulus $\tau$ stabilization have found that the VEV of $\tau$ often approaches specific fixed points, such as $i$, $e^{i2\pi/3}$ and $i\infty$\,\cite{Novichkov:2018ovf,Novichkov:2022wvg,Gonzalo:2018guu,Feruglio:2023uof}. 
We show that the scalar fields, including the $U(1)_X$-charged scalar fields and the $U(1)_X$-charged modulus, are stabilized, with the modulus $\tau$ settling near the modular fixed points, $\tau\approx i$ and $\tau\approx\pm1/2+i\sqrt{3}/2$. Although $SL(2,\mathbb{Z}$) is treated as an exact discrete gauge symmetry, loop corrections and small perturbations displace the modulus VEV from the modular fixed points, $\tau= i$ and $\tau=\pm1/2+i\sqrt{3}/2$, so that no non-trivial subgroup of the modular group survives in the low energy theory. In our phenomenological analysis, we select $\langle\tau\rangle\approx i$.
The flavored $U(1)_X$ plays a dual role in flavor physics and in the solution of the strong CP problem\,\cite{Ahn:2014gva,Ahn:2016hbn,Ahn:2018cau,Ahn:2023iqa}.
Upon spontaneous breaking of the $U(1)_X$ gauge symmetry and subsequent decoupling of the associated gauge boson, an effective global $U(1)_X$ symmetry emerges, which is protected against quantum gravitational effects\,\cite{Krauss:1988zc}. The resulting flavored QCD axion provides a dark matter candidate. 
Moreover, gauge invariance of the nonperturbative superpotential under $U(1)_X$ introduces quantized nonperturbative indices $N_i$ ($i=\alpha,\beta$), which play  roles in both $U(1)_X$ modulus stabilization and the resolution of the axion quality problem\,\footnote{For early discussions of the axion quality problem and the potential threat posed by Planck-scale or gravitational effects to the PQ mechanism, see Refs.\cite{Banks:1996ea,Svrcek:2006yi}}. The vacuum-stability requirement constrains the allowed values of $N_i$, thereby determining the corresponding nonperturbative instanton actions. We show that 
 anomaly cancellation and SM gauge coupling unification, as fundamental consistency conditions, constrain the framework in a parameter region where vacuum-stability is maintained and the string-induced gravitational contribution to the axion potential is effectively vanishing. This suppression is achieved through the normalized $U(1)_X$ gauge coupling at the flavored-GUT scale, thereby dynamically realizing a qualified QCD axion within the flavored-GUT framework. Consequently, the axion solution to the strong CP problem provides a compelling motivation for SM gauge-coupling unification within this framework. Moreover, we show that supersymmetry breaking is governed predominantly by the uplifting contribution associated with the Anti-de Sitter vacuum energy, with a scale consistent with that favored by renormalization-group running for SM gauge coupling unification.

The rest of this paper is organized as follows. In Sec.\,\ref{exam}, we present the flavored-GUT framework based on $G_{\rm SM}\times SL(2,\mathbb{Z}) \times U(1)_X\times U(1)_{B-L}$ and derive the modular, gauge, and gravitational anomaly cancellation conditions together with their implications for the chiral spectrum. In Sec.\,\ref{gcu}, we show that exact SM gauge coupling unification can be realized and determine the associated flavor dynamics scale.
In Sec.\,\ref{modvev}, we study the nonperturbative dynamics and QCD axion quality. In particular, Sec.\,\ref{modvev1} examines the scalar potential, vacuum stabilization, and supersymmetry breaking, while Sec.\,\ref{qual} demonstrates the dynamical realization of a qualified QCD axion within the flavored-GUT framework.
In Sec.\,\ref{visu}, we construct the quark and lepton superpotentials and discuss the resulting fermion mass hierarchies and mixing structures. And we analyze the phenomenological implications for quarks, leptons, neutrinos, and the flavored QCD axion. Finally, Sec.\,\ref{conc} summarizes our results and discusses their implications.

\section{setup: flavored Grand Unification Theory}
\label{exam}
In four-dimensional (4D) ${\cal N}=1$ string-derived supergravity in the Einstein frame, with chiral superfields $\Phi=(\varphi,\tau,...)$, the theory is characterized by a generalized K{\"a}hler function $G(\Phi,\bar{\Phi})=K(\Phi,\bar{\Phi})/M^2_P+\ln(|W(\Phi)|^2/M^6_P)$, together with an analytic gauge kinetic function $f(\Phi)$ and  a holomorphic gravitational kinetic function $f_{\rm grav}(\Phi)$. The most general action coupled to chiral and vector multiplets in curved superspace is given by\,\cite{Wess:1992cp}
{\begin{eqnarray}
 &&{\cal S}=\int d^4x d^2\theta\,2\varepsilon\Big\{\frac{3M^2_P}{8}(\bar{D}^2-8{\cal R})e^{-K(\Phi, \bar{\Phi}e^{2V})/3M^2_P}+W(\Phi)+\frac{f_{ab}(\Phi)}{4}{\cal W}^{\alpha a}{\cal W}^b_{\alpha}\nonumber\\
 &&\qquad\qquad\qquad\qquad+\frac{f_{\rm grav}(\Phi)}{8}{\cal W}^{\alpha\beta\gamma}{\cal W}_{\alpha\beta\gamma}\Big\}+{\rm h.c.}\,,
 \label{ac1}
\end{eqnarray}}
where $(2\varepsilon)$ is the chiral superspace density satisfying $(2\varepsilon|_{\theta=0}=\sqrt{-g})$ (spacetime measure), and $M_P=(8\pi G_N)^{-1/2}=2.435\times10^{18}$ GeV is the reduced Planck mass with Newtons's gravitational constant $G_N$. The operator $(\bar{D}^2-8{\cal R})$ is the chiral projection operator (where ${\cal R}$ denotes the chiral curvature superfield satisfying ${\cal R}|_{\theta^2}\supset-\frac{1}{12}R$, with Ricci scalar $R$. The combination $\frac{\bar{D}^2-8{\cal R}}{-8}$ produces the Einstein-Hilbert and gravitino terms. Here, $K(\Phi, \bar{\Phi}e^{2V})$ is a real gauge-invariant function of $\Phi$ and $\bar{\Phi}$, while $W(\Phi)$ is a holomorphic gauge-invariant function of $\Phi$. The chiral superfield $\Phi$ contains all chiral supermultiplets, both matter superfields $\varphi$ and moduli fields, including the K{\"a}hler modulus ($\tau$), complex-structure modulus ($U_X$), and dilaton ($S$). The gauge multiplet is denoted by  $V\equiv V^aT^a$, where $T^a$ are the gauge group generators. The chiral spinor superfield ${\cal W}_\alpha$ contains the Yang-Mills and Abelian gauge-field strengths, while ${\cal W}_{\alpha\beta\gamma}$ is the Weyl superfield. 

In type IIA intersecting D-brane models, which are $T$-dual to type IIB magnetized D-brane configurations, the low-energy effective theory should be invariant under the modular transformation\,\cite{string_book}
\begin{eqnarray}
\tau\rightarrow \frac{a\tau+b}{c\tau+d}=\gamma\tau\,,\qquad(a,b,c,d\in Z, ad-bc=1)\,,
\label{mo1}
\end{eqnarray}
acting on the modular group $SL(2,\mathbb{Z})$ of the complex modulus $\tau$, with ${\rm Im}(\tau)>0$. The transformation forms the modular group $SL(2,\mathbb{Z})$.
Assuming that modular forms are holomorphic everywhere in the $SL(2,\mathbb{Z})$ fundamental domain, including the cusp at infinity $\tau=i\infty$, and can be expressed as polynomials in the Eisenstein series $E_4$ and $E_6$, they form a finite-dimensional vector space at each modular weight\,\cite{Binetruy:1995nt,Feruglio:2023uof}. For modular weights $k\geq12$, the modular forms are generally not unique, but decompose into an Eisenstein-series amd cusp-form contributions. Since cusp forms vanish at the cusp, perturbative boundary conditions typically determine the modular forms uniquely up to an overall normalization.

Under the modular transformation Eq.(\ref{mo1}) and the gauged $U(1)$s, the action Eq.(\ref{ac1}) must remain invariant under
 {\begin{eqnarray}
&K(\Phi, \bar{\Phi}e^{2V})\rightarrow K(\Phi, \bar{\Phi}e^{2V})+\big(g(\tau)+g(\bar{\tau})\big)M^2_P\,,\nonumber\\
&W(\Phi)\rightarrow W(\Phi)e^{-g(\tau)}\,,\nonumber\\
&f(\Phi){\cal W}^\alpha{\cal W}_\alpha\rightarrow f(\Phi){\cal W}^\alpha{\cal W}_\alpha\,,\nonumber\\
&f_{\rm grav}(\Phi){\cal W}^{\alpha\beta\gamma}{\cal W}_{\alpha\beta\gamma}\rightarrow f_{\rm grav}(\Phi){\cal W}^{\alpha\beta\gamma}{\cal W}_{\alpha\beta\gamma}\,,
\label{tr1}
\end{eqnarray}}
where the K{\"a}hler transformation leaves the physical theory  invariant through the generalized K{\"a}hler  function.
At the quantum level, however, the symmetry $G_{\rm SM}\times SL(2,\mathbb{Z})\times U(1)_X\times U(1)_{B-L}$ can be violated by modular, gauge, and gravitational anomalies. These are classified as follows:
\begin{description}
\item[(i)] Modular anomalies associated with the nontrivial action of $SL(2,\mathbb{Z})$ on chiral fermions, leading to triangle diagrams of the form
 {\begin{eqnarray}
SL(2,\mathbb{Z})\times\{[SU(3)_C]^2, [SU(2)_L]^2, [U(1)_Y]^2, [U(1)_X]^2, [U(1)_{B-L}]^2\}\,.
\label{ano1}
\end{eqnarray}}
 Since the modular symmetry is exact in string theory and preserved under $T$-duality between type IIA intersecting and type IIB magnetized D-brane configurations, any apparent modular anomaly in the resulting low-energy supergravity must be cancelled (discrete gauge invariance).
\item[(ii)] Gauge anomalies involving the gauged Abelian symmetries, 
 {\begin{eqnarray}
U(1)_i\times\{[SU(3)_C]^2, [SU(2)_L]^2, [U(1)_Y]^2, [U(1)_X]^2, [U(1)_{B-L}]^2\}\,,\quad (i=X, B-L)\,,
\label{ano2}
\end{eqnarray}}
arising from triangle diagrams with external gauge bosons. The cancellation of these anomalies is required for quantum consistency and gauge invariance.
\item[(iii)] Mixed gravitational anomalies 
 {\begin{eqnarray}
\{SL(2,\mathbb{Z}), U(1)_X, U(1)_{B-L}\}\times[{\rm gravity}]^2\,,
\label{ano3}
\end{eqnarray}}
\end{description}
which must vanish to preserve general covariance (re-parametrization invariance) in the effective supergravity theory.

All such anomalies must cancel for the theory to be consistent. In particular, the SM fermion spectrum without three right-handed neutrinos carrying charge $-1$ under $U(1)_{B-L}$ is anomalous with respect to a gauged $U(1)_{B-L}$.  
The anomaly-cancellation conditions therefore impose stringent constraints on the allowed $U(1)_X$ charges, modular weights, and matter representations. Remarkably, when combined with the requirement of exact SM gauge coupling unification (see Sec.\ref{gcu}), they do not merely ensure quantum consistency but largely determine the flavor structure itself. As we will show, these consistency conditions correlate fermion mass hierarchies, axion physics, moduli-dependent threshold corrections, and the scales associated with $U(1)_X$ breaking and the seesaw mechanism\,\cite{Minkowski:1977sc}.
The resulting framework realizes a predictive flavored-GUT in which flavor and gauge unification originate from the same underlying structure.
  
First, for the action (\ref{ac1}) to be invariant under the modular group $SL(2,\mathbb{Z})$ and the gauged $U(1)_X$ and $U(1)_{B-L}$,
we consider a low-energy K{\"a}hler potential $K$,  superpotential $W$, gauge kinetic function $f_i(\Phi)$ (here $i$ labels the gauge group factor), and chiral gravitational kinetic function $f_{\rm grav}(\Phi)$ :
\begin{eqnarray}
&&K=-M^2_P\ln\Big\{(-i\tau+i\bar{\tau})^h \Big(S+\bar{S}-\frac{h}{16\pi^2}\ln(-i\tau+i\bar{\tau})\Big)\Big(U_X+\bar{U}_X-\frac{\delta^{\rm GS}_X}{8\pi^2}V_X\Big)^h\Big\}
\nonumber\\
&&\qquad+(-i\tau+i\bar{\tau})^{-k_i}|\varphi_i|^2+Z_X\varphi^\dag_X e^{-X 2V_X}\varphi_X+Z_{P}\varphi^\dag_{P} e^{-(B-L)2V_{P}}\varphi_{P}+...\,,\nonumber\\
&&W=\frac{1}{3}Y(\tau)\,\varphi_i\varphi_j\varphi_k+W(S,U_X,\tau)\,,
\label{kalher1}
\end{eqnarray}
with $h=3$ and subscript $P=B-L$. Here, $-k_i$ denotes the modular weight of the matter superfield $\varphi_i$, while $Z_P$ and $Z_X$ are normalization factors, with $Z_{X(P)}=1$. The dots denotes higher-dimensional operators suppressed by the ultraviolet cutoff $M_P$. 
$W(S,U_X,\tau)$ represents the moduli superpotential given by Eq.(\ref{mosu}), while the scalar-sector superpotential $W_v$ is given by Eq.(\ref{super_d}).
The Green-Schwarz (GS) parameter $\delta^{\rm GS}_X$ characterizes the coupling of the anomalous gauge boson to the closed string axion $\theta_X$. 
The $U(1)_X$ charged matter fields $\varphi_X$ and complex structure modulus $U_X$ and the vector superfield $V_X$ of the gauged $U(1)_X$ containing the gauge field $A^\mu_X$ participate in the 4D GS mechanism\,\cite{Green:1984sg}. 
The dilaton $S$, the $U(1)_X$ charged modulus $U_X$, and the $U(1)_X$ charged scalar field $\varphi_X$ can be decomposed as
\begin{eqnarray}
S=\frac{1}{g^2_{st}}\,,\qquad U_X=\sigma+i\theta_X\,,\qquad \varphi_X\big|_{\theta=\bar{\theta}=0}=\frac{1}{\sqrt{2}}e^{i\frac{A_X}{v_X}}(v_X+h_X)\,,
\label{kf1}
\end{eqnarray}
where
$\sigma=1/g^2_X$ with $g_X$ being the 4D gauge coupling of $U(1)_X$, and $A_X$, $v_X$, and $h_X$ are the NG mode, VEV, and Higgs boson of scalar components, respectively.

In order to cancel the $SL(2,\mathbb{Z})$ modular anomalies arising from gaugino and the gravitino loops, together with the mixed $U(1)_X$ gauge anomalies via the GS mechanism, the modular variations of the gauge and gravitational kinetic functions must be aligned. Since gauge and gravitational interactions originate from the same 10D (or 11D) underlying strings, we set the same chiral moduli $S$ and $U_X$ for all gauge and gravitational kinetic functions,
\begin{eqnarray}
f_{iab}(\Phi)=\delta_{ab}(\kappa_i\,S+\tilde{\kappa}_i\,U_X)\,,\qquad\quad
f_{\rm grav}(\Phi)=\kappa_R\,S+\tilde{\kappa}_R\,U_X\,,
\label{kifct1}
\end{eqnarray}
allowing\,\footnote{See Eq.(\ref{gac00}) for non-Abelian gauge group and Eq.(\ref{gac0}) for Abelian gauge group.} for different coefficient to account for the fact that gauge fields are localized on D-branes wrapping internal cycles, whereas gravity propagates in the bulk.

\subsection{$SL(2,\mathbb{Z})$ modular anomaly cancellation}
\label{ma01}
Under the modular transformation Eq.(\ref{mo1}), invariance of the 4D action Eq.(\ref{ac1}) requires that the matter fields $\varphi_i$ and the modulus $S$ transform as
\begin{eqnarray}
\varphi_i\rightarrow (c\tau+d)^{-k_i}\varphi_i\,,\qquad S\rightarrow S-\frac{1}{16\pi^2}\ln(c\tau+d)^h\,.
\label{mt0}
\end{eqnarray}
Under the modular transformation Eq.(\ref{mo1}) with Eq.(\ref{mt0}), the K{\"a}hler potential $K$ transforms as in Eq.(\ref{tr1}), yielding $g(\tau)=\ln(c\tau+d)^h$.
This redundancy in the K{\"a}hler transformation induces a modular anomaly\,\cite{Ferrara:1989bc,Derendinger:1991hq,Ibanez:1992hc,Ahn:2023iqa}, see below Eq.(\ref{cr01}).
For the superpotential $W(\Phi)$ to remain modular-invariant under the K{\"a}hler transformation in Eq.(\ref{tr1}), the modular form $Y(\tau)$ must transform as a modular form of weight $-k_Y$:
{\begin{eqnarray}
Y(\tau)\rightarrow Y(\gamma\tau)=(c\tau+d)^{-k_Y}Y(\tau)\,,
\label{mt2}
\end{eqnarray}}
where $k_Y=h-(k_i+k_j+k_k)$.
Canonically normalized fields $\hat{\varphi}$, defined by $\varphi_i=(K^{-1/2})_{ij}\hat{\varphi}_j$, ensures that the modular forms are normalized, leading to
\begin{eqnarray}
W=\frac{1}{3}\hat{Y}(\tau)\hat{\varphi}_i\hat{\varphi}_j\hat{\varphi}_k\quad\text{with}~\hat{Y}(\tau)=e^{K/2M^2_P}Y(\tau)(-i\tau+i\bar{\tau})^{(k_i+k_j+k_k)/2}\,.
\label{cn0}
\end{eqnarray}
Under the modular transformations of Eq.(\ref{mt0}), the canonically normalized fields and modular forms transform as
{\begin{eqnarray}
\hat{\varphi}_i\rightarrow\Big(\frac{c\tau+d}{c\bar{\tau}+d}\Big)^{-\frac{k_i}{2}}\hat{\varphi}_i\,,\qquad \hat{Y}(\tau)\rightarrow \Big(\frac{c\tau+d}{c\bar{\tau}+d}\Big)^{\frac{1}{2}(k_i+k_j+k_k-h)}\hat{Y}(\tau)\,.
\label{cn1}
\end{eqnarray}}
The kinetic and mass terms of the normalized SM fermions $\hat{\psi}$, gauginos $\hat{\lambda}$, and the gravitino $\hat{\xi}$ arise as functions determined by the K{\"a}hler potential $K$ and superpotential $W$. The relevant Lagrangian can be written in Weyl spinor notation as (see also Refs.\cite{Witten:1982hu,Wess:1992cp,Feruglio:2023uof})
\begin{eqnarray}
&&-\frac{1}{2}e^{\frac{K}{2M^2_P}}(K^{-1/2})^{~k}_i(K^{-1/2})^{~l}_j({\cal D}_kD_lW)\hat{\psi}^i\hat{\psi}^j-\frac{1}{4}(\text{Re}\,f)^{-1}_{ac}\,F^i\partial_if_{cb}\,\hat{\lambda}^a\hat{\lambda}^b\nonumber\\
&&-\frac{1}{2}m_{3/2}\hat{\xi}_\mu \sigma^{\mu\nu}\hat{\xi}_\nu+{\rm h.c.}+\Big(-\frac{i}{2}\bar{\hat{\psi}}^{\bar{i}}\bar{\sigma}^\mu\Gamma^i_{jk}\partial_\mu\phi^j\hat{\psi}^k+{\rm h.c.}\Big)\nonumber\\
&&-i\bar{\hat{\psi}}^{\bar{i}}\bar{\sigma}^\mu D_\mu\hat{\psi}^j-i\bar{\hat{\lambda}}\bar{\sigma}^\mu D_\mu\hat{\lambda}+\Big(-\frac{1}{2}\epsilon^{\mu\nu\rho\sigma}\bar{\hat{\xi}}_\mu\bar{\sigma}_\nu D_\rho\hat{\xi}_\sigma+{\rm h.c.}\Big)\,,
\label{sm0}
\end{eqnarray}
where $F^i=-e^{K/2M^2_P}K^{i\bar{j}}\bar{D}_{\bar{j}}\bar{W}$, $\sigma^\mu = (1, \sigma^k)$ and $\bar{\sigma}^\mu = (1, -\sigma^k)$ with $\sigma^k$ the Pauli matrices, $\Gamma^i_{jk} = K^{i\bar{\ell}}\partial_j K_{k\bar{\ell}}$ is the modular connection (K{\"a}hler Christoffel symbol), $K^{i\bar{j}} = (\partial_i \partial_{\bar{j}} K)^{-1}$ is the inverse K{\"a}hler metric, the K{\"a}hler-covariant derivative of the superpotential is $D_iW=W_i+\frac{K_i}{M^2_P}W$, the fully covariant second derivative\,\footnote{The covariant derivative ${\cal D}_i$ acting on an object $V_j$ with K{\"a}hler weight $(p,q)$, meaning it transforms as $V_j\rightarrow e^{-(p g + q \bar{g})} V_j$ under the K{\"a}hler transformation Eq.(\ref{tr1}), is given by ${\cal D}_iV_j=\partial_i V_j-\Gamma^k_{ij} V_k+\frac{p}{M^2_P}K_i V_j$. Here, under the K{\"a}hler transformation Eq.(\ref{tr1}), the object $D_j W$ transforms as $D_j W \rightarrow e^{-g(\Phi)} D_j W$.} is ${\cal D}_iD_jW=W_{ij}+\frac{K_{ij}}{M^2_P}W+\frac{K_i}{M^2_P}D_jW+\frac{K_j}{M^2_P}D_iW-\frac{K_iK_j}{M^4_P}W-\Gamma^k_{ij}D_kW$, and the mass parameter $m_{3/2}$ is given by
\begin{eqnarray}
m_{3/2}=e^{K/2M^2_P}\frac{W^\dag}{M^2_P}\,.
\label{mp32}
\end{eqnarray}
The spacetime covariant derivatives acting on SM fermions, gauginos, and the gravitino are given by
\begin{eqnarray}
D_\mu\hat{\psi}^j&=&\partial_\mu\hat{\psi}^j+iq_KK_\mu\hat{\psi}^j+...\,,\nonumber\\
D_\mu\hat{\lambda}^a&=&\partial_\mu\hat{\lambda}^a+f^{abc}A^b_\mu\hat{\lambda}^c+iq_KK_\mu\hat{\lambda}^a+...\,,\nonumber\\
D_\mu\hat{\xi}&=&\partial_\mu\hat{\xi}+iq_KK_\mu\hat{\xi}+...\,,
\label{kine}
\end{eqnarray}
for flat spacetime, where $K_\mu=-\frac{i}{2M^2_P}(K_i\partial_\mu\phi^i-K_{\bar{i}}\partial_\mu\bar{\phi}^{\bar{i}})$ is the K{\"a}hler connection\,\footnote{In the global supersymmetry (SUSY) limit $M_P\rightarrow\infty$, the K{\"a}hler connection vanishes. Consequently, the associated modular anomalies disappear (see below Eq.(\ref{mo00})), and the chiral transformations of Eq.(\ref{cr02}) effectively reduce to the case $h=0$. Meanwhile, the term containing the modular connection in Eq.(\ref{sm0}) is required to maintain covariance under SM fermion reparameterizations on the K{\"a}hler manifold. The quantum anomalies associated with the modular connection are benign in this limit, as it pertains to a spacetime-like symmetry rather than an internal gauge symmetry, and thus do not jeopardize the consistency of the quantum theory.} and the dots contain gauge connection and spacetime spin connection. For the gauginos $\hat{\lambda}^a$, which transform in the adjoint representation of the gauge group, the $f^{abc}$ are the totally antisymmetric structure constants.
The K{\"a}hler charge $q_K$ takes the values $-1/2$ for SM fermions, $+1/2$ for gauginos and the gravitino in Eqs.(\ref{sm0}) and (\ref{kine}). Under the K{\"a}hler transformation Eq.(\ref{tr1}) the K{\"a}hler connection transforms as
\begin{eqnarray}
K_\mu\rightarrow K_\mu+\frac{i}{2}\partial_\mu(g(\bar{\tau})-g(\tau))\,,
\label{kc01}
\end{eqnarray}
where $\partial_\mu g(\tau) = \frac{\partial g}{\partial \tau}\partial_\mu \tau$ and $g(\tau)=\ln(c\tau+d)^h$ , with the variation $\delta K_\mu=-i\frac{h}{2}\Big(\frac{c\partial\tau}{c\tau+d}-\frac{c\partial\bar{\tau}}{c\bar{\tau}+d}\Big)$.
The SM fermions, gauginos, and the gravitino then transform with their respective K{\"a}hler charge $q_k$ as
\begin{eqnarray}
\Psi\rightarrow e^{-q_K\frac{g-\bar{g}}{2}}\Psi\,,\qquad \bar{\Psi}\rightarrow e^{q_K\frac{g-\bar{g}}{2}}\bar{\Psi}\qquad\text{with}~\Psi=\hat{\psi}, \hat{\lambda}, \hat{\xi}\,.
\label{cr01}
\end{eqnarray}
In addition, under the modular transformation Eq.(\ref{mo1}) the term including the modular connection in Eq.(\ref{sm0}), that is, $\frac{1}{2}(\Gamma^\varphi_{\tau\varphi}\partial_\mu\tau-\Gamma^\varphi_{\bar{\tau}\varphi}\partial_\mu\bar{\tau})$ transforms as
\begin{eqnarray}
\frac{k_i}{2}\Big(\frac{c\partial\tau}{c\tau+d}-\frac{c\partial\bar{\tau}}{c\bar{\tau}+d}\Big)-\frac{k_i}{2}\Big(\frac{\partial_\mu\tau+\partial_\mu\bar{\tau}}{\tau-\bar{\tau}}\Big)\,,
\label{mc01}
\end{eqnarray}
where we have used that for a diagonal K{\"a}hler metric $\Gamma^i_{ji} =\partial_j\ln K_{i\bar{i}}$.
The second term in Eq.(\ref{mc01}) vanishes in the cusp limit $\tau\rightarrow i\infty$ and is therefore neglected. Including these contribution, the canonically normalized SM fermions $\hat{\psi}$, gauginos $\hat{\lambda}$, and the gravitino $\hat{\xi}$ transform as
\begin{eqnarray}
\hat{\psi}_i\rightarrow\Big(\frac{c\tau+d}{c\bar{\tau}+d}\Big)^{k_{\hat{\psi}_i}}\hat{\psi}_i\,,\qquad  \hat{\lambda}(\hat{\xi})\rightarrow \Big(\frac{c\tau+d}{c\bar{\tau}+d}\Big)^{-\frac{h}{4}}\hat{\lambda}(\hat{\xi})\,\qquad\text{with}\,\,k_{\hat{\psi}_i}=\frac{h}{4}-\frac{1}{2}k_i\,,
\label{cr02}
\end{eqnarray}
which ensures that the kinetic and mass terms of Eq.(\ref{sm0}) are invariant under the K{\"a}hler transformation in Eq.(\ref{kalher1}) and modular transformation Eq.(\ref{mo1}) (up to total derivative). However, these transformations Eq.(\ref{cr02}) induce chiral rotations in the fermionic path-integral measure, generating modular anomalies -- triangle anomalies analogous to the Adler-Bell-Jackiw anomaly\,\cite{Bell:1969ts}. At the quantum level, the anomalies, generated by K{\"a}hler connection and modular connection, appearing in the effective action 
 \begin{eqnarray}
SL(2,\mathbb{Z})\times \big\{[U(1)_X]^2,~[U(1)_{B-L}]^2,~ [U(1)_Y]^2,~ [SU(2)_L]^2,~ [SU(3)_C]^2,~[{\rm gravity}]^2\big\}\,.
\label{mo00}
\end{eqnarray}
should vanish.

\subsubsection{Discrete-gauge anomalies induced by Gauginos and the Gravitino}
\label{dgGG}
Under the dilaton transformation in Eq.(\ref{mt0}), the gauge and gravitational kinetic functions in Eq.(\ref{kifct1}) transform as $f_i\rightarrow f_i-\frac{\kappa_i}{16\pi^2}\ln(c\tau+d)^h$ and $f_{\rm grav}\rightarrow f_{\rm grav}-\frac{\kappa_R}{16\pi^2}\ln(c\tau+d)^h$. 
This induces the following variation of the action,
\begin{eqnarray}
\delta {\cal S}_S&=&\int d^4x\sqrt{-g} \Big(\frac{-\kappa_i}{64\pi^2}\Big)\Big\{-F^{\mu\nu}_iF_{i\mu\nu}\big(g(\tau)+g(\bar{\tau})\big)+iF^{\mu\nu}_i\tilde{F}_{i\mu\nu}\big(g(\tau)-g(\bar{\tau})\big)\Big\}\nonumber\\
&+&\int d^4x\sqrt{-g}\Big(\frac{-\kappa_R}{256\pi^2}\Big)\Big\{C^{\mu\nu\rho\sigma}C_{\mu\nu\rho\sigma}\big(g(\tau)+g(\bar{\tau})\big)-iR^{\mu\nu\rho\sigma}\tilde{R}_{\mu\nu\rho\sigma}\big(g(\tau)-g(\bar{\tau})\big)\Big\}
\label{mo0}
\end{eqnarray}
where the first term in the brackets corresponds to the gauge boson kinetic term and the gravity higher-derivative term\,\footnote{Here, $C^{\mu\nu\rho\sigma}C_{\mu\nu\rho\sigma}=R^{\mu\nu\rho\sigma}R_{\mu\nu\rho\sigma}-2R^{\mu\nu}R_{\mu\nu}+\frac{1}{3}R^2$, where $R^{\mu\nu\rho\sigma}$, $R^{\mu\nu}$, and $R$ denote the Riemann curvature tensor, Ricci tensor, and Ricci scalar, respectively.} in order, while the second terms, CP-odd terms involves the dual field strength $\tilde{F}_{i\mu\nu}=\frac{1}{2}\epsilon_{\mu\nu\rho\sigma}F^{\rho\sigma}_i$ and the dual Riemann curvature tensor. The gauge field strengths $F^{\mu\nu}_i$ are given by $i=\{G,W,Y,X,B-L\}$ for $SU(3)_C$, $SU(2)_L$, $U(1)_Y$, $U(1)_X$, and $U(1)_{B-L}$, respectively. 
Meanwhile, the gauginos and gravitino transformations of Eq.(\ref{cr02}) induce chiral rotations in the path-integral measure
\begin{eqnarray}
\delta {\cal S}_{P}&=&\int d^4x\sqrt{-g} \Big(\frac{\kappa_i}{64\pi^2}\Big)iF^{\mu\nu}_i\tilde{F}_{i\mu\nu}\big(g(\tau)-g(\bar{\tau})\big)\nonumber\\
&+&\int d^4x\sqrt{-g}\Big(\frac{-\kappa_R}{256\pi^2}\Big)iR^{\mu\nu\rho\sigma}\tilde{R}_{\mu\nu\rho\sigma}\big(g(\tau)-g(\bar{\tau})\big)
\label{mo10}
\end{eqnarray}
Therefore the total variation of the action, $\delta {\cal S}=\delta {\cal S}_S+\delta {\cal S}_P$, are exactly cancelled.

Moreover, the tree-level action including one-loop contribution from massless gauginos takes the form (see Ref.\cite{string_book}) ${\cal S}\supset\int d^4xd^2\theta\frac{1}{4}{\cal W}^\alpha{\cal W}_\alpha\big(f_i+\frac{C_i(G)}{16\pi^2}\frac{K}{M^2_P}\big)+{\rm h.c.}$, where $C_i(G)$ is the Dynkin index of the adjoint of each gauge group $G$, taking values $3,2,0$ for $SU(3)$, $SU(2)$, and $U(1)$, respectively. Then, under the K{\"a}hler transformation\,\footnote{Under the K{\"a}hler transformation, the modular anomaly manifests via the variation of the action $\delta {\cal S}=\tilde{c}\frac{1}{4}\int d^4xd^2\theta{\cal W}^\alpha{\cal W}_{\alpha}g(\tau)+{\rm h.c.}$\,\cite{Derendinger:1991hq}.} in Eq.(\ref{tr1}) the action varies as
\begin{eqnarray}
\delta{\cal S}_K=\int d^4x\sqrt{-g} \frac{C_i(G)}{16\pi^2}\Big\{-\frac{1}{8}F^{\mu\nu}_iF_{i\mu\nu}\big(g(\tau)+g(\bar{\tau})\big)+\frac{i}{8}F^{\mu\nu}_i\tilde{F}_{i\mu\nu}\big(g(\tau)-g(\bar{\tau})\big)\Big\}\,,
\label{mo0}
\end{eqnarray}
which exactly matches the anomaly induced by the chiral rotation of the gauginos in Eq.(\ref{cr02}).
Therefore the total variation of the action is $\delta{\cal S}=\delta{\cal S}_S+\delta{\cal S}_K$:
\begin{eqnarray}
\delta {\cal S}\supset\int d^4x\sqrt{-g} \frac{1}{16\pi^2}\big(C_i(G)-\kappa_i\big)\Big\{-\frac{1}{8}F^{\mu\nu}_iF_{i\mu\nu}\big(g(\tau)+g(\bar{\tau})\big)+\frac{i}{8}F^{\mu\nu}_i\tilde{F}_{i\mu\nu}\big(g(\tau)-g(\bar{\tau})\big)\Big\}\,.
\label{0th}
\end{eqnarray}
Hence the anomaly vanishes when $C_i(G)=\kappa_i$, which corresponds to the GS universality condition, see also Ref.\cite{Ibanez:1992hc}.  Therefore the modular anomalies by gauginos are cancelled consistently when the coefficient $\kappa_i$ equals the quadratic Casimir coefficient of each gauge group.

Under the K{\"a}hler transformation the gaugino masses and the gravitino mass parameters in Eq.(\ref{sm0}) transform as $M_{\hat{\lambda}}\rightarrow\Big(\frac{c\tau+d}{c\bar{\tau}+d}\Big)^{\frac{h}{2}}M_{\hat{\lambda}}$ and $m_{3/2}\rightarrow \Big(\frac{c\tau+d}{c\bar{\tau}+d}\Big)^{\frac{h}{2}}m_{3/2}$, respectively. And these mass terms are invariant by Eq.(\ref{cr02}). Since the gravitino and gauginos do not mix with SM fermions, their contribution to the modular anomalies cancels via the one-loop-corrected gravitational and gauge chiral functions, as discussed above.
In particular, (i) the gravitino contribution to the $SL(2,\mathbb{Z})$-mixed gravitational anomaly vanish (c.f. see Sec.-\ref{dgSM}, the SM fermion contribution to the $SL(2,\mathbb{Z})$-mixed gravitational anomaly). Meanwhile, (ii) the gluino contribution to the strong CP phase can be absorbed into the bare QCD vacuum angle after an anomalous chiral rotation renders the gluino mass real\,\footnote{The phases of the gaugino and gravitino mass parameters can be removed by appropriate field redefinitions. In particular, writing the gluino mass as $M_{\hat{\tilde{g}}}=|M_{\hat{\tilde{g}}}|e^{i\varphi_3}$, a chiral rotation can make it real. Because this rotation is anomalous under QCD, it shifts the bare QCD vacuum angle, and the gluino phase must be included in the definition of the effective vacuum angle.}. Consequently, after accounting for the gluino-induced shift, the effective strong CP phase can be expressed as
\begin{eqnarray}
 \vartheta_{\rm eff}= \vartheta_{\rm QCD}+{\cal A}_C\arg\Big(\frac{c\tau+d}{c\bar{\tau}+d}\Big)+\arg\big[\det (M_uM_d)\big]\,,
\label{qcdph}
\end{eqnarray}
where $\vartheta_{\rm QCD}$ is the bare QCD vacuum angle, with the gluino-induced shift included, and ${\cal A}_C$ denotes the sum of the modular weights of the quark fields.

\subsubsection{Discrete-gauge anomalies induced by SM fermions}
\label{dgSM}
Since the modular group $SL(2,\mathbb{Z})$ is treated as a discrete-gauge symmetry, see below Eq.(\ref{kine}), any anomalies generated in the effective action by the chiral rotation of SM fermions in Eq.(\ref{cr02}) that would break this invariance must be canceled. The anomaly coefficients ${\cal A}_i$ of the mixed $SL(2,\mathbb{Z})\times \big\{[SU(3)_C]^2,~[SU(2)_L]^2,~[U(1)_Y]^2,~[U(1)_{X}]^2,~ [U(1)_{B-L}]^2,~ [{\rm gravity}]^2\big\}$ are given, respectively, by
\begin{eqnarray}
{\cal A}_C&=&2{\rm Tr}[k_\psi\,T^2_{SU(3)_C}]=\sum_{i=1}^3\big(2k_{\hat{Q}_i}+k_{\hat{U}^c_i}+k_{\hat{D}^c_i}\big)\,,\nonumber\\
{\cal A}_L&=&2{\rm Tr}[k_\psi\,T^2_{SU(2)}]=\sum^3_{i=1}\big(k_{\hat{L}_i}+3k_{\hat{Q}_i}\big)\,,\nonumber\\
{\cal A}_Y&=&2{\rm Tr}[k_\psi\,c_YY^2]=\frac{c_Y}{3}\sum^3_{i=1}\big(k_{\hat{Q}_i}+8k_{\hat{U}^c_i}+2k_{\hat{D}^c_i}+3k_{\hat{L}_i}+6k_{\hat{\ell}^c_i}\big)\,,\nonumber
\end{eqnarray}
\begin{eqnarray}
{\cal A}_{X}&=&2{\rm Tr}[k_\psi\,X^2_\psi]=2\sum_{i=1}^3\Big\{3k_{\hat{Q}_i}\big(2X^2_{Q_i}-X^2_{U^c_i}-X^2_{D^c_i}\big)+k_{\hat{L}_i}\big(2X^2_{L_i}-X^2_{\ell^c_i}\big)+k_{\hat{N}^c_i}X^2_{N^c_i}\Big\}\,,
\nonumber\\
{\cal A}_{B-L}&=&2{\rm Tr}[k_\psi\,(B-L)^2_i]=2\sum^3_{i=1}\big(k_{\hat{L}_i}+k_{\hat{N}^c_i}\big)\,,\nonumber\\
{\cal A}_{\rm grav}&=&2{\rm Tr}[k_\psi]=2\sum^3_{i=1}\big(6k_{\hat{Q}_i}+3k_{\hat{U}^c_i}+3k_{\hat{D}^c_i}+2k_{\hat{L}_i}+k_{\hat{\ell}^c_i}+k_{\hat{N}^c_i}\big)\,,
\label{cr04}
\end{eqnarray}
where\,\footnote{For simplicity, we consider a minimal realization in which only one right-handed neutrino carries a $U(1)_X$ charge (see Table-\ref{reps_l}), which is sufficient for the low-energy phenomenology and anomaly cancellation discussed in this work. In the flavored-GUT framework, the hypercharge normalization factor $c_Y$ is fixed by the SM gauge coupling unification (see Eq.({\ref{th03}}) below).} the trace is over all fermions $\psi$ carrying modular weights $k_\psi$, $T_{SU(3)}$, $T_{SU(2)}$ are gauge group generators, $Y$ is the hypercharge operator with normalization factor $c_Y$, and $X_\psi$ denotes the $U(1)_X$ charge of $\psi$. 
Here $k_{\hat{Q}_i}$, $k_{\hat{U}^c_i} (k_{\hat{D}^c_i}$), $k_{\hat{L}_i}$, $k_{\hat{\ell}^c_i}$, and $k_{\hat{N}^c_i}$ denote the weights for the normalized left-handed quarks, right-handed up (down)-type quarks, left-handed leptons, right-handed leptons, and right-handed neutrinos, respectively, and $X_{Q_i}$ ($X_{L_i}$) represent the $U(1)_X$ charges of the left-handed quark (lepton) doublets, $X_{D^c_i}$ ($X_{U^c_i}$) represent the charges of the gauge singlet right-handed down (up)-type quarks, and $X_{\ell^c_i}$ ($X_{N^c_i}$) represent the charges of the gauge singlet right-handed charged-leptons (neutrinos). Note that at least one right-handed neutrino is charged under $U(1)_X$, see the above ${\cal A}_{X}$.
The $U(n)$ generators ($n\geq2$) are normalized to ${\rm Tr}[T^aT^b]=\delta^{ab}/2$. For convenience, ${\cal A}_{Y, X, B-L}$ are defined as above for $U(1)$ and ${\cal A}_R$ for gravity. Similarly, the electromagnetic anomaly coefficient for $SL(2,\mathbb{Z})\times[U(1)_{\rm EM}]^2$ is given by ${\cal A}_E=2{\rm Tr}[k_i\,(Q^{\rm em}_i)^2]$:
\begin{eqnarray}
{\cal A}_E=\frac{2}{3}\sum_{i=1}^3\big\{3(k_{\hat{L}_i}+k_{\hat{\ell}^c_i})+5k_{\hat{Q}_i}+4k_{\hat{U}^c_i}+k_{\hat{D}^c_i}\big\}\,.
\label{cr05}
\end{eqnarray}
Due to the modular- and SM gauge-invariant structure of the superpotential with non-negative weight modular forms, from Eqs.(\ref{cr04}) and (\ref{cr05}) we obtain
\begin{eqnarray}
\sum_{i=1}^3(k_{\hat{Q}_i}+k_{\hat{U}^c_i})=0\,,\qquad \sum_{i=1}^3(k_{\hat{Q}_i}+k_{\hat{D}^c_i})=0\,,\qquad \sum_{i=1}^3(k_{\hat{L}_i}+k_{\hat{\ell}^c_i})=0\,,
\label{cr000a}
\end{eqnarray}
which induce ${\cal A}_C={\cal A}_E=0$ and ${\cal A}_L=-{\cal A}_Y/c_Y$.
If the GS counterterms are introduced to cancel non-zero anomalies (that is, ${\cal A}_L=-{\cal A}_Y/c_Y$, ${\cal A}_X$, ${\cal A}_{B-L}$, and ${\cal A}_{\rm grav}$), certain anomalies that were already cancelled are reintroduced, and additional anomalous contributions arise. So, ${\cal A}_L=-{\cal A}_Y/c_Y$ should be zero:
\begin{eqnarray}
\sum^3_{i=1}\big(k_{\hat{L}_i}+3k_{\hat{Q}_i}\big)=0\,.
\label{cr0500}
\end{eqnarray}
${\cal A}_X$ must also vanish, constraining the modular weight $k_{\hat{N}^c_3}$ and $U(1)_X$ charge $X_{N^c_3}$ of right-handed neutrino $N^c_3$.
Using Eq.(\ref{cr000a}) ${\cal A}_{\rm grav}$ reduces to
\begin{eqnarray}
{\cal A}_{\rm grav}=2\sum^3_{i=1}\big(k_{\hat{L}_i}+k_{\hat{N}^c_i}\big)\,,
\label{SU01}
\end{eqnarray}
which is equivalent to ${\cal A}_{B-L}$. In the presence of $U(1)_{B-L}$ gauge field and gravity as backgrounds, their mixed anomalies appearing in the effective action that should be free impose an identical constraint on the modular weight, ${\cal A}_{B-L}={\cal A}_{\rm grav}=0$ (analogously, $\delta^{B-L}_{B-L}=\delta_{B-L}^{\rm grav}=0$, see Eq.(\ref{blA}) and below).

\subsection{Gauged $U(1)_X$ and $U(1)_{B-L}$ anomaly cancellation}
\label{NGm}
The 4D action of Eq.(\ref{ac1}), combined with the K{\"a}hler potential Eq.(\ref{kalher1}), must remain $U(1)_X$ and $U(1)_{B-L}$ gauge invariant. Under the $U(1)_X$ gauge transformation $2V_X\rightarrow 2V_X+i(\Lambda_X-\bar{\Lambda}_X)$, the matter superfields $\varphi_X$ and the complex structure modulus transform as $\varphi_X\rightarrow e^{iX\Lambda_X}\varphi_X$ and $U_X\rightarrow U_X+i\frac{\delta^{\rm GS}_X}{16\pi^2}\Lambda_X$, respectively, while under the $U(1)_{B-L}$ gauge transformation $2V_{P}\rightarrow 2V_{P}+i(\Lambda_{P}-\bar{\Lambda}_{P})$ (where $P=B-L$), only the matter superfields $\varphi_{P}$ transform as $\varphi_{P}\rightarrow e^{iP_i\Lambda_{P}}\varphi_{P}$, where $\Lambda_{X(P)}(\bar{\Lambda}_{X(P)})$
 are (anti)chiral supefields parametrizing $U(1)_{X(B-L)}$ transformation on the superspace. For the anomalous $U(1)_X$, there are the axionic modulus $\theta_X$ (from $U_X$) and axion $A_X$ (from $\varphi_X$) which have shift symmetries
\begin{eqnarray}
\theta_X\rightarrow\theta_X+\frac{\delta^{\rm GS}_X}{16\pi^2}\xi_X\,,\qquad A_X\rightarrow A_X+\frac{f_X}{\tilde{\kappa}_{i(R)}}\xi_X\,,
\label{ss01}
\end{eqnarray}
where $\xi_X=-{\rm Re}\Lambda_X|_{\theta=\bar{\theta}=0}$, $f_X=Xv_X$ is the $U(1)_X$ breaking scale, and $\tilde{\kappa}_i$ ($\tilde{\kappa}_R$) are given by Eq.(\ref{kifct1}). Then, the $U(1)_X$ gauge field $A^\mu_X$ transforms as\,\footnote{See also Eq.(\ref{ag01}) for $U(1)_X$ and Eq.(\ref{gare}) for $U(1)_{B-L}$.}
\begin{eqnarray}
A^\mu_X\rightarrow A^\mu_X-\partial^\mu\xi_X\,.
\label{ss02}
\end{eqnarray}
For the non-anomalous $U(1)_{B-L}$, the gauge field $A^\mu_{P}$ transforms as
\begin{eqnarray}
A^\mu_{P}\rightarrow A^\mu_{P}-\partial^\mu\xi_{P}\,,
\label{ps02}
\end{eqnarray}
 where $\xi_{P}=-{\rm Re}\Lambda_{P}|_{\theta=\bar{\theta}=0}$. Since the gauged $U(1)_X$ is anomalous, the axion $A_X$ and axionic modulus $\theta_X$ couple to the (non-)Abelian Chern-Pontryagin densities of the SM gauge group, $U(1)_X$, $U(1)_{B-L}$, and {\it gravity} in the compactified theory.
Under the $U(1)_X$ gauge transformations (see above Eq.(\ref{ss01})), the gauge and gravitational kinetic functions in Eq.(\ref{kifct1}) transform as $f_i\rightarrow f_i+i\tilde{\kappa}_i\frac{\delta^{\rm GS}_X}{16\pi^2}\Lambda_X$ and $f_{\rm grav}\rightarrow f_{\rm grav}+i\tilde{\kappa}_R\frac{\delta^{\rm GS}_X}{16\pi^2}\Lambda_X$. As a result, the action varies as $\delta {\cal S}_{U_X}=\int d^4x\sqrt{-g}\big\{\frac{1}{4}F^{\mu\nu}_i\tilde{F}_{i\mu\nu}\big(\tilde{\kappa}_i\frac{\delta^{\rm GS}_X}{16\pi^2}\xi_X\big)-\frac{1}{8}R^{\mu\nu\rho\sigma}\tilde{R}_{\mu\nu\rho\sigma}\big(\tilde{\kappa}_R\frac{\delta^{\rm GS}_X}{16\pi^2}\xi_X\big)\big\}$. Meanwhile, the chiral rotation of the SM fermions induces anomaly contributions through the path-integral measure,
$\delta{\cal S}_{U(1)_X}=\int d^4x\sqrt{-g}\Big\{\frac{1}{4}F^{\mu\nu}_i\tilde{F}_{i\mu\nu}\big(-\frac{\delta^i_X}{16\pi^2}\xi_X\big)+\frac{1}{8}R^{\mu\nu\rho\sigma}\tilde{R}_{\mu\nu\rho\sigma}\big(\frac{\delta^R_X}{16\pi^2}\xi_X\big)\Big\}$.
Combining both effects, the total variation of the action is
\begin{eqnarray}
\delta{\cal S}\supset\int d^4x\sqrt{-g}\frac{\xi_X}{16\pi^2}\Big\{-\frac{1}{4}F^{\mu\nu}_i\tilde{F}_{i\mu\nu}\Big(\delta^i_X-\tilde{\kappa}_i\delta^{\rm GS}_X\Big)+\frac{1}{8}R^{\mu\nu\rho\sigma}\tilde{R}_{\mu\nu\rho\sigma}\Big(\delta^R_X-\tilde{\kappa}_R\delta^{\rm GS}_X\Big)\Big\}\,.
\label{1th}
\end{eqnarray}
Therefore, the mixed gauge and gravitational anomalies are cancelled when $\delta^i_X=\tilde{\kappa}_i\delta^{\rm GS}_X$ and $\delta^R_X=\tilde{\kappa}_R\delta^{\rm GS}_X$, which is the GS anomaly cancellation condition and is exactly equivalent to Eq.(\ref{gt02}).

Then the 4D gauge-invariant effective action for the axions $\theta_X$ and $A_X$, and the $U(1)_X$ and $U(1)_{B-L}$ gauge fields $A^\mu_{X(P)}$ reads (see also Refs.\cite{Ahn:2017dpf, Ahn:2016typ})
\begin{eqnarray}
 &&K_{U_X\bar{U}_X}\Big(\partial^\mu\theta_X+\frac{\delta^{\rm GS}_X}{16\pi^2}A^\mu_X\Big)^2-\frac{1}{4g^2_i}F^{\mu\nu}_iF_{i\mu\nu}+\tilde{g}_X\xi^{\rm FI}_XD_X-D_X\tilde{g}_XX|\varphi_X|^2+\frac{1}{2}D^2_X\nonumber\\
 &&+|D_\mu\varphi_X|^2+|D_\mu\varphi_{P}|^2-\tilde{\kappa}_i\frac{\theta_X}{2}{\rm Tr}(F^{\mu\nu}_i\tilde{F}_{i\mu\nu})+\tilde{\kappa}_i\frac{A_X}{f_X}\frac{\delta^i_X}{32\pi^2}{\rm Tr}(F^{\mu\nu}_i\tilde{F}_{i\mu\nu})\nonumber\\
 &&+\tilde{\kappa}_R\frac{\theta_X}{8}R^{\mu\nu\rho\sigma}\tilde{R}_{\mu\nu\rho\sigma}-\tilde{\kappa}_R\frac{A_X}{f_X}\frac{\delta^R_X}{128\pi^2}R^{\mu\nu\rho\sigma}\tilde{R}_{\mu\nu\rho\sigma}
 \label{act1}
\end{eqnarray}
where $F^{\mu\nu}_i$ with $i=G, W, Y, X, B-L$ denote the gauge field strengths for $SU(3)_C$, $SU(2)_L$, $U(1)_Y$, $U(1)_X$, and $U(1)_{B-L}$, respectively, with gauge couplings absorbed into their definitions. From Eqs.(\ref{kalher1}, \ref{kf1}) the gauge coupling constants at a string scale $M_{st}$ are given by 
\begin{eqnarray}
\frac{1}{g^2_i(M_{st})}=\frac{\kappa_i}{g^2_{st}}+\frac{\tilde{\kappa}_i}{g^2_X}\,, 
\label{gac00}
\end{eqnarray}
which suggests that the physical gauge couplings unify at the scale $M_{st}$, see Sec.\ref{gcu}. 
The coefficients $\kappa_i$ and $\tilde{\kappa}_i$ are fixed by GS universality conditions (Eqs.(\ref{0th}) and (\ref{1th}) and see Sec.\ref{dgGG}).  For the $U(1)_X$ gauge group this gives
\begin{eqnarray}
\frac{1}{\tilde{g}^2_X(M_{st})}=\frac{\tilde{\kappa}_X}{g^2_X}\,, 
\label{gac0}
\end{eqnarray}
where $1/g^2_X\equiv \sigma^{\rm comp}_0$, see Eq.(\ref{d0aa}), since $C_X(G)=0$ for an Abelian gauge group, consistent with the GS cancellation condition, see below Eq.(\ref{mo0}).
Similarly, the $U(1)_{B-L}$ gauge coupling can be written as $1/g^2_{B-L}(M_{st})=1/g^2_{P}$.

In the effective action (\ref{act1}), the first, third, fourth, sixth, and seventh terms result from expanding the K{\"a}hler potential Eq.(\ref{kalher1}), and the second, fifth, eighth, ninth, tenth, and eleventh terms result from the gauge interaction term in Eq.(\ref{ac1}).
$F^{\mu\nu}_X=\partial^\mu A^\nu_X-\partial^\nu A^\mu_X$ is the $U(1)_X$ gauge field strength. The $U(1)_X$ gauge covariant derivative $D^\mu\varphi_X=\partial^\mu\varphi_X-iXA^\mu_X\varphi_X$ governs the coupling of the scalar component of $\varphi_X$ to the $U(1)_X$ gauge boson, where the gauge coupling $g_X$ is absorbed into $A^\mu_X$.  The corresponding expression for $U(1)_{B-L}$ is identical, obtained by replacing $X$ with $B-L$.
The coefficients $\delta^{i}_X$ of the mixed $U(1)_X\times[SU(3)_C]^2$, $U(1)_X\times[SU(2)_L]^2$, $U(1)_X\times[U(1)_Y]^2$, $[U(1)_X]^3$, and $U(1)_X\times[U(1)_{B-L}]^2$ anomalies are given, respectively, by
\begin{eqnarray}
&&\delta^G_X=2{\rm Tr}[X_{\psi}T^2_{SU(3)}]=\sum_{i=1}^3\big(2X_{Q_i}+X_{D^c_i}+X_{U^c_i}\big)\,,\nonumber\\
&&\delta^W_X=2{\rm Tr}[X_{\psi}T^2_{SU(2)}]=\sum_{i=1}^3\big(X_{L_i}+3X_{Q_i}\big)\,,\nonumber\\
&&\delta^Y_X=2{\rm Tr}[X_{\psi}\,c_YY^2]=\sum_{i=1}^3\Big(\frac{1}{3}X_{Q_i}+\frac{8}{3}X_{U^c_i}+\frac{2}{3}X_{D^c_i}+X_{L_i}+2X_{\ell^c_i}\Big)c_Y\,,\nonumber\\
&&\delta^{X}_X=2{\rm Tr}[X^3_{\psi}]=2\sum_{i=1}^3\big(6X^3_{Q_i}+3X^3_{D^c_i}+3X^3_{U^c_i}+2X^3_{L_i}+X^3_{\ell^c_i}+X^3_{N^c_i}\big)\,,\nonumber\\
&&\delta^{B-L}_X=2{\rm Tr}[X_{\psi}(B-L)^2_i]=\frac{2}{3}\sum_{i=1}^3\big(2X_{Q_i}+X_{D^c_i}+X_{U^c_i}+6X_{L_i}+3X_{\ell^c_i}\big)+2\sum_{i=1}^3X_{N^c_i}\,.
\label{SAo}
\end{eqnarray}
And the coefficient $\delta^R_{X}$ of the mixed $U(1)_X\times[{\rm gravity}]^2$ anomaly is given by
\begin{eqnarray}
\delta^{R}_X=2{\rm Tr}[X_{\psi}]=2\Big\{3\delta^G_X+\sum_{i=1}^3(2X_{L_i}+X_{\ell^c_i}+X_{N^c_i})\Big\}\,.
\label{grA}
\end{eqnarray}
The absence of GS counterterms for the $U(1)_{B-L}$ implies that all associated gauge anomalies must cancel explicitly. Therefore, the mixed anomalies $U(1)_{B-L}\times \big\{[SU(3)_C]^2,~[SU(2)_L]^2,~[U(1)_Y]^2,~[U(1)_{X}]^2,~ [U(1)_{B-L}]^2,~ [{\rm gravity}]^2\big\}$ can not appear in the effective action. While the anomaly coefficients vanish automatically
\begin{eqnarray}
\delta^{G}_{B-L}=\delta^{W}_{B-L}=\delta^{Y}_{B-L}=\delta^{B-L}_{B-L}=0\,,
\label{blA}
\end{eqnarray}
where $(B-L)_{Q_i}=1/3$, $(B-L)_{D^c_i}=(B-L)_{U^c_i}=-1/3$, $(B-L)_{L_i}=-1$, $(B-L)_{\ell^c_i}=1$, and $(B-L)_{N^c_i}=1$, the condition $\delta^{X}_{B-L}=0$ imposes a non-trivial constraint on the flavor-dependent $U(1)_X$ charges:
\begin{eqnarray}
\delta^{X}_{B-L}=2\sum_{i=1}^3\big(2X^2_{Q_i}-X^2_{D^c_i}-X^2_{U^c_i}-2X^2_{L_i}+X^2_{\ell^c_i}+X^2_{N^c_i}\big)=0\,.
\label{SAo1}
\end{eqnarray}
Furthermore, the mixed $U(1)_{B-L}\times[{\rm gravity}]^2$ anomaly vanishes automatically, $\delta_{B-L}^{R}=0$.

The Fayet-Iliopoulos (FI) term ${\cal L}^{\rm FI}_X=\xi^{\rm FI}_{X}\int d^2\theta d^2\bar{\theta}\,2V_X=\xi^{\rm FI}_{X}\,\tilde{g}_XD_X$ with $D_X=\tilde{g}_X(-\xi^{\rm FI}_{X}+X|\varphi_X|^2)$ leads to D-term potential for the anomalous $U(1)_X$,
\begin{eqnarray}
V_D=\frac{\tilde{g}^2_X}{2}(-\xi^{\rm FI}_{X}+X|\varphi_X|^2)^2\,,
\label{DV01}
\end{eqnarray}
where $\tilde{g}_X$ given by Eq.(\ref{gac0}) depends only on  the closed string modulus ${\rm Re}[U_X]=\sigma$, and $\xi^{\rm FI}_X$ is the FI factor $\xi^{\rm FI}_X=\frac{\partial K}{\partial V_X}|_{V_X=0, \sigma=\sigma_0}\Delta\sigma$ produced by expanding the K{\"a}hler potential Eq.(\ref{kalher1}) in components linear in $V_X$ and depends on ${\rm Re}[U_X]=\sigma$: $\xi^{\rm FI}_X=hM^2_P\frac{\delta^{\rm GS}_X}{16\pi^2}\frac{\Delta\sigma}{\sigma_0}$,
where $\Delta\sigma=\sigma-\sigma_0$. If $U_X$ is stabilized at $\langle U_X\rangle$, see Eq.(\ref{susy_0}), $\xi^{\rm FI}_X$ becomes a constant. Since the FI term is controlled by the string coupling, it may not be zero. Then, for $\delta^{\rm GS}_X\neq0$, see Sec.\,\ref{modvev}, the restabilization of VEVs by $\varphi_X$ necessarily implies spontaneous breaking of the anomalous $U(1)_X$. For the non-anomalous $U(1)_{B-L}$, no such FI term is present. Consequently, its D-flatness condition is satisfied with a vanishing potential, $V^{B-L}_D=0$.

Under the $U(1)_X$ transformations Eqs.(\ref{ss01}) and (\ref{ss02}), the followings in Eq.(\ref{act1}) are gauge invariant 
\begin{eqnarray}
&&{\cal L}^{\rm int}_X=-A^\mu_X J^X_\mu+\tilde{\kappa}_i\frac{A_X}{f_X}\frac{\delta^i_X}{32\pi^2}{\rm Tr}(F^{\mu\nu}_i\tilde{F}_{i\mu\nu})\,,\quad {\cal L}^{\rm int}_\theta=-A^\mu_\theta J^\theta_\mu-\tilde{\kappa}_i\frac{\theta_X}{2}{\rm Tr}(F^{\mu\nu}_i\tilde{F}_{i\mu\nu})\,,\nonumber\\
&&{\cal L}^{\rm grav}_X=-A^\mu_X J^X_\mu-\tilde{\kappa}_R\frac{\delta^R_X}{128\pi^2}R^{\mu\nu\rho\sigma}\tilde{R}_{\mu\nu\rho\sigma}\,,\qquad {\cal L}^{\rm grav}_\theta=-A^\mu_\theta J^\theta_\mu+\tilde{\kappa}_R\frac{\theta_X}{8}R^{\mu\nu\rho\sigma}\tilde{R}_{\mu\nu\rho\sigma}\,.
\label{gi01}
\end{eqnarray}
They require
\begin{eqnarray}
&&\delta^{\rm GS}_X=\frac{\delta^i_X}{\tilde{\kappa}_i}=\frac{\delta^R_X}{\tilde{\kappa}_R}\,,
\label{gt02}
\end{eqnarray}
and 
\begin{eqnarray}
&&\partial_\mu J^\mu_X=\frac{\delta^{i}_X}{32\pi^2}{\rm Tr}(F^{\mu\nu}_i\tilde{F}_{i\mu\nu})=-\partial_\mu J^\mu_\theta\,,\qquad \partial_\mu J^\mu_X=-\frac{\delta^{R}_X}{128\pi^2}R^{\mu\nu\rho\sigma}\tilde{R}_{\mu\nu\rho\sigma}=-\partial_\mu J^\mu_\theta\,,
\label{gt03}
\end{eqnarray}
where the anomalous current $J^\mu_X$ and $J^\mu_\theta$ couplings to $A^\mu_X$ are represented by $J^\theta_\mu=-K_{U_X\bar{U}_X}\frac{\delta^{\rm GS}_X}{8\pi^2}\partial_\mu\theta_X$ and $J^X_\mu=-iX\varphi_X^\dag\overleftrightarrow{\partial_\mu}\varphi_X+\frac{1}{2}\sum_\psi X_\psi\bar{\psi}\gamma_\mu\gamma_5\psi$ with $\psi$ being all $U(1)_X$ charged Dirac fermions. On the other hand, under the $U(1)_{B-L}$ transformation, below Eq.(\ref{ss02}),
\begin{eqnarray}
&&{\cal L}^{\rm int}_{B-L}=-A^\mu_{B-L} J^{B-L}_\mu\,,\qquad \partial_\mu J^\mu_{B-L}=0\,,
\label{gt04}
\end{eqnarray}
with the non-anomalous current $J^{B-L}_\mu=-i(B-L)_i\varphi_{B-L}^\dag\overleftrightarrow{\partial_\mu}\varphi_{B-L}$.

The effective action (\ref{act1}), after canonical normalization Eq.(\ref{act2}) with $\theta_X=a_\theta/8\pi^2 f_\theta$ where $f_\theta=\sqrt{2K_{U_X\bar{U}_X}}/8\pi^2$, and incorporating the gauge kinetic function, yields kinetic terms for both axions along with their couplings to the topological terms ${\rm Tr}(F^{\mu\nu}_i\tilde{F}_{i\mu\nu})$ and $R^{\mu\nu\rho\sigma}\tilde{R}_{\mu\nu\rho\sigma}$. The $U(1)_X$ and $U(1)_{B-L}$ gauge bosons acquire masses 
\begin{eqnarray}
M_X=|\tilde{g}_X|\sqrt{2K_{U_X\bar{U}_X}(\delta^{\rm GS}_X/16\pi^2)^2+2f^2_X}\,,\qquad M_{B-L}=|g_{P}|\sqrt{2(B-L)^2_\varphi|\langle\varphi_{B-L}\rangle|^2}\,,
 \label{gbm01}
\end{eqnarray}
through the St{\"u}ckelberg mechanism for $U(1)_X$ and the ordinary Higgs mechanism for $U(1)_{B-L}$, while the D-term potentials remain. Within the flavored-GUT, the $U(1)_X$ gauge boson mass $M_X$ is identified with the Froggatt-Nielsen (FN) cutoff scale:
\begin{eqnarray}
M_X=\text{\it flavor dynamics scale}\,,
 \label{FN01}
\end{eqnarray}
which is given, to a good approximation, as $2\times10^{15}$ GeV by the SM gauge coupling unification (see Eq.(\ref{gbm01m}) and Eq.(\ref{axide1})).  At this scale, the gauged $U(1)_X$ symmetry effectively decouples and the $U(1)_X$ gauge boson is integrated out. Below this scale $M_X$ (see Eqs.(\ref{gbm01}) and (\ref{gbm01m})), the gauge boson decouples, leaving an anomalous global $U(1)_X$, and the low-energy effective theory is described by FN higher-dimensional operators that generate the hierarchical Yukawa structures (see Eqs.(\ref{lagrangian_q}) and (\ref{lagrangian_l})).

The open string axion $A_X$ (with its decay constant $f_X$) is mixed linearly with the closed string axion $a_\theta$ (with its decay constant $f_\theta$)\,\cite{Ahn:2023iqa, Ahn:2016typ}
\begin{eqnarray}
\tilde{A}=\frac{A_X\frac{\delta^{\rm GS}_X}{2}f_\theta-a_\theta\,f_X}{\sqrt{f^2_X+(\frac{\delta^{\rm GS}_X}{2}f_\theta)^2}}\,,\qquad\qquad G=\frac{A_X\,f_X+a_\theta\,\frac{\delta^{\rm GS}_X}{2}f_\theta}{\sqrt{f^2_X+(\frac{\delta^{\rm GS}_X}{2}f_\theta)^2}}\,,
 \label{gt05}
\end{eqnarray}
such that the orthogonal combinations $G$ (NG mode) and $\tilde{A}$ (pseudo-NG mode) emerge. 
The gauged $U(1)_X$ absorbs one linear combination of $A_X$ and $a_\theta$, denoted $G$, giving it a string scale mass through the $U(1)_X$ gauge boson, while the other combination, approximated as $\tilde{A}\approx A_X$ when 
\begin{eqnarray}
 f_\theta\frac{|\delta^{\rm GS}_X|}{2}=\frac{g^2_X|\delta^{\rm GS}_X|}{16\pi^2}\sqrt{\frac{3}{2}}M_P\gg f_X\,,
 \label{gt06}
\end{eqnarray}
remains at low energies and contributes to the QCD axion. The quantities $g_X$ and $\delta^{\rm GS}_X$ are not free parameters, see Eqs.(\ref{gac0},\ref{gt02}): Here $\delta^i_X$ is fixed by flavor structure, see Eq.(\ref{SAo}), while $\tilde{\kappa}_i$ and $g_X$ are fixed by the gauge coupling unification, see Eq.(\ref{gc01}), together with the measured values of $\sin^2\theta_W$ and $\alpha_3$, see Eq.(\ref{th0301}).
 After the St{\"u}ckelberg mechanism, since the NG mode $G$ in Eq.(\ref{gt05}) is eaten by the $U(1)_X$ gauge boson, gauge fixing $G=0$ in unitary gauge provides
\begin{eqnarray}
a_\theta=-2\frac{A_X}{\delta^{\rm GS}_X}\frac{f_X}{f_\theta}\,,
 \label{gt07}
\end{eqnarray}
and the field $G$ no longer appears. 
Since the $U(1)_X$ gauge transformation that shifts the phases of the charged fields also shifts $G$, the gauge fixing $G=0$ fixes the corresponding gauge freedom.
So, in the nonperturbative  string-induced gravitational potential, see Eqs.(\ref{sup_01},\ref{gr_01}), the closed string axion $a_\theta$ can be expressed in terms of the low-energy axion $A_X$. The resulting nonperturbative effects therefore generate a periodic potential for the physical axion. Such a contribution could, in principle, spoil the QCD axion solution to the strong CP problem. However, in Sec.\,\ref{qual} we demonstrate that the potentially dangerous string-induced gravitational contributions can be removed, ensuring that the PQ mechanism remains intact. Furthermore, as we show in Sec.\,\ref{gcu}, the axion decay constant $f_X$ is fixed by the consistency conditions associated with SM gauge coupling unification.

There is one comment regarding this section. When the mixed anomaly coefficients vanish, $\delta^i_X=0$, the GS parameter satisfies $\delta^{\rm GS}_X=0$, eliminating the anomalous GS St{\"u}ckelberg coupling. The $U(1)_X$ gauge symmetry is then broken solely through the Higgs mechanism, and the corresponding gauge boson decouples below the symmetry-breaking scale. The flavor-dependent charge assignments satisfying $\delta^i_X=0$ cancel all mixed gauge anomalies, rendering both the $U(1)_X$ current and the associated axionic shift current exactly conserved, {\it i.e.} $\partial_\mu J^\mu_X=0$ and $\partial_\mu J^\mu_\theta=0$. In addition, $\delta^{\rm GS}_X=0$ removes the FI term ($\xi^{\rm FI}_X=0$), allowing the vacuum to satisfy the D-flatness condition and stabilizing the scalar potential.
The superpotential $W(U_X,\tau)$ (for $\alpha=0$ in Eq.(\ref{mosu})) generates a mass for the axionic mode $a_\theta$, while the NG mode $A_X$ is absorbed as the longitudinal component of the massive $U(1)_X$ gauge boson. Consequently, only an effective non-anomalous global $U(1)_X$ remains below the $U(1)_X$ breaking scale. A particularly simple realization corresponds to baryon-lepton number. 
This symmetry exactly reproduces $U(1)_{B-L}$ for the charge assignment $X_{Q_i}=1/3$, $X_{D^c_i}=X_{U^c_i}=-1/3$, $X_{L_i}=-1$, $X_{\ell^c_i}=1$, and $X_{N^c_i}=1$, yielding $\delta^G_X=\delta^W_X=\delta^Y_X=\delta^{X}_X=\delta^{R}_X=0$ and ensuring complete anomaly cancellation.

\section{Gauge coupling unification}
\label{gcu}
In this section, we show that the SM gauge coupling unification can be realized naturally within the flavored-GUT framework. 
Unlike the conventional GUTs based on a simple unified gauge group\,\cite{cGUT, cGUT1, cGUT2}, the unification is controlled by  anomaly coefficients, Green-Schwarz contributions, Abelian kinetic mixing effects\,\cite{bob,gcuk0}, and flavored $U(1)$ gauge sectors associated with the underlying flavor structure. As a result, the three SM gauge couplings unify precisely at a common scale while remaining consistent with the experimentally measured low energy values of the Weinberg angle $\sin^2\theta_W(M_Z)$ and the strong coupling constant $\alpha_3(M_Z)$(see the current status in Refs.\cite{PDG, BelleII}). This provides a qualitatively different mechanism from those of conventional GUT models\,\cite{cGUT, cGUT1, cGUT2}, establishing a direct connection between flavor physics and exact SM gauge coupling unification.   

We investigate gauge coupling unification in a flavored-GUT framework in which the flavor structure of the SM is constrained by an underlying symmetry $G_{\rm SM}\times SL(2,\mathbb{Z})\times U(1)_X\times U(1)_{B-L}$ motivated by type IIA string theory. In this setup, the gauge kinetic term in Eq.(\ref{act1}) is generalized to 
\begin{eqnarray}
-{\cal L}_{\rm kin}=\frac{1}{4}F^{\mu\nu}_i(G^{-2})_{ij}F_{j\mu\nu}\,,
 \label{gk01}
\end{eqnarray}
which is gauge invariant. Here, $G^{-2}$ denotes the gauge kinetic matrix encoding both gauge couplings and kinetic mixing among Abelian factors. Refs.\cite{cGUT1,Giunti:1991ta} established that ${\cal N}=1$ supersymmetry provides a framework for gauge coupling unification through the supersymmetric contributions to the RG running of the gauge couplings.
In 4D ${\cal N}=1$ type IIA string-derived supergravity, the running gauge coupling at a renormalization scale $\mu$ is given by
\begin{eqnarray}
(G^{-2})_{ij}(\mu)=(G^{-2})_{ij}(M_{st})-\frac{b_{ij}}{16\pi^2}\ln\Big(\frac{\mu}{M_{st}}\Big)^2+\Delta^{\rm thres}_{ij}\,,
 \label{gc01}
\end{eqnarray}
where $M_{st}$ denotes the string scale.
The quantity $(G^{-2})_{ij}(M_{st})$ represents the tree-level gauge kinetic function at $M_{st}$, determined by the underlying string compactification (see Eq.(\ref{gac00})), and is independent of the renormalization scale $\mu$. The matrix-valued one-loop $\beta$-function coefficients $b_{ij}$ are defined via the RG equation $d(G^{-2})_{ij}/d\ln\mu=-b_{ij}/8\pi^2$ and is given by $b_{ij}=-3C_{ij}(G)+\sum_{\text{chiral multiplets}}T_{ij}(R)$, where $C_{ij}(G)$ is the quadratic Casimir of the adjoint representation of $G$ (non-zero only for non-Abelian groups, and diagonal in gauge indices), $T_{ij}(R)$ generalizes the Dynkin index of the representation $R$ to include Abelian charges, with $T_{ij}(R)=Q_iQ_j$ for $U(1)$ factors. For a single gauge factor, the $\beta$-function reduces to $\beta(g_i)=dg_i/d\ln\mu=b_i\,g^3_i/16\pi^2$; thus for $b_i<0$ for asymptotically free theory. Our normalization conventions are ${\rm Tr}[T^aT^b]=T_i(R)\delta^{ab}$ with $T(N)=1/2$ for $SU(N)$ and $T_Y(R)=c_YY^2$ for $U(1)_Y$ (see below Eq.(\ref{cr04}))\,\footnote{The hypercharge generator used in the gauge kinetic term is not $Y$, but a rescaled one $\tilde{Y}=\sqrt{c_Y}Y$. So the hypercharge coupling is rescaled as $g_{0Y}\rightarrow g_{0Y}/\sqrt{c_Y}\equiv g_Y$ (effective $U(1)_Y$ coupling) and the covariant derivative $D^\mu=\partial^\mu-ig_{0Y}YA^\mu_Y$ is invariant.}. For instance, with minimal matter content including right-handed neutrinos and flavons, the $\beta$-function coefficients are 
\begin{eqnarray}
&&(b^{\rm SUSY}_3,\; b^{\rm SUSY}_2,\; b^{\rm SUSY}_Y) = (-3,\; 1,\; c_Y\,11) ~ \text{with}~ c_Y = \frac{30}{23}\;\;~\text{for} \; SU(3)_C,\; SU(2)_L,\; U(1)_Y,\nonumber\\
&&b_P = 16,\qquad b_X = 8398 \quad \text{for} \; U(1)_{B-L},\; U(1)_X,\nonumber\\
&&b_{YP} = 8\sqrt{c_Y},\qquad b_{YX} = -129\sqrt{c_Y},\qquad b_{XP} = -222 \quad \text{(mixed coefficients)}\,,
 \label{bfc}
\end{eqnarray}
where $c_Y$ denotes the normalization factor of the hypercharge gauge coupling (see below Eq.(\ref{th03}) with Table-\ref{reps_q} and -\ref{reps_l}, and the discussion above Eq.(\ref{super_d})). 
Finally, $\Delta^{\rm thres}_{ij}$ denotes the threshold corrections arising from integrating out heavy states at symmetry-breaking scale, including massive gauge sectors associated with broken $U(1)$ symmetries breaking, massive string states, Kaluza-Klein and winding modes, and GS-induced moduli-dependent effects\,\cite{string_book}.

In orbifold compactifications containing unrotated complex planes, the gauge coupling constants (see Eq.(\ref{gc01})) receive moduli-dependent threshold corrections from (i) massive momentum and winding states in ${\cal N}=2$ subsectors associated with the K{\"a}hler modulus\,\cite{Ibanez:1992hc}, and (ii) the GS St{\"u}ckelberg structure induced by an anomalous $U(1)_X$.
In order to get the moduli-dependent 1-loop coefficients $\Delta^{\rm thres}_i$, we consider the loop anomalies and GS anomaly cancellations (see Secs.\ref{dgGG} and \ref{NGm}).
As shown in Sec.\ref{ma01}, the $SL(2,\mathbb{Z})\times[G_i]^2$ anomalies induced by the chiral rotations of SM fermions cancel owing to the SM field content, and therefore no anomaly-induced threshold corrections are requires. Meanwhile, there exist the $SL(2,\mathbb{Z})\times[G_i]^2$ anomalies induced by the chiral rotations of gauginos, and therefore the $SL(2,\mathbb{Z})$ GS-induced threshold correction is required.
And as shown in Sec.\ref{NGm}, for $U(1)_X$ GS-induced threshold correction, the gauge kinetic function (see Eq.(\ref{kifct1})) transforms nontrivially under the $U(1)_X$ gauge transformation, while the chiral rotations of the SM fermions also induce $U(1)_X\times[G_i]^2$ anomalies through the path-integral measure. Consequently, the total variation is proportional to $(\delta^i_X-\tilde{\kappa}_i\delta^{\rm GS}_X)$, see Eq.(\ref{1th}), which is cancelled by the corresponding GS-induced threshold correction.
The corresponding modular- and gauge-invariant threshold corrections then take the form
\begin{eqnarray}
&&\Delta^{(\tau)}_i=+\frac{h}{16\pi^2}(C_i(G)-\kappa_i)\ln(-i\tau+i\bar{\tau})|\eta(\tau)|^4\,,\nonumber\\
&&\Delta^{(U_X)}_i=-\frac{h}{16\pi^2}(\delta^i_X-\tilde{\kappa}_i\delta^{\rm GS}_X)\ln\Big(U_X+\bar{U}_X-\frac{\delta^{\rm GS}_X}{8\pi^2}V_X\Big)\,,
 \label{th01}
\end{eqnarray}
where $C_i(G)$ and $\delta^i_X$ arise from one-loop anomalies of the massless spectrum, while $\kappa_i$ and $\tilde{\kappa}_i\delta^{\rm GS}_X$ originate from GS anomaly cancellation (see Sec.\ref{dgGG} and \ref{NGm}). Gauge coupling unification requires that all moduli-dependent terms be universal across gauge sectors, imposing $C_i(G)=\kappa_i$ and $\delta^i_X=\tilde{\kappa}_i\delta^{\rm GS}_X$, independent of the gauge group $G$. Consequently, in orbifold compactifiocations with unrotated complex plane, gauge coupling unification becomes a non-trivial consistency condition relating {\it geometry} (moduli dependence)$\leftrightarrow${\it spectrum} (anomaly coefficients)$\leftrightarrow${\it GS couplings}. This interplay severely restricts the allowed massless spectrum, making such compactifications highly constrained from the viewpoint of phenomenologically viable unification.  

At a gauge coupling unification scale $\mu=\Lambda_{\rm fGUT}$, the SM gauge coupling unification is defined by $g_3(\Lambda_{\rm fGUT})=g_2(\Lambda_{\rm fGUT})=g_Y(\Lambda_{\rm fGUT})$, that is, $G^{-2}(\Lambda_{\rm fGUT})=g^{-2}_i(\Lambda_{\rm fGUT})\times{\rm diag}(1,1,1)$. Equivalently, at $\Lambda_{\rm fGUT}$ any two gauge couplings satisfy $1/g^2_i(\Lambda_{\rm fGUT})=1/g^2_j(\Lambda_{\rm fGUT})$. Using the one-loop running with string threshold corrections Eq.(\ref{gc01}) with Eq.(\ref{th01}), this condition leads to 
\begin{eqnarray}
&&\frac{\Lambda_{\rm fGUT}}{M_{st}}=e^{\frac{2\pi}{b_i-b_j}\{\frac{1}{\alpha_i(M_{st})}-\frac{1}{\alpha_j(M_{st})}\}}\nonumber\\
&&\qquad\times\Big[(-i\tau+i\bar{\tau})|\eta(\tau)|^4\Big]^{h\frac{C_i(G)-\kappa_i-C_j(G)+\kappa_j}{2(b_i-b_j)}}\Big[U_X+\bar{U}_X-\frac{\delta^{\rm GS}_X}{8\pi^2}V_X\Big]^{h\frac{\delta^j_X-\tilde{\kappa}_j\delta^{\rm GS}_X-\delta^i_X+\tilde{\kappa}_i\delta^{\rm GS}_X}{2(b_i-b_j)}}_{V_X=0}\,,
 \label{th02}
\end{eqnarray}
where $\alpha_i\equiv g^2_i/4\pi$ with $\alpha_3$ for $SU(3)_C$, $\alpha_2$ for $SU(2)_L$, $\alpha_Y$ for $U(1)_Y$.
Since gauge coupling unification depends only on differences of gauge couplings, the universal GS contribution cancels out and plays no role in determining the unification scale $\Lambda_{\rm fGUT}$. The unification scale $\Lambda_{\rm fGUT}$ becomes independent of the moduli $\tau$ and $U_X$ whenever $C_i(G)=\kappa_i$ and $\delta^i_X=\tilde{\kappa}_i\delta^{\rm GS}_X$, so that the moduli-dependent threshold corrections are identical for all gauge sectors. Thus for the flavored-GUT it immediately follows that $\Lambda_{\rm fGUT}=M_{st}\,exp\{{\frac{2\pi}{b_i-b_j}(\frac{1}{\alpha_i(M_{st})}-\frac{1}{\alpha_j(M_{st})})}\}$.
Although three gauge couplings are involved, they can meet at a single point because they descend from a common high-energy gauge kinetic function (see Eq.(\ref{kifct1})). Only two independent matching conditions are required. The three couplings intersect at one scale if and only if $(i)$ $\alpha_i(M_{st})=\alpha_j(M_{st})$ or $(ii)$ $\frac{1}{b_Y-b_2}\big(\frac{1}{\alpha_Y(M_{st})}-\frac{1}{\alpha_2(M_{st})}\big)=\frac{1}{b_2-b_3}\big(\frac{1}{\alpha_2(M_{st})}-\frac{1}{\alpha_3(M_{st})}\big)$; otherwise, gauge coupling unification does not occur. Using Eqs.(\ref{gac00}) and (\ref{gt02}), these can be interpreted in terms of the $U(1)_X$-mixed anomaly coefficients and the $U(1)_X$ gauge coupling (which is proportional to $\alpha_{st}/\delta^{\rm GS}_X$) as 
\begin{eqnarray}
&&(i)\quad\kappa_G\delta^W_X=\kappa_W\delta^G_X+(\kappa_G-\kappa_W)\delta^Y_X\quad\text{and}\quad \alpha_X\delta^{\rm GS}_X=\alpha_{st}\frac{\delta^Y_X-\delta^W_X}{\kappa_W}\,,\nonumber\\
&&(ii)\quad\frac{1}{\alpha_X\,\delta^{\rm GS}_X}\Big(\frac{\delta^Y_X-\delta^W_X}{b_Y-b_2}-\frac{\delta^W_X-\delta^G_X}{b_2-b_3}\Big)=\frac{1}{\alpha_s}\Big(\frac{\kappa_W-\kappa_G}{b_2-b_3}+\frac{\kappa_W}{b_Y-b_2}\Big)\,,
 \label{th03}
\end{eqnarray}
where $\alpha_{st}=g^2_{st}/4\pi$ and $\alpha_X=g^2_X/4\pi$.
In case $(i)$, gauge coupling unification occurs at the string scale $\Lambda_{\rm fGUT}=M_{st}$, referred to as flavored-GUT scale (see Eq.(\ref{gac00})), for $\delta^i_X\neq\delta^j_X$, whereas in case $(ii)$ it takes place at a scale $\Lambda_{\rm fGUT}<M_{st}$ ({\it mirage unification}). The hypercharge normalization factor $c_Y$ (see Eqs.(\ref{cr04}) and (\ref{SAo})) is then determined by the gauge coupling unification condition in Eq.(\ref{th03}).
In this work, we focus on case $(i)$, which reduces to
\begin{eqnarray}
-\alpha_X\,\delta^{\rm GS}_X=\alpha_{st}\frac{\delta^i_X-\delta^j_X}{\kappa_i-\kappa_j}\,,\qquad(i\neq j=G, W,Y)\,,
 \label{th0003}
\end{eqnarray}
under which exact SM gauge coupling unification is realized at the string scale through the GS-induced threshold corrections.

\begin{figure}[t]
\begin{minipage}[h]{12.0cm}
\epsfig{figure=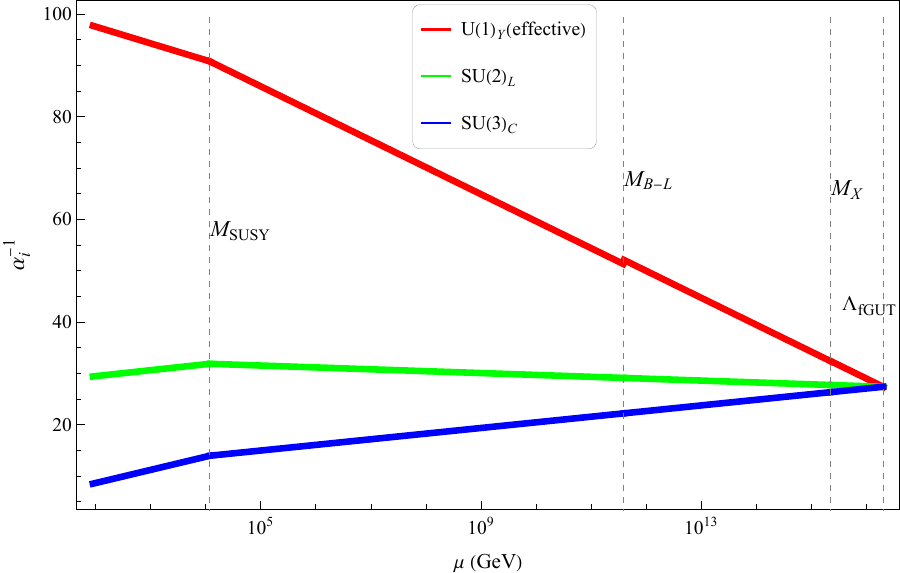,width=12.0cm,angle=0}
\end{minipage}
\caption{\label{Fig2} Plot for running of the inverse gauge couplings $\alpha^{-1}_i(\mu)$ as a function of the renormalization scale $\mu$ (in GeV). The solid blue, green, and red curves represent the $SU(3)_C$, $SU(2)_L$, and effective $U(1)_Y$ couplings, respectively. The vertical dashed lines indicate the energy thresholds $M_X=2.228\times10^{15}$ GeV (QCD axion decay constant $F_a=1.625\times10^{14}$ GeV), $M_{B-L}=3.897\times10^{11}$ GeV, $M_{\rm SUSY}=12$ TeV, and the effective flavored-GUT scale $\Lambda_{\rm fGUT}=2\times10^{16}$ GeV, listed in Table-\ref{bmp} (where $g_{B-L}=0.28$ and $g_X=0.2917$ with $g_X/g_{B-L}\sim1$). The three SM gauge couplings unify at $\Lambda_{\rm fGUT}$ with the value $\alpha^{-1}_{\rm fGUT}=27.355$, consistent with the experimental low‑energy constraints $\alpha_3(M_Z)=0.1179$ and $\sin^2\theta_W(M_Z)=0.23122$ in Eq.(\ref{th0301}).}
\end{figure} 
At the flavored-GUT scale $\Lambda_{\rm fGUT}=M_{st}$, case $(i)$ of Eq.(\ref{th03}) imposes a stringent constraint on the flavor structure through the mixed $U(1)_X$ anomaly coefficients, while simultaneously realizing the SM gauge coupling unification $g_3(\Lambda_{\rm fGUT})=g_2(\Lambda_{\rm fGUT})=g_Y(\Lambda_{\rm fGUT})=g_{st}$. In contrast, the anomalous $U(1)_X$ gauge coupling is not determined by ordinary gauge unification, but instead by the unification condition of Eq.(\ref{th03}): $\tilde{g}_X(M_{st})=g_{st}\sqrt{(\delta^Y_X-\delta^W_X)/\kappa_W\delta^X_X}$ as follows from Eqs.(\ref{gac0}) and (\ref{gt02}).
On the other hand, since the $U(1)_{B-L}$ is non-anomalous, its coupling $g_{B-L}$ is not constrained by the unification condition Eq.(\ref{th03}).
Because both $U(1)_X$ and $U(1)_{B-L}$ originate from the Abelian sector of the same string compactification, their gauge couplings are expected to  be of comparable magnitude at the string scale
\begin{eqnarray}
g_{B-L}(M_{st})\sim g_X(M_{st})\quad\text{up to}~{\cal O}(1)\,,
 \label{th04}
\end{eqnarray}
where $g_X(M_{st})$ is obtained from Eq.(\ref{th0003}).
Moreover, there is no kinetic mixing at $\Lambda_{\rm fGUT}$. While the $SU(3)_C$ and $SU(2)_L$ couplings continue to evolve according to their respective $\beta$-functions, radiative corrections generated in the running from $\Lambda_{\rm fGUT}$ down to $M_{X}$ (see Eq.(\ref{gbm01}))\,\footnote{Since the $U(1)_X$ gauge symmetry is broken via a GS-induced St{\"u}ckelberg mechanism, the NG mode in Eq.(\ref{gt05}) (including the axionic component of a complex structure modulus $U_X$) is absorbed by the gauge boson, which thereby acquires a mass of order the string scale (see Sec.\ref{NGm}). We therefore assume $M_X>M_{B-L}$ (see Eq.(\ref{gbm01})).} induce kinetic mixing\,\cite{bob,gcuk0} among $U(1)_Y\times U(1)_X\times U(1)_{B-L}$ gauge sectors:
  \begin{eqnarray}
  G^{-2}(\mu)={\left(\begin{array}{ccc}
 g^{-2}_Y(\mu) &  -\frac{b_{YX}}{8\pi^2}\ln(\mu/\Lambda_{\rm fGUT}) & -\frac{b_{YP}}{8\pi^2}\ln(\mu/\Lambda_{\rm fGUT}) \\
-\frac{b_{YX}}{8\pi^2}\ln(\mu/\Lambda_{\rm fGUT})  &   \tilde{g}^{-2}_X(\mu) & -\frac{b_{XP}}{8\pi^2}\ln(\mu/\Lambda_{\rm fGUT}) \\
-\frac{b_{YP}}{8\pi^2}\ln(\mu/\Lambda_{\rm fGUT}) & -\frac{b_{XP}}{8\pi^2}\ln(\mu/\Lambda_{\rm fGUT})  &   g^{-2}_P(\mu) 
 \end{array}\right)}\,,
 \label{gth04}
  \end{eqnarray}
where the diagonal entries $1/g^2_i(\mu)=1/g^2_i(\Lambda_{\rm fGUT})-\frac{b_i}{8\pi^2}\ln(\mu/\Lambda_{\rm fGUT})$ with $i=Y, P$ ($P=B-L$) and $1/\tilde{g}^2_X(\mu)=1/\tilde{g}^2_X(\Lambda_{\rm fGUT})-\frac{b_X}{8\pi^2}\ln(\mu/\Lambda_{\rm fGUT})$. The off-diagonal entries of Eq.(\ref{gth04}) show that gauge kinetic mixing among $U(1)_Y$, $U(1)_X$ and $U(1)_{B-L}$ is generated radiatively, even if the mixing terms vanish at $\Lambda_{\rm fGUT}$. 
At the $U(1)_X$ breaking scale $M_X$, the kinetic mixing terms can be removed by the gauge field redefinition\,\footnote{Note that this gauge field redefinition also affects the covariant derivative $D^\mu=\partial^\mu-iYA^\mu_Y-iXA^\mu_X-i(B-L)A^\mu_P$ as $D^\mu\rightarrow D^\mu-iXA^\mu_Y\,\tilde{g}^2_X(M_X)\frac{b_{YX}}{8\pi^2}\ln(M_X/\Lambda_{\rm fGUT})-iXA^\mu_P\,\tilde{g}^2_X(M_X)\frac{b_{XP}}{8\pi^2}\ln(M_X/\Lambda_{\rm fGUT})$, where the gauge couplings are absorbed into the gauge fields. However, this redundancy can be removed by the gauge transformation $\varphi_X\rightarrow\varphi_X\,e^{iX\Lambda_X}$ (see Eq.(\ref{ss02}) and above Eq.(\ref{ss01})). A similar argument applies to Eq.(\ref{gare}) (see Eq.(\ref{ps02})).} (see Eq.(\ref{ss02})) 
\begin{eqnarray}
A^\mu_X\rightarrow A^\mu_X+A^\mu_Y\,\tilde{g}^2_X(\mu)\,\frac{b_{YX}}{8\pi^2}\ln\Big(\frac{\mu}{\Lambda_{\rm fGUT}}\Big)+A^\mu_P\,\tilde{g}^2_X(\mu)\,\frac{b_{XP}}{8\pi^2}\ln\Big(\frac{\mu}{\Lambda_{\rm fGUT}}\Big)\qquad\text{at}~\mu=M_X\,,
\label{ag01}
\end{eqnarray}
 after which the mixed kinetic mixing terms $F_{Y\mu\nu}F^{\mu\nu}_X$ and $F_{X\mu\nu}F^{\mu\nu}_P$ are eliminated. 
The scale $M_X$ in Eqs.(\ref{gbm01}) and (\ref{FN01}) can be expressed, using Eq.(\ref{gac0}) together with the case $(i)$ of Eq.(\ref{th03}), as 
\begin{eqnarray}
M_X&=&g_{st}\Big(2\frac{\delta^Y_X-\delta^W_X}{\delta^X_X\kappa_W}\Big)^{\frac{1}{2}}\Big[g^4_{st}\frac{3}{4}\Big(\frac{\delta^Y_X-\delta^W_X}{\kappa_W}\Big)^2\Big(\frac{M_P}{16\pi^2}\Big)^2+f^2_X\Big]^{\frac{1}{2}}\nonumber\\
&=&g^3_{st}\frac{M_P}{16\pi^2}\sqrt{\frac{3}{2|\delta^X_X|}}\Big|\frac{\delta^Y_X-\delta^W_X}{\kappa_W}\Big|^{\frac{3}{2}} \Big(1-8\Delta^2_\chi\,g^2_{st}\frac{\delta^Y_X-\delta^W_X}{\delta^X_X\kappa_W}\Big)^{-\frac{1}{2}}\,,
 \label{gbm01m}
\end{eqnarray}
where $\tilde{\kappa}_X=\delta^X_X/\delta^{\rm GS}_X$ (see Eq.(\ref{gt02})) and $K_{U_X\bar{U}_X}=3g^4_XM^2_P/4$ have been used in the first equality, while Eqs.(\ref{AFN1}) and (\ref{NGboson}) together with $f_X\equiv f_A$ have been used in the second equality. Note that, apart from $g_{st}$, all  parameters appearing in Eq.(\ref{gbm01m}) are fixed by the flavor sector: For instance, $\delta^X_X=-64572$, $\delta^Y_X=-80$, $\delta^W_X=-38$ (obtained from Table-\ref{reps_q} and -\ref{reps_l}) and $\Delta_\chi=0.62$ (from Eq.(\ref{delchi})).
Once the coupling $g_{st}$ is fixed so as to satisfy the measured strong coupling constant (see Eq.(\ref{th0301})), the flavor dynamics scale $M_X$ of Eq.(\ref{FN01}) and the QCD axion decay constant $F_a$ defined in Eq.(\ref{stcp}) via Eq.(\ref{AFN1}) are predicted, to a good approximation, as\,\footnote{See the details in Table-\ref{bmp} and Fig.\,\ref{Fig4}.}
\begin{eqnarray}
M_X\simeq2\times10^{15}\,{\rm GeV}\,,\qquad \quad F_a=2M_X\frac{\Delta_\chi}{|\delta^G_X|}\simeq1.6\times10^{14}\,{\rm GeV}
 \label{axide1}
\end{eqnarray}
with $v_\chi\simeq2\times10^{15}$ GeV, where in $F_a$ we have used Eqs.(\ref{dGi}) and (\ref{delchi}).
At the scale $\mu=M_X$, the matching conditions are
\begin{eqnarray}
&&\frac{1}{g'^2_Y(M^-_X)}=\frac{1}{g^2_Y(M^+_X)}-\tilde{g}^2_X(M_X)\Big(\frac{b_{YX}}{8\pi^2}\ln\Big(\frac{M_X}{\Lambda_{\rm fGUT}}\Big)\Big)^2\,,\nonumber\\
&&\frac{1}{g'^2_P(M^-_X)}=\frac{1}{g^2_P(M^+_X)}-\tilde{g}^2_X(M_X)\Big(\frac{b_{XP}}{8\pi^2}\ln\Big(\frac{M_X}{\Lambda_{\rm fGUT}}\Big)\Big)^2\,,
\label{th1}
\end{eqnarray}
where $M^+_X$ ($M^-_X$) denotes the scale just above (below) the $U(1)_X$ breaking threshold. 

In the running from $M_{X}$ down to $M_{B-L}$, radiative corrections induce kinetic mixing in the $U(1)_Y\times U(1)_{B-L}$ sector:
  \begin{eqnarray}
  G^{-2}(\mu)={\left(\begin{array}{cc}
 1/g'^2_Y(M^-_X)-\frac{b_Y}{8\pi^2}\ln(\mu/M_X)&  -\frac{b_{YP}}{8\pi^2}\ln(\mu/M_X)  \\
 -\frac{b_{YP}}{8\pi^2}\ln(\mu/M_X) &   1/g'^2_P(M^-_X)-\frac{b_P}{8\pi^2}\ln(\mu/M_X) 
 \end{array}\right)}\,,
 \label{gth04p}
  \end{eqnarray}
  where the off-diagonal terms represent the induced mixing.
At the $U(1)_{B-L}$ breaking scale $M_{B-L}$, the mixed kinetic term can be removed by the gauge field redefinition (see Eq.(\ref{ps02}))  
\begin{eqnarray}
A^\mu_P\rightarrow A^\mu_P+A^\mu_Y\,g^{2}_{P}(\mu)\,\frac{b_{YP}}{8\pi^2}\ln\Big(\frac{\mu}{M_X}\Big)\qquad\text{at}~\mu=M_{B-L}\,,
\label{gare}
\end{eqnarray}
which eliminates the kinetic mixing term $F_{Y\mu\nu}F^{\mu\nu}_P$. The matching condition at $\mu=M_{B-L}$ is then given by 
\begin{eqnarray}
\frac{1}{g'^2_Y(M^-_{B-L})}=\frac{1}{g^2_Y(M^+_{B-L})}-g^2_{P}(M_{B-L})\,\Big(\frac{b_{YP}}{8\pi^2}\ln\Big(\frac{M_{B-L}}{M_X}\Big)\Big)^2\,,
\label{th2}
\end{eqnarray}
 where $M^+_{B-L}\,(M^+_{B-L})$ denotes the scale just above (below) the $U(1)_{B-L}$ breaking threshold. 
Eqs.(\ref{th1}) and (\ref{th2}) show that the kinetic mixing modifies the running of $g_Y$, while the threshold corrections at $M_X$ and $M_{B-L}$ further affect the RG evolution. 
Consequently, the scales $M_X$ and $M_{B-L}\sim g_P\langle\varphi_{B-L}\rangle$ with the coupling $g_P(\Lambda_{\rm fGUT})$ in Eq.(\ref{gbm01}) can be constrained by requiring the measured value of the Weinberg angle, $\sin^2\theta_W(M_Z)$. In particular, once $M_X$ (or equivalently $g_{st}$ or $F_a$) is fixed by the measured value of the strong coupling constant $\alpha_3(M_Z)$, $M_{B-L}$ with $g_P(\Lambda_{\rm fGUT})$ exhibits a higher sensitivity to this constraint than $M_X$ does. 
Unlike the $U(1)_X$, the $U(1)_{B-L}$ does not acquire its breaking scale through the GS mechanism.  Instead, the scale $M_{B-L}$ must be determined by the VEV of a scalar field through the scalar potential in Eq.(\ref{super_d}), together with the heavy-neutrino Yukawa sector in Eq.(\ref{lagrangian_l}) responsible for neutrino mass generation. For instance, in the illustrative model presented in Sec.\ref{visu}, viable mass scales with $v_\rho>10^6$ GeV are obtained, depending on the modular weight (see Eq.(\ref{AxionLag2}) above), where $\langle\varphi_{B-L}\rangle=v_\rho/\sqrt{2}$.
\begin{figure}[t]
\begin{minipage}[h]{10.0cm}
\epsfig{figure=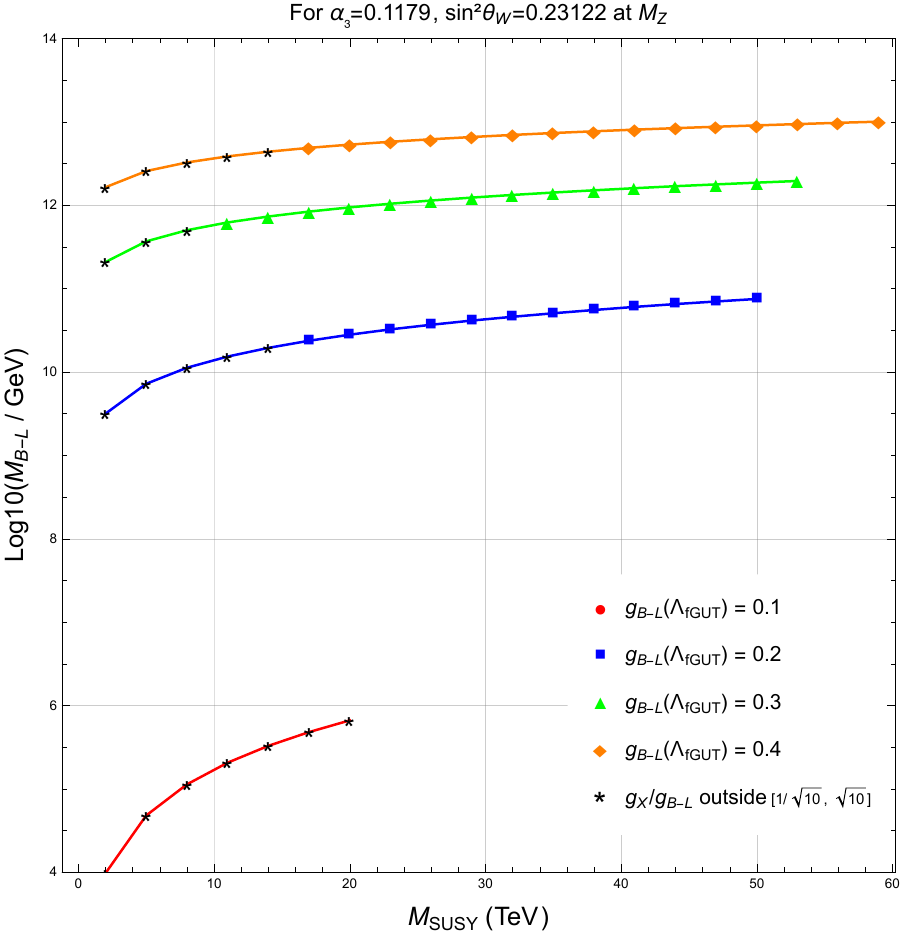,width=10.0cm,angle=0}
\end{minipage}
\caption{\label{Fig3} The $U(1)_{B-L}$ breaking scale $M_{B-L}$ as a function of the SUSY breaking scale $M_{\rm SUSY}$, satisfying the central experimental values of $\sin^2\theta_W(M_Z)$ and $\alpha_3(M_Z)$ in Eq.(\ref{th0301}).
Black circles indicate points where $\frac{g_{B-L}}{g_X}|_{\Lambda_{\rm fGUT}}$ lies outside $[1/\sqrt{10},\sqrt{10}]$. Some representative numerical values are listed in Table-\ref{bmp}.}
\end{figure}

Below the $U(1)_{B-L}$ breaking scale, the gauge couplings evolve according to the particle content of the effective theory. 
Between the $U(1)_{B-L}$ breaking scale and the SUSY breaking scale $M_{\rm SUSY}$, $U(1)_Y$ coupling runs as
  \begin{eqnarray}
  \frac{1}{g^2_Y(\mu)}=\frac{1}{g'^{2}_Y(M_{B-L})}-\frac{b_Y}{8\pi^2}\ln\frac{\mu}{M_{B-L}}
 \label{gth04pp}
  \end{eqnarray}
and similar for $1/g^2_2(\mu)$ and $1/g^2_3(\mu)$, with the corresponding supersymmetric $\beta$-function coefficients. At $\mu=M_{\rm SUSY}$, the matching conditions are $1/g^2_i(M^-_{\rm SUSY})=1/g^2_i(M^+_{\rm SUSY})-\Delta^{\rm SUSY}_i$, where $\Delta^{\rm SUSY}_i$ denotes the one-loop threshold corrections from heavy supersymmetric particles (in the present analysis, these threshold corrections are neglected\,\footnote{Even when non-degenerate superpartner masses are taken into account, the resulting SUSY threshold corrections can be incorporated consistently and do not spoil SM gauge coupling unification within the flavored-GUT framework.}, {\it i.e.} $\Delta^{\rm SUSY}_i=0$, by assuming that all superpartners are almost degenerate at the scale $M_{\rm SUSY}$).  
This implies that the scale $M_{B-L}$ (and through its dependence on the gauge coupling $g_{B-L}$) is correlated with gauge coupling unification and is therefore constrained by the supersymmetry breaking scale $M_{\rm SUSY}$, see Fig.\ref{Fig3}.

Below $M_{\rm SUSY}$, the supersymmetric particles decouple, and the gauge couplings evolve according to the SM particle content down to the electroweak scale $M_Z$.
The corresponding one-loop $\beta$-function coefficients are 
\begin{eqnarray}
(b^{\rm SM}_3,b^{\rm SM}_2,b^{\rm SM}_Y)=\Big(-7,\,-\frac{19}{6},\,c_Y\frac{41}{6}\Big)\,.
 \label{bfc1}
\end{eqnarray}

\begin{table}[htbp]
\centering
\caption{\label{bmp} Benchmark points satisfying the central values of Eq.(\ref{th0301}) from the numerical analysis.}
\begin{tabular}{ccccccc}
\hline
${M_{\rm SUSY} ({\rm TeV})}$ & ${\delta^{\rm GS}_X}$ & ${M_X/10^{15}\,{\rm GeV}}$ & ${M_{B-L}\,({\rm GeV}})$ & ${\alpha^{-1}_{\rm fGUT}}$ & ${g_X}$ & $g_{B-L}$ \\
\hline
5    & $-1.9437\times10^{4}$ & 2.298 & $4.79\times10^{4}$  & 26.798 & 0.0225 & 0.1 \\
14   & $-1.6477\times10^{5}$ & 2.216 & $3.28\times10^{5}$  & 27.453 & 0.00764 & 0.1 \\
20   & $-1.2794\times10^{2}$ & 2.189 & $2.79\times10^{10}$ & 27.680 & 0.273  & 0.2 \\
23   & $-2.5294\times10^{2}$ & 2.178 & $3.24\times10^{10}$ & 27.769 & 0.194  & 0.2 \\
29   & $-3.8341\times10^{2}$ & 2.161 & $4.13\times10^{10}$ & 27.917 & 0.157  & 0.2 \\
41   & $-1.2152\times10^{2}$ & 2.135 & $6.02\times10^{10}$ & 28.137 & 0.278  & 0.2 \\
50   & $-2.8000\times10^{1}$ & 2.121 & $7.50\times10^{10}$ & 28.263 & 0.577  & 0.2 \\
12   & $-1.1338\times10^{2}$ & 2.228 & $3.90\times10^{11}$ & 28.263 & 0.292  & 0.28 \\
11   & $-1.7928\times10^{2}$ & 2.234 & $6.20\times10^{11}$ & 27.300 & 0.232  & 0.3 \\
17   & $-4.9644\times10^{1}$ & 2.201 & $8.36\times10^{11}$ & 27.577 & 0.439  & 0.3 \\
32   & $-1.3168\times10^{1}$ & 2.153 & $1.32\times10^{12}$ & 27.979 & 0.846  & 0.3 \\
47   & $-2.7686\times10^{1}$ & 2.126 & $1.77\times10^{12}$ & 28.224 & 0.581  & 0.3 \\
17   & $-7.1080\times10^{1}$ & 2.201 & $4.86\times10^{12}$ & 27.577 & 0.367  & 0.4 \\
29   & $-1.9404\times10^{1}$ & 2.161 & $6.55\times10^{12}$ & 27.917 & 0.698  & 0.4 \\
41   & $-1.2188\times10^{1}$ & 2.135 & $8.04\times10^{12}$ & 28.137 & 0.877  & 0.4 \\
50   & $-1.4385\times10^{1}$ & 2.121 & $9.04\times10^{12}$ & 28.263 & 0.806  & 0.4 \\
\hline
\end{tabular}
\end{table}
The predictions for the Weinberg angle $\sin^2\theta_W$ and the strong coupling constant $\alpha_3$ at $M_Z$ depend on the $\beta$-function coefficients $b_i$, the string-derived gauge kinetic coefficients $\kappa_i$, $\tilde{\kappa}_i$ (or equivalently the $U(1)_X$-mixed anomaly coefficients $\delta^i_X$ and the GS parameter $\delta^{\rm GS}_X$), the RG logarithms, and the $U(1)_{X}$ and $U(1)_{B-L}$ breaking threshold effects, as well as SUSY breaking threshold effect.
With experimental measurements, the Weinberg angle $\sin^2\theta_W(\mu)=\alpha_Y(\mu)/(\alpha_Y(\mu)+\alpha_2(\mu))$ and the strong coupling constant $\alpha_3(\mu)$ at $M_Z$ scale\cite{PDG}, respectively,
{\begin{eqnarray}
\alpha_3(M_Z)=0.1179\pm0.0009\,,\qquad \sin^2\theta_W(M_Z)=0.23122\pm0.00015\,.
 \label{th0301}
\end{eqnarray}}
In the flavored-GUT framework, all detailed moduli dependence cancels, rendering gauge coupling unification predictive and directly testable against low-energy data Eq.(\ref{th0301}).
Taking $\Lambda_{\rm fGUT}=2\times10^{16}$ GeV, $M_Z=91.1876$ GeV, and Eqs.(\ref{bfc}) and (\ref{bfc1}) (fixed by the quantum anomaly-free conditions in Sec.-\ref{ma01} and -\ref{NGm} with the quantum numbers of Table-\ref{reps_q} and -\ref{reps_l}), the $U(1)_X$ gauge symmetry breaking scale $M_X$ Eq.(\ref{gbm01m}) and the QCD axion decay constant $F_a$ Eq.(\ref{axide1}) are determined, as well as both the scale $M_{\rm SUSY}$ and $M_{B-L}$ can be determined depending on the couplings $g_X$ (equivalently $\delta^{\rm GS}_X$) and $g_{B-L}$ respectively, in a way that the gauge coupling unification is realized with the low-energy data Eq.(\ref{th0301}). 
For instance, see Fig.\ref{Fig3} and a list of data in Table-\ref{bmp}, which indicates $M_{\rm SUSY}\sim{\cal O}(10)$ TeV for $g_{B-L}/g_X|_{\Lambda_{\rm fGUT}}\sim{\cal O}(1)$.

\section{Nonperturbative Dynamics and the Qualified QCD Axion}
\label{modvev}
In the present flavored-GUT framework\,\footnote{We do not introduce an asymptotically-free hidden Yang-Mills sector.}, the same moduli $S$ and $U_X$ appearing in both the gravitational and visible-sector gauge kinetic functions also govern the string-induced nonperturbative dynamics. Consequently, SM gauge-coupling unification, moduli stabilization, supersymmetry breaking, the flavor sector, and the axion sector become correlated through a common holomorphic structure. Since the $U(1)_X$ charged complex structure modulus $U_X$ couples to curvature and gauge fields through terms $\theta_X R\tilde{R}$ and $\theta_X F_i\tilde{F}_i$ in Eq.(\ref{act1}), the string-induced nonperturbative contribution carries gravitational (or GS) charge, see Eq.(\ref{mosu}). 

Motivated by the universal holomorphic structure of string nonperturbative effects, to determine Yukawa and gauge couplings, the supersymmetry-breaking scale, and cosmological constant, while also addressing the axion quality problem, we consider a simple modular- and gauge-invariant superpotential $W(S,U_X,\tau; \varphi_X)$ in terms of the Dedekind $\eta$-function\,\footnote{It is a modular form of weight $1/2$, $\eta(\tau)\rightarrow(c\tau+d)^{1/2}\eta(\tau)$. Here, for simplicity, the Dedekind multiplier $e^{i\epsilon(a,b,c,d)}$ is omitted, where $\epsilon(a,b,c,d)$ is a moduli-independent phase, which can depend on the $SL(2,\mathbb{Z})$ transformation. Under the modular transformation Eq.(\ref{mo1}), in general, the Dedekind eta function $\eta(\tau)$ transforms as $\eta(\tau)\rightarrow e^{i\epsilon}(c\tau+d)^{1/2}\eta(\tau)$ and the superpotential $W$ transforms as $W\rightarrow e^{-g(\tau)-i6\epsilon} W$ where the matter fields $\varphi_i$ transform as: $\varphi_i\rightarrow e^{-i\epsilon_{i}}(c\tau+d)^{-k_i}\varphi_i$ with the condition $\sum_{i}\epsilon_i=\epsilon$, and the dilaton transforms as $S\rightarrow S-\frac{1}{16\pi^2}\{g(\tau)+i6\epsilon\}$.} and nonperturbative effects:
\begin{eqnarray}
W(S,U_X,\tau;\varphi_X)= \frac{M^3_P}{[\eta(\tau)]^{6}}\Big\{C_0+A(\varphi_X)\frac{e^{-\alpha f_R}}{[\eta(\tau)]^{3\alpha \kappa_R/8\pi^2}}-B (\varphi_X)\frac{e^{-\beta f_R}}{[\eta(\tau)]^{3\beta \kappa_R/8\pi^2}}\Big\}\,,
\label{mosu}
\end{eqnarray}
with $f_R\equiv\kappa_R S+\tilde{\kappa}_R U_X= f_{\rm grav}(\Phi)$ in Eq.(\ref{kifct1}), where $C_0$ is, in general, a complex constant representing a flux- or vacuum-induced contribution, whose value is determined by the underlying compactification and, after SUSY breaking, may be adjusted to reproduce the observed vacuum energy. The parameters $\alpha$ and $\beta$ are real coefficients characterizing the underlying string-induced nonperturbative dynamics. The modular invariance of the generalized K{\"a}hler potential requires the superpotential to transform as $W\rightarrow We^{-g(\tau)}$ with Eq.(\ref{mt0}) under the modular transformation Eq.(\ref{mo1}).
The holomorphic function $A(\varphi_X)$ and $B(\varphi_X)$ of $\varphi_X=\{\chi,\tilde{\chi}\}$ are determined by the modular weights and $U(1)_X$ charges of the flavon fields and transform as
\begin{eqnarray}
A(\varphi_X)\rightarrow A(\varphi_X)\,e^{i\frac{\alpha\,\tilde{\kappa}_R}{16\pi^2}\delta^{\rm GS}_X\Lambda_X}\,,\qquad B(\varphi_X)\rightarrow B(\varphi_X)\,e^{i\frac{\beta\,\tilde{\kappa}_R}{16\pi^2}\delta^{\rm GS}_X\Lambda_X}\,.
\label{u1t}
\end{eqnarray}
The $A(\varphi_X)$ and $B(\varphi_X)$ can be parametrized as 
\begin{eqnarray}
A(\varphi_X)&=&A_0\Big(\frac{\chi}{M_X}\Big)^{N_\alpha}\Big(1+c_1\frac{\chi\tilde{\chi}}{M^2_X}+c_2\Big(\frac{\chi\tilde{\chi}}{M^2_X}\Big)^2+...\Big)\,,\nonumber\\
B(\varphi_X)&=&B_0\Big(\frac{\chi}{M_X}\Big)^{N_\beta}\Big(1+d_1\frac{\chi\tilde{\chi}}{M^2_X}+d_2\Big(\frac{\chi\tilde{\chi}}{M^2_X}\Big)^2+...\Big)\,,
\label{u1t01}
\end{eqnarray}
where $N_i>0$ ($i=\alpha,\beta$), and the coefficients $A_0,B_0,c_i, d_i$ are generally complex numbers of unit magnitude. Here, for $N_i<0$, the corresponding powers of $\chi$ are replaced by powers of $\tilde{\chi}$, according to the $U(1)_X$ charge assignments. Recall that $M_X$ denotes the flavor-dynamics scale associated with the $U(1)_X$-breaking sector, see Eq.(\ref{gbm01m}). 
Under the $U(1)_X$ gauge transformations (see Eq.(\ref{ss01}) and the text above it), gauge invariance of the nonperturbative superpotential Eq.(\ref{mosu}) requires  
\begin{eqnarray}
N_\alpha=\frac{\alpha\tilde{\kappa}^2_R\,\delta^{\rm GS}_X}{16\pi^2}\in\mathbb{Z}\,,\qquad N_\beta=\frac{\beta\tilde{\kappa}^2_R\,\delta^{\rm GS}_X}{16\pi^2}\in\mathbb{Z}\,,
\label{u1t001}
\end{eqnarray}
indicating that the $U(1)_X$ gauge invariance quantizes the allowed combinations $N_\alpha$ and $N_{\beta}$, which we refer to as the quantized nonperturbative indices.
Using the GS anomaly-cancellation condition Eq.(\ref{gt02}), these can equivalently be written as $N_\alpha=\alpha(\delta^R_X)^2/(16\pi^2\,\delta^{\rm GS}_X)$ and $N_\beta=\beta(\delta^R_X)^2/(16\pi^2\,\delta^{\rm GS}_X)$.
$N_i$ characterizes discrete nonperturbative sectors in a manner analogous to the instanton number in Yang-Mills theory. In the present effective description, an identification of $N_i$ with a topological charge would require an explicit realization of the underlying stringy or quantum-gravitational nonperturbative configuration. Rather than identifying $N_i$ with a topological charge, we will show that $N_i$ is closely related to the resolution of axion quality problem in the next subsection. 
After $U(1)_X$ breaking, Eq.(\ref{u1t01}) can be effectively written, when Eq.(\ref{bac00b}) is satisfied, as
\begin{eqnarray}
A(\langle\varphi_X\rangle)=A_0\,\Delta^{N_\alpha}_\chi\,e^{i\frac{A_X}{f_A}N_\alpha}\Big|_{h_\chi=0=h_{\tilde{\chi}}}\,, \qquad B(\langle\varphi_X\rangle)=B_0\,\Delta^{N_\beta}_\chi\,e^{i\frac{A_X}{f_A}N_\beta}\Big|_{h_\chi=0=h_{\tilde{\chi}}}\,,
\label{u1t002}
\end{eqnarray}
with $\Delta_\chi=\langle\chi\rangle/M_X$ or $\Delta_\chi=\langle\tilde{\chi}\rangle/M_X$, see Eq.(\ref{AFN1}), where Eq.(\ref{kf1}) (or see Eq.(\ref{NGboson})) is used, and then state directly
\begin{eqnarray}
1-\frac{\Delta^2_\chi}{1-\Delta^2_\chi}\leq|A_0|,|B_0|\leq 1+\frac{\Delta^2_\chi}{1-\Delta^2_\chi}\,.
\label{u1t003}
\end{eqnarray}

After the moduli $S$ and $U_X$ are stabilized by the nonperturbative superpotential, the gauge couplings at the flavored-GUT scale $\Lambda_{\rm fGUT}=M_{st}$ are determined by Eq.(\ref{gac00}). Together with the gauge-coupling unification condition Eq.(\ref{th0003}) at $\Lambda_{\rm fGUT}$, these relations determine the gauge coupling parameters subject to the anomaly and unification constraints. The coefficient $\tilde{\kappa}_R$ is correlated with the coefficients $\tilde{\kappa}_i$ through the GS anomaly-cancellation condition Eq.(\ref{gt02}). In contrast, $\kappa_R$ is not fixed by either the gauge coupling unification or the GS anomaly-cancellation conditions. It therefore remains a dynamical, model-dependent parameter associated with the nonperturbative sector in Eq.(\ref{mosu}). In particular, the QCD gauge coupling is determined by $f_G=\kappa_G S+\tilde{\kappa}_G U_X$ via Eq.(\ref{gac00}). Subsequent RG running determines the QCD dynamical scale $\Lambda_{\rm QCD}$, below which QCD becomes strongly coupled and nonperturbative effects generate the QCD axion potential through QCD instantons. At the same time, because the gravitational kinetic function $f_R=\kappa_R S+\tilde{\kappa}_R U_X$ contains the axionic component of $U_X$, string-induced nonperturbative effects explicitly break the perturbative PQ shift symmetry and generate an additional periodic axion potential. The physical axion potential therefore receives contributions from both QCD instantons and string-induced quantum-gravitational effects.
The validity of the flavored-GUT framework therefore requires the string-induced contribution to be sufficiently suppressed, or appropriately aligned in field space, so that the QCD instanton contribution remains dominant and the strong CP problem is resolved. We will show that, within the flavored-GUT framework, the string-induced gravitational potential can be naturally suppressed by  the anomaly cancellation and SM gauge coupling unification, which acts as fundamental consistency conditions.

For the matter sector, we introduce a minimal set of superfields responsible for $U(1)_X$ and $U(1)_{B-L}$ breaking. These fields also determine the vacuum values of the  $A(\varphi_X)$ and $B(\varphi_X)$ appearing in the nonperturbative superpotential Eq.(\ref{mosu}). Their stabilization must therefore be analyzed simultaneously with the moduli sector.
 The field content consists of a gauge-singlet field\,\footnote{The field $\chi_0$ may also play the role of the inflaton\,\cite{Ahn:2017dpf}.} $\chi_0$ with modular weight $3$, a gauge-singlet field $\rho_0$ with modular weight $7/2$, a pair of field $\varphi_X=\{\chi, \tilde{\chi}\}$ responsible for $U(1)_X$ breaking, a pair $\varphi_{B-L}=\{\rho, \tilde{\rho}\}$ responsible for $U(1)_{B-L}$ breaking, and two Higgs doublets $H_{u(d)}$ that trigger electroweak symmetry breaking. All symmetry-breaking fields are taken to have vanishing modular weight. The fields $\chi$ and $\tilde{\chi}$ carry $U(1)_X$ charges $+1$ and $-1$, respectively, while $\rho$ and $\tilde{\rho}$ carry $U(1)_{B-L}$ charges $-2$ and $+2$. These charge assignments are enforced by the extended $U(1)$ gauge symmetries and are compatible with holomorphy of the superpotential. The fields $\chi_0$ and $\rho_0$, with modular weight $3$ and $7/2$, respectively, play a special role: their modular weights compensate the modular weights required for modular invariance of the corresponding superpotential terms, allowing the modular forms $Y(\tau)$ appearing in these terms to be treated as  $\tau$-independent constants\,\cite{Ahn:2023iqa}. The leading-order superpotential invariant under $SL(2,\mathbb{Z})\times U(1)_X\times U(1)_{B-L}$ is given by
\begin{eqnarray}
 W_v&=&g_{\chi_0}\chi_0\,H_uH_d+\chi_0(g_\chi\,\chi\tilde{\chi}-\mu^2_\chi)+\rho_0(g_\rho\,\rho\tilde{\rho}-\mu^2_\rho)\,\eta(\tau)\,\,,
  \label{super_d}
\end{eqnarray}
where the coefficients $g_{\chi_0}$, $g_{\chi}$, $g_{\rho}$ are taken to be complex-unity (after K{\"a}hler-rescaling $W\rightarrow e^{K/2M^2_P}W$, see Eq.(\ref{cn0})), but receive corrections from higher-dimensional operators (see Eq.(\ref{AFN1})).
The parameters $\mu_\chi$ and $\mu_\rho$ set the scales of spontaneous $U(1)_X$ and $U(1)_{B-L}$ breaking, respectively.

\subsection{Scalar Potential, Vacuum Stabilization, and Supersymmetry Breaking}
\label{modvev1}
First, we set the SM matter fields $\{q^c,\ell, H_u,...\}$ to be zero and focus on the moduli and symmetry-breaking sectors. Assuming an approximately vanishing cosmological constant, the scalar potential satisfies $V=|F|^2-3m^2_{3/2}M^2_P+\frac{1}{2}D^2\approx0$. In the supersymmetric Minkowski limit considered below, the vacuum satisfies $W=0$, $D_IW=0$, $D_a=0$, so that the gravitino mass vanishes, $m_{3/2}\rightarrow0$, and both the $F$- and $D$-term contributions vanish at the vacuum. Small departures from this supersymmetric Minkowski vacuum, generated for example by nonvanishing\,\footnote{In the limit $\kappa_R\rightarrow0$, the superpotential (\ref{mosu}) reduces to the Kallosh-Linde (KL)-type racetrack superpotential\,\cite{Kachru:2003aw}.} $\kappa_R$, uplifting effects, or additional superpotential contributions, induce supersymmetry breaking, implying that the $F$-terms scale proportionally to $m_{3/2}$ near the minimum when the $D$-terms vanish.
The $F$-term scalar potential is $V_F=e^{K/M^2_P}\big\{K^{I\bar{J}}D_IW\bar{D}_{\bar{J}}\bar{W}-\frac{3}{M^2_P}|W|^2\big\}$ with $W=W(S,U_X,\tau; \varphi_X)+W_v$,
where $I,J$ run over the moduli and symmetry-breaking fields, $\{U_X, \tau, S; \chi_0, \chi(\tilde{\chi}); \rho_0, \rho(\tilde{\rho})\}$. 
For the moduli sector, the $F$-term equations are expressed as
\begin{eqnarray}
&&D_{U_X} W=\frac{M^3_P}{[\eta(\tau)]^{6}}\tilde{\kappa}_R\,\Omega-\frac{3W}{U_X+\bar{U}_X}\,,\nonumber\\
&&D_\tau W=-3W\Big[\frac{1-1/(16\pi^2y)}{\tau-\bar{\tau}}+2\frac{\eta'(\tau)}{\eta(\tau)}\Big]+\kappa_R\frac{3}{8\pi^2}\frac{M^3_P}{[\eta(\tau)]^{6}}\frac{\eta'(\tau)}{\eta(\tau)}\,\Omega\nonumber\\
&&\qquad\quad\quad-\rho_0(g_\rho\rho\tilde{\rho}-\mu^2_\rho)\eta'(\tau)-\frac{W}{M^2_P}\frac{k_i}{\tau-\bar{\tau}}(-i\tau+i\bar{\tau})^{-k_i}|\varphi_i|^2\,,\nonumber\\
&&D_S W=-\frac{W}{y}+\kappa_R\,\Omega\frac{M^3_P}{[\eta(\tau)]^{6}}\,,  
\label{susy_1}
\end{eqnarray}
where, with $\tilde{x}=\frac{3}{8\pi^2}\ln\eta(\tau)$,
\begin{eqnarray}
y&=&S+\bar{S}-\frac{3}{16\pi^2}\ln(-i\tau+i\bar{\tau})\,,\nonumber\\
\Omega&=&-\alpha\,A(\varphi_X)\,e^{-\alpha\,\tilde{\kappa}_RU_X}\,e^{-\alpha\kappa_R(S+\tilde{x})}+\beta\,B(\varphi_X)\,e^{-\beta\,\tilde{\kappa}_RU_X}\,e^{-\beta\kappa_R(S+\tilde{x})}\,.
\label{susy_a1}
\end{eqnarray}
For the matter fields, the corresponding the $F$-term equations are
\begin{eqnarray}
&&D_{\chi_0}W=g_\chi \chi\tilde{\chi}-\mu^2_\chi+\frac{W}{M^2_P}(-i\tau+i\bar{\tau})^{-3}\bar{\chi}_0\,\nonumber\\
&&D_\chi W=g_\chi \chi_0\tilde{\chi}+\frac{M^3_P}{[\eta(\tau)]^6}\Big(\frac{\partial A}{\partial \chi}\frac{e^{-\alpha\,\tilde{\kappa}_R U_X}}{e^{\alpha\kappa_R(S+\tilde{x})}}-\frac{\partial B}{\partial \chi}\frac{e^{-\beta\,\tilde{\kappa}_R U_X}}{e^{\beta\kappa_R(S+\tilde{x})}}\Big)+\frac{W}{M^2_P}\bar{\chi}\,,\nonumber\\
&&D_{\tilde{\chi}}W=g_\chi \chi_0\chi+\frac{M^3_P}{[\eta(\tau)]^6}\Big(\frac{\partial A}{\partial\tilde{\chi}}\frac{e^{-\alpha\,\tilde{\kappa}_R U_X}}{e^{\alpha\kappa_R(S+\tilde{x})}}-\frac{\partial B}{\partial\tilde{\chi}}\frac{e^{-\beta\,\tilde{\kappa}_R U_X}}{e^{\beta\kappa_R(S+\tilde{x})}}\Big)+\frac{W}{M^2_P}\bar{\tilde{\chi}}\,,\nonumber\\
&&D_{\rho_0}W=(g_\rho\,\rho\tilde{\rho}-\mu^2_\rho)\eta(\tau)+\frac{W}{M^2_P}(-i\tau+i\bar{\tau})^{-\frac{7}{2}}\bar{\rho}_0\,\nonumber\\
&&D_\rho W=g_\rho \chi_0\tilde{\rho}+\frac{W}{M^2_P}\tilde{\rho}\,,\qquad\qquad D_{\tilde{\rho}}W=g_\rho \chi_0\rho+\frac{W}{M^2_P}\rho\,.
\label{super_01}
\end{eqnarray}

Temporarily neglecting the exponential terms as well as small supergravity mixings, from the above Eqs.(\ref{susy_1}) and (\ref{super_01}), at leading order we obtain
\begin{eqnarray}
\langle\chi_0\rangle=0\,,\qquad \langle\chi\rangle=\langle\tilde{\chi}\rangle=\frac{v_\chi}{\sqrt{2}}\,,\qquad \langle\rho\rangle=\langle\tilde{\rho}\rangle=\frac{v_\rho}{\sqrt{2}}\,,
  \label{super_02}
\end{eqnarray}
with $\mu_\chi=v_\chi\sqrt{g_\chi/2}$ and $\mu_\rho=v_\rho\sqrt{g_\rho/2}$. 
 In this approximation, the F-flatness conditions reduce to $g_\chi\langle\chi\tilde{\chi}\rangle\simeq\mu^2_\chi$, $g_\rho\langle\rho\tilde{\rho}\rangle\simeq\mu^2_\rho$, $\langle\chi_0\rangle\simeq0$, while D-flatness (from Eq.(\ref{DV01}) and below Eq.(\ref{DV01})) implies $|\langle\chi\rangle|=|\langle\tilde{\chi}\rangle|$ (see below Eq.(\ref{Ux_pot})) and $|\langle\rho\rangle|=|\langle\tilde{\rho}\rangle|$. Expanding the $\chi$ and $\tilde{\chi}$ fields around the vacuum\,\cite{Ahn:2014gva, Ahn:2016hbn, Ahn:2018cau} in Eq.(\ref{super_02})
 \begin{eqnarray}
\chi=\frac{v_{\chi}}{\sqrt{2}}e^{i\frac{A_X}{f_{A}}}\left(1+\frac{h_{\chi}}{f_{A}}\right)\,,\quad\,\tilde{\chi}=\frac{v_{\tilde{\chi}}}{\sqrt{2}}e^{-i\frac{A_X}{f_{A}}}\left(1+\frac{h_{\tilde{\chi}}}{f_{A}}\right)\qquad\text{with}~f_{A}=\sqrt{v^2_{\chi}+v^2_{\tilde{\chi}}}\,,
  \label{NGboson}
 \end{eqnarray}
with $v_\chi=v_{\tilde{\chi}}$ and $h_{\chi}=h_{\tilde{\chi}}$ in supersymmetric limit, the radial and perpendicular modes are given by $h_+=(h_\chi+h_{\tilde{\chi}})/2\sqrt{2}$ and $h_-=(h_\chi-h_{\tilde{\chi}})/2\sqrt{2}$, respectively.
After the canonical normalization of Eq.(\ref{cn0}), the F-term potential for $\chi$, $\tilde{\chi}$, and $\chi_0$ is then given by $V_F\supset(2 {\rm Im}\,\tau_0)^3|g_\chi\chi\tilde{\chi}-\mu^2_\chi|^2+g^2_\chi|\chi_0|^2(|\chi|^2+|\tilde{\chi}|^2)$, where the K{\"a}hler-rescaling $W\rightarrow e^{K/2M^2_P}W$ has been used. And the D-term potential from Eq.(\ref{DV01}) is given by  $V_D=\frac{1}{2}(-\xi^{\rm FI}_{X}+|\chi|^2-|\tilde{\chi}|^2)^2\,g^2_X/\tilde{\kappa}^2_X$. The leading scalar masses are given by
\begin{eqnarray}
m^2_\chi\simeq16g_\chi\,\mu^2_\chi\big(1+\frac{\tilde{g}^2_X}{4g^2_\chi}\big)\,,\qquad\qquad m^2_{\chi_0}\simeq2g_\chi\,\mu^2_\chi\,,
  \label{chma}
\end{eqnarray}
where the second term in parentheses can be neglected for  $|g_\chi|\gg\tilde{g}_X|$.

Temporarily, for the moduli sector, in the limit $\kappa_R\rightarrow0$ with $A(\varphi_X)$ and $B(\varphi_X)$ replaced by the constants $|A_0|\Delta^{N_\alpha}_\chi$ and  $|B_0|\Delta^{N_\beta}_\chi$ (see Eq.(\ref{u1t002}), after integrating out the heavy fields $\varphi_X$), the scalar potential for the moduli $\sigma, \tau$ has a local minimum at $\sigma_0$, $\tau_0$. The vacuum is supersymmetric and Minkowski, {\it i.e.}, 
\begin{eqnarray}
W(\sigma_0,\tau_0)=0\,,\qquad D_I W(\sigma_0,\tau_0)=0\,,\qquad  V(\sigma_0,\tau_0)=0\,.
  \label{susy_0}
\end{eqnarray}
Note that, in this limit, the F-term equation for $\tau$ in Eq.(\ref{susy_1}) is automatically satisfied at the exact Minkowski point and therefore does not, by itself\,\footnote{In this limit $\tau_0$ may lie within the fundamental domain of the modular group, see Eq.(\ref{do_0}).}, determine $\tau$.
The parameters $|C_0|$ and $\sigma_0$ are determined by the conditions $V^{(0)}_F(\sigma_0)=0$ and $\partial V^{(0)}_F/\partial\sigma|_{\sigma=\sigma_0}=0$ : in terms of the quantized nonperturbative indices $N_{i}$ of Eq.(\ref{u1t001}) with Eq.(\ref{gt02}),
\begin{eqnarray}
&&\sigma_0=\frac{\delta^R_X}{16\pi^2(N_\alpha-N_\beta)}\ln\Big(\frac{N_\alpha|A_0|}{N_\beta|B_0|}\Delta^{N_\alpha-N_\beta}_\chi\Big)\equiv\sigma^{\rm comp}_0\,,\nonumber\\
&&|C_0|=-|A_0|\Big(\frac{N_\alpha|A_0|}{N_\beta|B_0|}\Big)^{-\frac{N_\alpha}{N_\alpha-N_\beta}}+|B_0|\Big(\frac{N_\alpha|A_0|}{N_\beta|B_0|}\Big)^{-\frac{N_\beta}{N_\alpha-N_\beta}}\,,
\label{d0}
\end{eqnarray}
where $\sigma^{\rm comp}_0$ denotes the modulus value determined during string compactification.
Using Eqs.(\ref{gt02}) and (\ref{u1t001}), we define the nonperturbative actions to be positive, without loss of generality, as
\begin{eqnarray}
S_i=\frac{16\pi^2}{\delta^R_X}N_i\,\sigma\,,\quad (i=\alpha,\beta)\,.
\label{d0ac}
\end{eqnarray}

As shown in Eq.(\ref{susy_0}), in the limit $\kappa_R\rightarrow0$ ({\it i.e.} $V_F\rightarrow V^{(0)}_F$), the potential for $U_X$ is given by $V^{(0)}_F=V^{(0)}_F(\sigma)+V^{(0)}_F(\theta_X)$. After the K{\"a}hler-rescaling $W\rightarrow e^{K/2M^2_P}W$, the coefficients $A_0, B_0, C_0$ are correspondingly rescaled, the scalar potential $V^{(0)}_F$ can be written as
\begin{eqnarray}
V^{(0)}_F(\sigma)&=&\frac{M^4_P}{|\eta(\tau)|^{12}}\frac{4}{3}\Big\{\big(|A_0|\Delta^{N_\alpha}_\chi\,S_\alpha\,e^{-S_\alpha}-|B_0|\Delta^{N_\beta}_\chi\,S_\beta\,e^{-S_\beta}\big)^2\nonumber\\
&+&3\big(|C_0|+|A_0|\Delta^{N_\alpha}_\chi\,e^{-S_\alpha}-|B_0|\Delta^{N_\beta}_\chi\,e^{-S_\beta}\big)\big(|A_0|\Delta^{N_\alpha}_\chi\,S_\alpha\,e^{-S_\alpha}-|B_0|\Delta^{N_\beta}_\chi\,S_\beta\,e^{-S_\beta}\big)\Big\}\nonumber\\
V^{(0)}_F(\theta_X)&=&\frac{M^4_P}{|\eta(\tau)|^{12}}\frac{8}{3}\Big\{|A_0B_0|\Delta^{N_\beta+N_\beta}_\chi\,e^{-S_\alpha-S_\beta}\big(2S_\alpha S_\beta+3S_\alpha+3S_\beta\big)\sin^2\frac{\xi_\alpha-\xi_\beta}{2}\nonumber\\
&-&3|C_0|\Big(|A_0|\Delta^{N_\alpha}_\chi\,S_\alpha\,e^{-S_\alpha}\sin^2\frac{\xi_\alpha}{2}-|B_0|\Delta^{N_\beta}_\chi\,S_\beta\,e^{-S_\beta}\sin^2\frac{\xi_\beta}{2}\Big)\Big\}\,,
  \label{sup_01}
\end{eqnarray}
where $A_0=|A_0|e^{i\varphi_\alpha}$, $B_0=|B_0|e^{i\varphi_\beta}$, and $C_0=|C_0|e^{i\varphi_c}$ have been used, and 
\begin{eqnarray}
\xi_i=N_i\frac{A_X}{f_A}-\frac{16\pi^2\,N_i}{\delta^R_X}\theta_X+\varphi_{ic}\,,\qquad(i=\alpha,\beta)\quad\text{with}~\varphi_{ic}=\varphi_i-\varphi_c\,.
  \label{sup_00a1}
\end{eqnarray}
In the supersymmetric Minkowski limit, the gravitino mass vanishes, while the mass squared of the field $\sigma$ at the local minimum is given by $m^2_\sigma=\frac{1}{2}K^{U_X\bar{U}_X}\partial^2_{\sigma}V_F|_{\theta_X=0, \sigma^{\rm comp}_0,\tau_0}$:
\begin{eqnarray}
m^2_\sigma=\frac{8}{9}\frac{M^2_P}{|\eta(\tau_0)|^{12}}\Big|(S^{\rm comp}_\alpha)^2|A_0|\Big(\frac{N_\alpha}{N_\beta}\frac{|A_0|}{|B_0|}\Big)^{-\frac{N_\alpha}{N_\alpha-N_\beta}}-(S^{\rm comp}_\beta)^2|B_0|\Big(\frac{N_\alpha}{N_\beta}\frac{|A_0|}{|B_0|}\Big)^{-\frac{N_\beta}{N_\alpha-N_\beta}}\Big|^2\,.
  \label{sig_01}
\end{eqnarray}
where we chose $\tau_0\approx i$, see Eq.(\ref{Dta_01}), and $|\eta(i)|\approx0.768$, and $S^{\rm comp}_i=(16\pi^2/\delta^R_X)N_i\sigma^{\rm comp}_0$.

Next, we examine the backreaction of $\varphi_X$ on $U_X$ since the dependence of $A(\varphi_X)$ and $B(\varphi_X)$ on $\varphi_X=\{\chi,\tilde{\chi}\}$ induces mixing between the $\varphi_X$ sector and the modulus $U_X$. We find that the corrections generated by integrating out the heavy $\varphi_X$ fields are parametrically suppressed. Consequently, one may consistently integrate out $\varphi_X$ at its leading-order vacuum Eq.(\ref{super_02}) and use Eq.(\ref{u1t002}) effectively.
In the limit $\kappa_R\rightarrow0$, from Eqs.(\ref{susy_1}) and (\ref{super_01}) the relevant f-term potential for $\chi, \tilde{\chi}$, $\chi_0$, and $U_X$, after the canonical normalization and the K{\"a}hler-rescaling, see Eq.(\ref{cn0}), takes form
\begin{eqnarray}
V_F&\supset&\Big|(-i\tau+i\bar{\tau})^{\frac{3}{2}}(g_\chi\chi\tilde{\chi}-\mu^2_\chi)+\frac{W}{M^2_P}\bar{\chi}_0\Big|^2\nonumber\\
&+&\Big|(-i\tau+i\bar{\tau})^{\frac{3}{2}}g_\chi\chi_0\tilde{\chi}+\frac{M^3_P}{[\eta(\tau)]^{6}}\Big(\frac{\partial A}{\partial\chi}e^{-\alpha\tilde{\kappa}_RU_X}-\frac{\partial B}{\partial\chi}e^{-\beta\tilde{\kappa}_RU_X}\Big)+\frac{W}{M^2_P}\bar{\chi}\Big|^2\nonumber\\
&+&\Big|(-i\tau+i\bar{\tau})^{\frac{3}{2}}g_\chi\chi_0\chi+\frac{M^3_P}{[\eta(\tau)]^{6}}\Big(\frac{\partial A}{\partial\tilde{\chi}}e^{-\alpha\tilde{\kappa}_RU_X}-\frac{\partial B}{\partial\tilde{\chi}}e^{-\beta\tilde{\kappa}_RU_X}\Big)+\frac{W}{M^2_P}\bar{\tilde{\chi}}\Big|^2\nonumber\\
&+&\frac{(U_X+\bar{U}_X)^2}{3|\eta(\tau)|^{12}}M^4_P\big|-\alpha\tilde{\kappa}_RA\,e^{-\alpha\tilde{\kappa}_RU_X}+\beta\tilde{\kappa}_RB\,e^{-\beta\tilde{\kappa}_RU_X}\big|^2\nonumber\\
&-&(U_X+\bar{U}_X)\Big\{\frac{M^3_P}{[\eta(\tau)]^{6}}(-\alpha\tilde{\kappa}_RA\,e^{-\alpha\tilde{\kappa}_RU_X}+\beta\tilde{\kappa}_RB\,e^{-\beta\tilde{\kappa}_RU_X})\bar{W}+\text{h.c.}\Big\}
  \label{Fp_01}
\end{eqnarray}
Expanding around the vacuum Eqs.(\ref{super_02}) and (\ref{susy_0}), $\chi=\langle\chi\rangle+\delta\chi$ and $U_X=\langle\sigma_0\rangle+\delta U_X$, the linearized equation of motion $\partial V/\partial\chi=0$ gives $\delta\chi=-\frac{1}{m^2_\chi}\frac{\partial^2V}{\partial\chi\partial U_X}\delta U_X$. Using Eq.(\ref{sig_01}), this induced shift $\delta\chi$ generated by fluctuations of $U_X$ around the racetrack minimum Eq.(\ref{d0}) can be obtained as
\begin{eqnarray}
\frac{\delta\chi}{\langle\chi\rangle}\simeq2\frac{m^2_{\sigma}}{m^2_\chi}\frac{M_P\{\alpha\tilde{\kappa}_R(\partial_\chi A)e^{-\alpha\tilde{\kappa}_R\sigma_0}-\beta\tilde{\kappa}_R(\partial_\chi B)e^{-\beta\tilde{\kappa}_R\sigma_0}\}}{\alpha^2\tilde{\kappa}^2_RA\,e^{-\alpha\tilde{\kappa}_R\sigma_0}-\beta^2\tilde{\kappa}^2_RB\,e^{-\beta\tilde{\kappa}_R\sigma_0}}\Big|_{\langle\chi\rangle,\langle\tilde{\chi}\rangle}\Big(\frac{M_P}{\langle\chi\rangle}\Big)\delta U_X\,.
  \label{bac0}
\end{eqnarray}
Using $\partial A/\partial\chi\big|_{\langle\chi\rangle,\langle\tilde{\chi}\rangle}\simeq (AN_\alpha/\chi)\big|_{\langle\chi\rangle,\langle\tilde{\chi}\rangle}$ and $\partial B/\partial\chi\big|_{\langle\chi\rangle,\langle\tilde{\chi}\rangle}\simeq (BN_\beta/\chi)\big|_{\langle\chi\rangle,\langle\tilde{\chi}\rangle}$ from Eq.(\ref{u1t01}), together with Eqs.(\ref{gt02}) and (\ref{u1t001}), the shift is simplified to 
\begin{eqnarray}
\frac{\delta\chi}{\langle\chi\rangle}\simeq2\frac{m^2_{\sigma}}{m^2_\chi}\Big(\frac{\delta^R_X}{16\pi^2}\Big)^2\Big(\frac{M_P}{\langle\chi\rangle}\Big)^2\delta U_X\,,
  \label{bac0a}
\end{eqnarray}
indicating that the deviation of $\chi$-fields from its leading vacuum Eq.(\ref{super_02}) is controlled by the ratio of the modulus mass to the $\varphi_X$-sector mass, as well as by the low-energy axion decay constant $f_A=2\langle\chi\rangle$ and the mixed gravitational-$U(1)_X$ anomaly coefficient, which is determined by the flavor sector. Therefore, Eq.(\ref{u1t002}) is justified only when 
\begin{eqnarray}
\delta R\equiv2\frac{m^2_{\sigma}}{m^2_\chi}\Big(\frac{\delta^R_X}{8\pi^2}\Big)^2\Big(\frac{M_P}{f_A}\Big)^2\ll1\,,
  \label{bac00b}
\end{eqnarray}
is satisfied.

As in the KL model\,\cite{Kachru:2003aw}, in the limit $\kappa_R\rightarrow0$ there exists a deeper supersymmetric Anti-de Sitter (AdS) vacuum along the $U_X$ direction. When a small weak-scale perturbation $\Delta W$ is introduced along $U_X$, the potential minimum shifts from zero to a slightly negative value $V_{\rm AdS}<0$ at ($\tilde{\sigma}_0,\tau_0,s_0$), with $\tilde{\sigma}_0=\sigma_0+\delta\sigma$. 
At this shifted minimum, supersymmetry is preserved, that is $D_{U_X}W(\sigma_0+\delta\sigma,\tau_0,s_0)=0$. This implies $W_{U_X}(\sigma_0,\tau_0)=0$, and the minimum shifts by $\delta\sigma\simeq3\Delta W/(2\sigma_0 W_{U_XU_X}(\sigma_0,\tau_0))$. 
The potential at this minimum, expressed in terms of $W(\sigma_0+\delta\sigma)=\Delta W+{\cal O}(\Delta W)^2$, becomes $V_{\rm AdS}(\Delta W)=-e^{\langle K/M^2_P\rangle}\frac{3}{M^2_P}\big|\langle W\rangle_{\rm AdS}\big|^2$, where $\Delta W=\langle W\rangle_{\rm AdS}$ is the value of the superpotential at the AdS minimum. In terms of the K{\"a}hler-rescaled superpotenial by $W\rightarrow e^{K/2M^2_P}W$, the AdS vacuum energy takes
\begin{eqnarray}
V_{\rm AdS}(\Delta W)=-\frac{3}{M^2_P}\big|\langle W\rangle_{\rm AdS}\big|^2\,.
  \label{Ux_pot}
\end{eqnarray}
The $U(1)_X$ D-flatness condition is $|\chi|^2-|\tilde{\chi}|^2-\xi^{\rm FI}_X(U_X)=0$ from Eq.(\ref{DV01}), where $X|\varphi_X|^2=|\chi|^2-|\tilde{\chi}|^2$. Thus, a non-zero  FI term generally requires a splitting between the VEVs of $\chi$ and $\tilde{\chi}$. In particular, $\langle\chi\rangle=\langle\tilde{\chi}\rangle$ in Eq.(\ref{super_02}) is compatible with D-flatness only when the FI term contribution ($\xi^{\rm FI}_X$) is sufficiently small\,\cite{Burgess:2003ic}. If a small residual FI term remains at the supersymmetric vacuum, the $U(1)_X$ D-term generates a positive contribution to the scalar potential, $\Delta V=\frac{1}{2}(\xi^{\rm FI}_X)^2\tilde{g}^2_X$, which can be treated as a small uplifting perturbation of the supersymmetric AdS vacuum. 
 Thus uplift shifts the minimum slightly along the $\sigma={\rm Re}[U_X]$ and simultaneously breaks supersymmetry, converting the AdS minimum into a de Sitter (dS) minimum\,\cite{Kachru:2003sx,Burgess:2003ic}.
Solving the minimization condition $\partial_\sigma(V+\Delta V)|_{\sigma_0+\delta\sigma}=0$ for the uplifted potential yields a small displacement $\Delta\sigma=\sigma-\sigma_0$, see below Eq.(\ref{DV01}), which is proportional to the AdS vacuum energy and the inversely proportional to the modulus mass squared.
 For the case of a D-term induced uplift parametrized as $\Delta V=\frac{1}{2}(\xi^{\rm FI}_X)^2\tilde{g}^2_X\approx|V_{\rm AdS}|(\sigma_0/\sigma)^3$\,\cite{Burgess:2003ic}, and using the approximation $(D_\sigma W)_\sigma\approx W_{\sigma\sigma}|_{\sigma_0}$, we obtain the displacement
$\Delta\sigma\simeq18M^2_P|V_{\rm AdS}|/(W_{\sigma\sigma}|_{\sigma_0})^2(2\sigma_0)^3$. 
This leads to the SUSY-breaking $F$-term in the uplifted minimum
 $D_\sigma W\simeq(D_\sigma W)_\sigma\Delta\sigma\simeq W_{\sigma\sigma}(\sigma_0)\Delta\sigma$:
\begin{eqnarray}
D_\sigma W\simeq\frac{3|V_{\rm AdS}|}{\sigma_0\,m_\sigma}\,,
  \label{Ux_1}
\end{eqnarray}
where the suppression factor 
\begin{eqnarray}
m^2_\sigma\gg\sqrt{|V_{\rm AdS}|}
  \label{Ux_100}
\end{eqnarray}
is characteristic of the KL model framework\,\cite{Kachru:2003aw}. As a result, the shift of the vacuum is strongly suppressed, and the generated F-term remains naturally small. This hierarchy together with Eq.(\ref{bac00b}) ensures that the $U_X$ modulus stabilization is essentially unaffected by the uplift while providing a controlled source of supersymmetry breaking.

The spontaneous breaking of modular symmetry is governed by the VEV of the modulus $\tau$ $({\rm Im}\,\tau>0)$, which can always be constrained to lie within the fundamental domain ${\cal D}$ of the modular group. This domain is defined as
\begin{eqnarray}
 {\cal D}\equiv\Big\{\tau\in{\cal H}:-\frac{1}{2}\leq{\rm Re}\,\tau<\frac{1}{2}, |\tau|>1\Big\}\cup\Big\{\tau\in{\cal H}:-\frac{1}{2}<{\rm Re}\,\tau\leq0, |\tau|=1\Big\}\,\,,
  \label{do_0}
\end{eqnarray}
where ${\cal H}$ denotes the upper half-plane of complex numbers ${\cal H}\equiv \{\tau\in\mathbb{C}| {\rm Im}(\tau)>0\}$.
While no specific value of $\tau$ preserves the full modular symmetry, partial modular symmetries are retained at special symmetric points such as $\tau=i, i\infty, e^{i2\pi/3}$\,\cite{Novichkov:2018ovf,Gonzalo:2018guu}. As shown in Ref.\cite{Cvetic:1991qm}, all extrema of the potential $V(\tau,\bar{\tau})$ must lie either on the boundary of the fundamental domain ${\cal D}$ or on the imaginary axis. 
For the introduction of a small perturbation $\Delta W$, or for a finite but small $\kappa_R$, the supersymmetric vacuum is shifted away from the exact KL Minkowski vacuum Eq.(\ref{susy_0}). In the shifted vacuum, $W\neq0$, and the F-term equation for $\tau$ in Eq.(\ref{susy_1}) becomes nontrivial. Considering the F-term equation for $\tau$ evaluated at $(\tilde{\sigma}_0, s_0)$
\begin{eqnarray}
D_\tau W&\simeq&-3W\Big[\frac{1-1/(16\pi^2y)}{\tau-\bar{\tau}}+2\frac{\eta'(\tau)}{\eta(\tau)}\Big]\nonumber\\
&+&\Big(\frac{\kappa_R\langle S\rangle}{\tilde{\kappa}_R \langle\sigma\rangle}\Big)\frac{3}{\langle S\rangle8\pi^2}\frac{M^3_P}{[\eta(\tau)]^{6}}\frac{\eta'(\tau)}{\eta(\tau)}\Big(|A_0|\Delta^{N_\alpha}_\chi S^2_\alpha\,e^{-S_\alpha}-|B_0|\Delta^{N_\beta}_\chi S^2_\beta\,e^{-S_\beta}\Big)+...
  \label{Dta_0}
\end{eqnarray}
The resulting modular-stabilization condition can then determine $\tau$ near the special fixed points,
\begin{eqnarray}
\tau_0\approx i\,,\qquad \pm\frac{1}{2}+i\frac{\sqrt{3}}{2}\,,
  \label{Dta_01}
\end{eqnarray}
with deviations controlled by the small perturbation $\Delta W$, finite $\kappa_R$, and the $1/(16\pi^2\langle y\rangle)$ corrections.
While the $\kappa_R$ dependent term in the superpotential, see Eq.(\ref{Dta_0}), modifies the scalar potential and shifts the vacuum value of $\tau$, its effect remains perturbative for sufficiently small $\kappa_R$. Moreover, since the vacuum-stability condition of Eq.(\ref{bac00b}) requires $N_i\in[18,47]$, see Eq.(\ref{fppe}), together with $g_X\in[0.13,0.34]$, see Eq.(\ref{fppe100}), the shape of the potential for $\tau$ is not significantly affected, and the minimum remains close to Eq.(\ref{Dta_01}).

\begin{figure}[t]
\begin{minipage}[h]{8.0cm}
\epsfig{figure=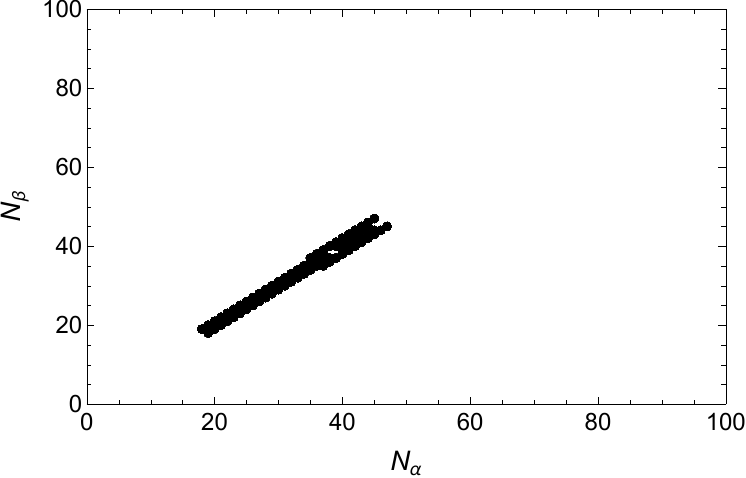,width=8.0cm,angle=0}
\end{minipage}
\begin{minipage}[h]{8.0cm}
\epsfig{figure=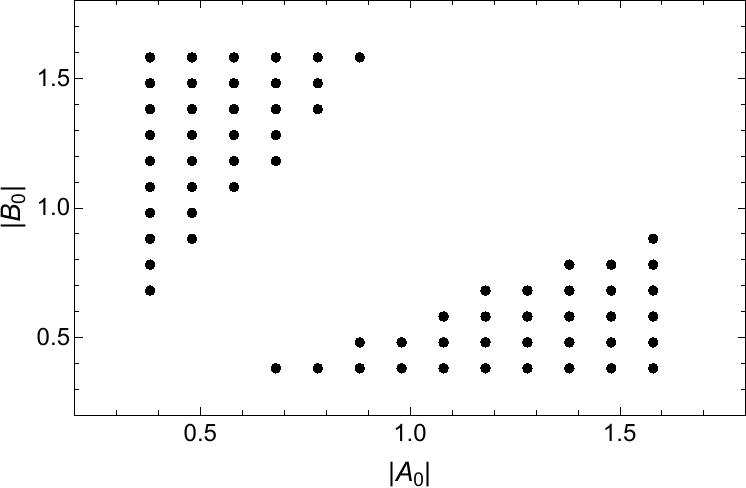,width=8.0cm,angle=0}
\end{minipage}\\
\begin{minipage}[h]{8.0cm}
\epsfig{figure=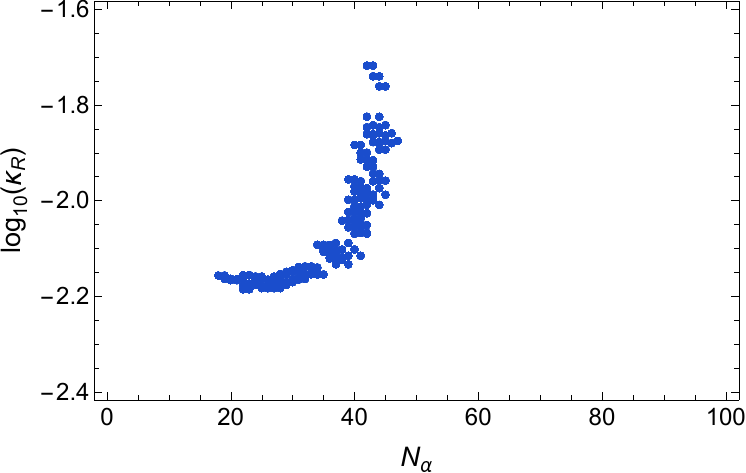,width=8.0cm,angle=0}
\end{minipage}
\begin{minipage}[h]{8.0cm}
\epsfig{figure=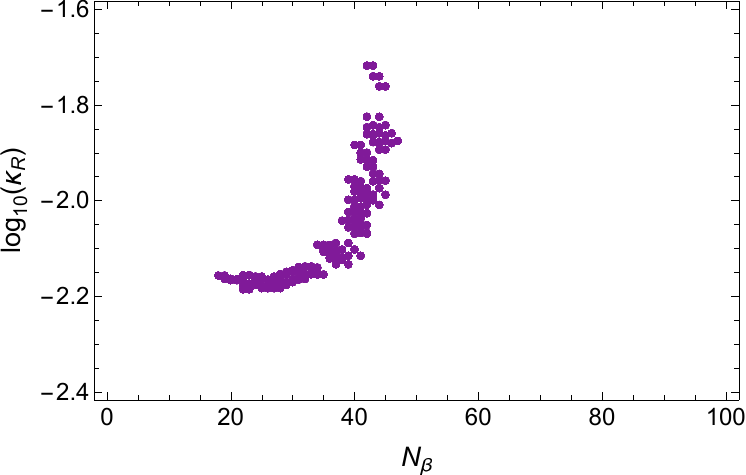,width=8.0cm,angle=0}
\end{minipage}\\
\begin{minipage}[h]{8.0cm}
\epsfig{figure=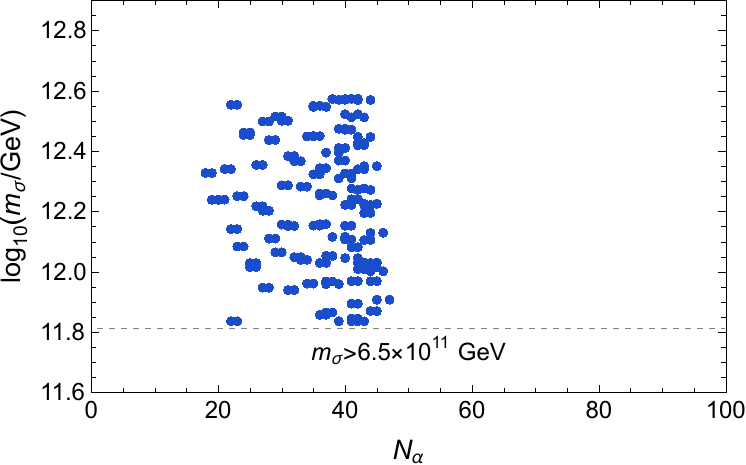,width=8.0cm,angle=0}
\end{minipage}
\begin{minipage}[h]{8.0cm}
\epsfig{figure=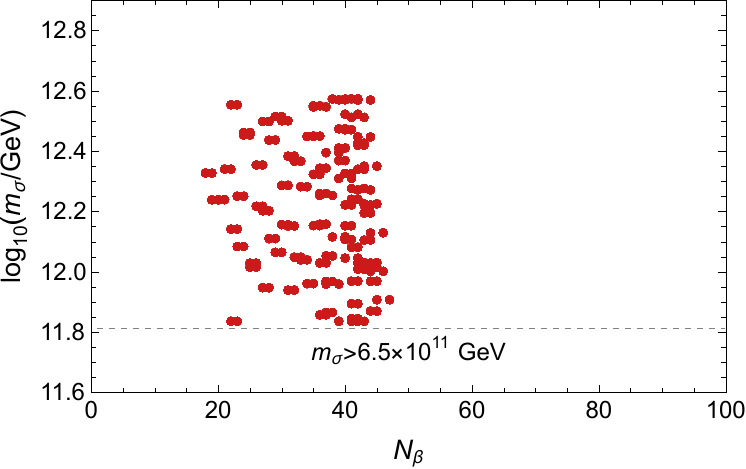,width=8.0cm,angle=0}
\end{minipage}
\caption{\label{FigB1} Allowed parameter spaces in the $(N_\alpha,N_\beta)$ plane (upper left panel) and the $(|A_0|,|B_0|)$ plane (upper right panel), subject to $\delta R<0.01$, $m_\sigma>6.5\times10^{11}$ GeV, and the constraints from SM gauge-coupling unification and the flavor structure discussed in Secs.\ref{gcu} and \ref{visu}. The middle panels show the allowed ranges of $\kappa_R$ as functions of $N_i$. The lower panels show the allowed modulus mass $m_\sigma$ as a function of $N_\alpha$ and $N_\beta$.}
\end{figure}

In the shifted vacuum, $W\neq0$, and the F-term equation for $S$ in Eq.(\ref{susy_1}) becomes nontrivial. Evaluated at $(\sigma_0+\delta\sigma, \tau_0)$, the F-term equation for $S$ takes the form
\begin{eqnarray}
&&D_S W\simeq\frac{M^3_P}{[\eta(\tau_0)]^{6}}\alpha A\,e^{-\alpha\tilde{\kappa}_R\sigma_0}\sum^{\infty}_{n=2}(-1)^n(\alpha^{n-1}-\beta^{n-1})\kappa^n_R\Big\{\frac{(S+\tilde{x})^{n-1}}{(n-1)!}-\frac{(S+\tilde{x})^{n}}{n!(2S+\tilde{y})}\Big\}\,,
  \label{Dta_1}
\end{eqnarray}
where $\alpha\,A\,e^{-\alpha\tilde{\kappa}_R \sigma_0}=\beta\,B\,e^{-\beta\tilde{\kappa}_R \sigma_0}$ is used in a good approximation due to small $\delta\sigma$, $\tilde{y}=\frac{3}{16\pi^2}\ln(2\,{\rm Im}\tau_0)$ and $\tilde{x}=\frac{3}{8\pi^2}\ln\eta(\tau_0)$.
The resulting equation can be solved in a good approximation for $S$: 
\begin{eqnarray}
\kappa_R\big(2S+\tilde{y}\big)\big(e^{-T_\beta}-e^{-T_\alpha}\big)=\frac{(\delta^R_X)^2}{16\pi^2|\delta^{\rm GS}_X|}\Big\{\frac{1}{N_\beta}\big(1-e^{-T_\beta}\big)-\frac{1}{N_\alpha}\big(1-e^{-T_\alpha}\big)\Big\}
  \label{Dta_2}
\end{eqnarray}
where, similar to $S_i$ of Eq.(\ref{d0ac}), the nonperturbative instanton actions are defined as
 \begin{eqnarray}
T_i=N_i\kappa_R\Big(S+\frac{3}{8\pi^2}\ln\eta(\tau)\Big)\frac{16\pi^2|\delta^{\rm GS}_X|}{(\delta^R_X)^2}\,,\quad(i=\alpha,\beta)\,.
  \label{inst01}
\end{eqnarray}
The parameter $\kappa_R$ and the VEV of $S$ can be fixed by the vacuum stability conditions discussed above as well as by requiring consistency with SM gauge coupling unification through RG running, see Fig.\ref{FigB1}.

\begin{figure}[t]
\begin{minipage}[h]{8.0cm}
\epsfig{figure=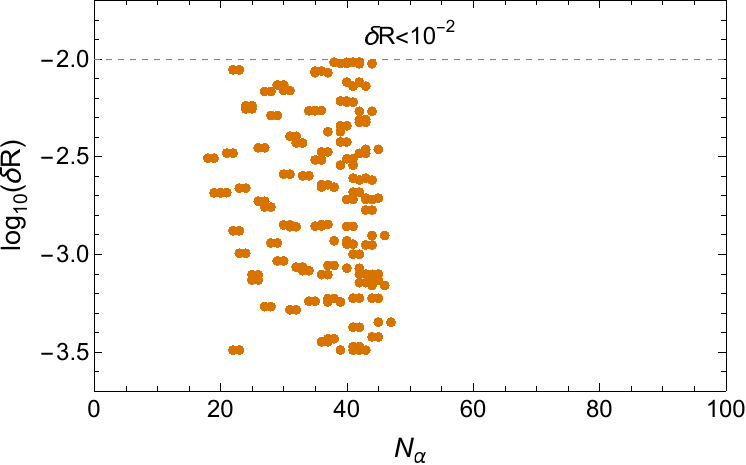,width=9.0cm,angle=0}
\end{minipage}
\caption{\label{FigB2} Vacuum-stability constraint $\delta R$ as a function of $N_\alpha$, with $\delta R<10^{-2}$ ({\it i.e.} less than a $1\%$ deviation) imposed. The plot for $N_\beta$ is similar and is not shown.}
\end{figure}

For $\kappa_R\neq0$, non-vanishing F-terms may not be generated in the $\tau$ and $S$ directions, before the uplift is included. 
At the shifted minimum $\tilde{\sigma}_0,\tau_0,s_0$, Eqs.(\ref{susy_1}) and (\ref{Ux_1}) imply
\begin{eqnarray}
\langle D_\tau W\rangle\neq0\,,\qquad \langle D_S W\rangle\neq0\,,\qquad\langle D_{U_X} W\rangle\neq0\,.
  \label{Sb_1}
\end{eqnarray}
This implies that supersymmetry is broken by the modulus $U_X$.
When including the uplifting contribution $\Delta V$, the gravitino mass Eq.(\ref{mp32}) becomes
\begin{eqnarray}
m^2_{3/2}\approx\frac{|V_{\rm AdS}|}{3M^2_P}=\frac{|\Delta W|^2}{M^4_P}\,,
  \label{m32_m}
\end{eqnarray}
in agreement with Ref.\cite{Kachru:2003aw}, where $\Delta W$ is the K{\"a}hler-rescaled quantity.
Requiring $m_{3/2}\gtrsim1$ TeV therefore implies $|\Delta W|\gtrsim4.1\times10^{-16}\,M^3_P$.
The condition Eq.(\ref{Ux_100}), $m^2_{\sigma}\gg\sqrt{3}m_{3/2}M_P$ where $m_{\sigma}$ is evaluated at the flavored-GUT normalized value $\sigma_0(\Lambda_{\rm fGUT})$ of Eq.(\ref{d0aa}), places a lower bound on the modulus mass $m_\sigma$. For $m_{3/2}\sim1,10,10^2$, and $10^3$ TeV, respectively, it requires $m_\sigma\gg6.5\times10^{10}$ GeV, $m_\sigma\gg2.0\times10^{11}$ GeV, $m_\sigma\gg 6.5\times10^{11}$ GeV, $m_\sigma\gg2.0\times10^{12}$ GeV. On the other hand, the vacuum stability condition Eq.(\ref{bac00b}), illustrated in Fig.\ref{FigB2}, yields an upper bound $m_\sigma\lesssim3.8\times10^{12}$ GeV in Eq.(\ref{fppe1}), given the values of $\Delta_\chi$ and $M_X$ fixed by the flavor structure and the RG running associated with SM gauge-coupling unification. Combining these two conditions, Eqs.(\ref{bac00b}) and (\ref{Ux_100}), yields the approximate allowed ranges
\begin{eqnarray}
6.5\times10^{10}\,{\rm GeV}\ll m_\sigma\lesssim3.8\times10^{12}\,{\rm GeV}\,,\qquad\quad 1\,{\rm TeV}\lesssim m_{3/2}\ll10^{2}\,{\rm TeV}\,.
  \label{m32_m2}
\end{eqnarray}
Thus, the two conditions jointly constrain the modulus mass $m_\sigma$ and the gravitino mass $m_{3/2}$. In our framework, the gravitino mass $m_{3/2}$ characterizes the supersymmetry-breaking mass scale, whereas the fundamental F-term scale is parametrically $\sqrt{|F|}\sim\sqrt{m_{3/2}M_P}$. The resulting allowed range of $m_{3/2}$ in Eq.(\ref{m32_m2}) is consistent with the supersymmetry-breaking scale inferred independently from the RG evolution associated with SM gauge-coupling unification, as illustrated in Fig.\ref{Fig3} and Table-\ref{bmp}. Although $|\Delta W|$ is not specified explicitly in our analysis, the supersymmetry-breaking scale can therefore be constrained independently through the RG running associated with SM gauge-coupling unification, as discussed in the previous section, and through the vacuum-stability condition Eq.(\ref{bac00b}) and the condition Eq.(\ref{Ux_100}).

In the flavored-GUT framework, once the anomaly coefficient $\delta^R_X$ (see Eqs.(\ref{grA}) and (\ref{gt02})) is fixed by the flavor sector, while the gauge coupling $g_X(\Lambda_{\rm fGUT})$ (see Eqs.(\ref{gac0}) and (\ref{d0aa}) and Fig.\ref{Fig2}) and the flavor-dynamics scale $M_X$ (see Eqs.(\ref{gbm01}) and (\ref{gbm01m})) are determined by RG running and the SM gauge-coupling unification conditions. Consequently, the QCD axion decay constant $f_X=f_A$ (Eqs.(\ref{axide1}) and (\ref{AFN1})) and the mass $m_\chi$ in Eq.(\ref{chma}) are also determined. Then, the vacuum-stability condition Eq.(\ref{bac00b}) is governed by the remaining five\,\footnote{For $\kappa_R\neq0$, the superpotential Eq.(\ref{mosu}) can be written as $W=\frac{M^3_P}{[\eta(\tau)]^6}\big(C_0+A\,e^{-\alpha\tilde{\kappa}_R U_X}\,e^{-\alpha\kappa_R(S+\tilde{x})}-B\,e^{-\beta\tilde{\kappa}_R U_X}\,e^{-\beta\kappa_R(S+\tilde{x})}\big)$. Then, the modulus mass $m_\sigma$ of Eq.(\ref{sig_01}) receives the correction $m^2_\sigma=\frac{8}{9}\frac{M^2_P}{|\eta(\tau_0)|^{12}}\big|S^2_\alpha|A_0|e^{-T_\alpha}\big(\frac{N_\alpha}{N_\beta}\frac{|A_0|}{|B_0|}e^{-T_\alpha+T_\beta}\big)^{-\frac{N_\alpha}{N_\alpha-N_\beta}}-S^2_\beta|B_0|e^{-T_\beta}\big(\frac{N_\alpha}{N_\beta}\frac{|A_0|}{|B_0|}e^{-T_\alpha+T_\beta}\big)^{-\frac{N_\beta}{N_\alpha-N_\beta}}\big|^2$.} parameters
\begin{eqnarray}
N_\alpha\,,\quad N_\beta~(\text{equivalently } S_\alpha\,,~S_\beta)\,,\quad\kappa_R\,,\quad |A_0|\,,\quad |B_0|\,.
  \label{fpp}
\end{eqnarray}
which correspond to the five degree of freedoms $\alpha$, $\beta$, $\kappa_R$, $A$, and $B$ in the superpotential (\ref{mosu}). Note that, in our numerical calculations, the parameters $g_{\chi_0}$, $g_\chi$ appearing in the superpotential (\ref{super_d}) were set to unity.
In particular, the corrected modulus mass $m_\sigma$ in footnote-25, evaluated at the compactification minimum $\sigma_0^{\rm comp}$ in Eq.(\ref{d0}), depends on these parameters. Recall that $|A_0|$ and $|B_0|$ are constrained by Eq.(\ref{u1t003}), whereas $|C_0|$ is fixed by Eq.(\ref{d0}).

\begin{figure}[t]
\begin{minipage}[h]{8.0cm}
\epsfig{figure=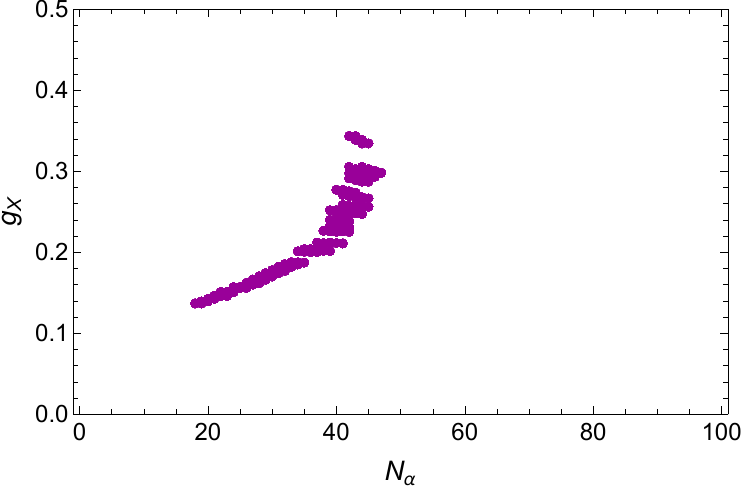,width=8.0cm,angle=0}
\end{minipage}
\begin{minipage}[h]{8.0cm}
\epsfig{figure=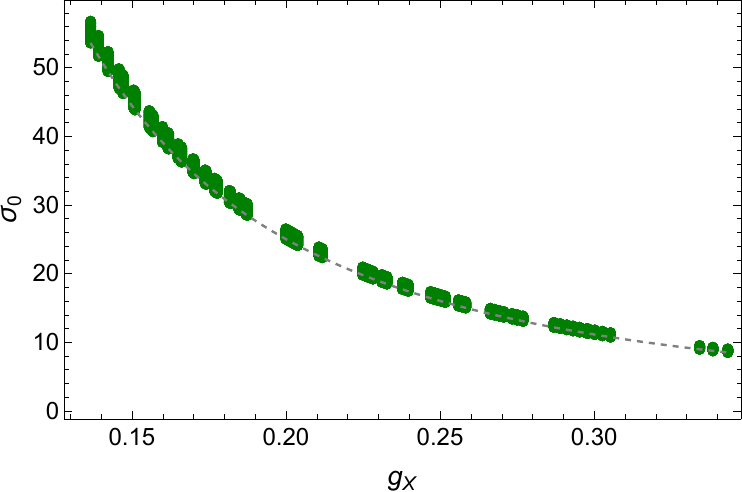,width=8.0cm,angle=0}
\end{minipage}
\caption{\label{FigB3}  Plots of the gauge coupling $g_X$ as a function of $N_\alpha$ (left) and $\sigma_0\equiv\sigma^{\rm comp}_0$ (right). The plot for $N_\beta$ is similar and is not shown.}
\end{figure}

As a numerical illustration, we consider the input parameters $|A_0|,|B_0|\in[19/50,79/50]$ (see Eqs.(\ref{u1t003}) and (\ref{delchi})), $g_X\in[0.001,2\pi]$, $\delta R<0.01$ ({\it i.e.}, less than $1\%$), as required by Eq.(\ref{bac00b}), $\alpha^{-1}_{\rm fGUT}=4\pi\langle S\rangle\in[26.70, 28.27]$ as in Table-\ref{bmp}, and $m_\sigma>6.5\times10^{11}$ GeV, as required by Eq.(\ref{m32_m2}). We take $v_\chi=1.9532\times10^{15}$ GeV and $M_X=2.2277\times10^{15}$ GeV, as specified in Eq.(\ref{axide1}), and set $\delta^R_X=-84$ from Table-\ref{reps_q} and -\ref{reps_l}. The upper panels of Fig.\ref{FigB1} show the allowed regions in the $(N_\alpha,N_\beta)$
and $(|A_0|,|B_0|)$ planes. The viable parameter regions are characterized by
\begin{eqnarray}
N_i\in[18,47]\,,(N_\alpha\neq N_\beta)\,;\quad |A_0|,|B_0|\in[19/50,79/50]\,, ( |A_0|\neq |B_0|)
  \label{fppe}
\end{eqnarray}
which yields 
\begin{eqnarray}
6.5\times10^{11}< m_\sigma[{\rm GeV}]\lesssim3.7\times10^{12}\,,\quad 3.2\times10^{-4}\lesssim\delta R\lesssim9.6\times10^{-3}\,.
  \label{fppe1}
\end{eqnarray}
The allowed modulus mass range is shown in the lowest panels of Fig.\ref{FigB1}. The middle panels of Fig.\ref{FigB1} show allowed parameter space for $\kappa_R$  as a function of $N_i$, yielding a bound on $\kappa_R$:
\begin{eqnarray}
6.5\times10^{-3}\lesssim\kappa_R\lesssim1.9\times10^{-2}\,.
  \label{kR1}
\end{eqnarray}
Fig.\ref{FigB2} illustrates the dependence of $\delta R$ on $N_\alpha$; the corresponding plot for $N_\beta$ is similar and is not shown.
 As illustrated in Fig.\ref{FigB3}, the viable values of $N_i$ ($i=\alpha,\beta$), see Fig.\ref{FigB1}, constrains the allowed range of $g_X(\Lambda_{\rm fGUT})$, through Eq.(\ref{d0aa}), to
\begin{eqnarray}
0.13\lesssim g_X\lesssim0.34\,.
  \label{fppe100}
\end{eqnarray}
This range is consistent with the values obtained from SM gauge coupling unification, as shown in Table-\ref{bmp} and discussed in connection with Eq.(\ref{th04}).
These results demonstrate that the vacuum-stability requirement selects the allowed quantized nonperturbative indices $N_\alpha$ and $N_\beta$; see also Fig.\ref{FigC1}. 
A parameter set satisfying these conditions can then be used to test
the axion-quality constraint Eq.(121).

\subsection{Dynamical realization of a Qualified QCD axion}
\label{qual}
The axionic shift symmetry (see Eq.(\ref{ss01})) is broken only by non-perturbative effects. Additional non-perturbative contributions beyond the QCD one can, in general, spoil the axion solution to the strong CP problem, leading to the well-known axion quality problem unless they are sufficiently suppressed\,\cite{axiQu,Banks:1996ea,Svrcek:2006yi}. Here, we show how the flavored-GUT framework dynamically suppresses these unwanted contributions, thereby resolving the axion quality problem.

The physical QCD axion potential, with the vacuum energy at the minimum subtracted, is defined\,\footnote{Here $V(A_X)$ is given by $V(A_X)=-\mu f^2_\pi\Big\{m_u\cos\frac{1}{1+z+\omega}\Big(\frac{A_X}{F_a}+\vartheta_{\rm eff}\Big)+m_d\cos\frac{z}{1+z+\omega}\Big(\frac{A_X}{F_a}+\vartheta_{\rm eff}\Big)+m_s\cos\frac{\omega}{1+z+\omega}\Big(\frac{A_X}{F_a}+\vartheta_{\rm eff}\Big)\Big\}$, see for an example Ref.\cite{Ahn:2014gva}.} as $V_{\rm QCD}(A_X)=V(A_X)-V(A^{\rm min}_X)$ with $V(A^{\rm min}_X)=-\mu f^2_\pi(m_u+m_d+m_s)$ at the minimum $\langle A_X\rangle=-F_a\vartheta_{\rm eff}$ with $A_X=\langle A_X\rangle+a_X$. It is given by
\begin{eqnarray}
&&V_{\rm QCD}(A_X)=\mu f^2_\pi\Big\{m_u\Big(1-\cos\frac{A_X/F_a+\vartheta_{\rm eff}}{1+z+\omega}\Big)\nonumber\\
&&\qquad\qquad\qquad+m_d\Big(1-\cos\frac{z(A_X/F_a+\vartheta_{\rm eff})}{1+z+\omega}\Big)+m_s\Big(1-\cos\frac{\omega(A_X/F_a+\vartheta_{\rm eff})}{1+z+\omega}\Big)\Big\}\,,
  \label{qcd_01}
\end{eqnarray}
where $z=m_u/m_d$ and $\omega=m_u/m_s$, which is equivalent to the form $V_{\rm QCD}(A_X)=m^2_aF^2_a\big[1-\cos(A_X/F_a+\vartheta_{\rm eff})\big]$ with
\begin{eqnarray}
m^2_a=\frac{f^2_\pi}{F^2_a}\frac{\mu m_u}{1+z+\omega}\sim \frac{\Lambda^4_{\rm QCD}}{F^2_a}\,,\quad\text{with}~F_a=\frac{f_A}{|\delta^G_X|}\,.
  \label{qcd_001}
\end{eqnarray}
 This QCD axion potential comes from the QCD instanton as well as the chiral rotation of SM quarks. 

On the other hand, the non-perturbative superpotential Eq.(\ref{mosu}) originates from a common consistency condition: the requirement that the moduli simultaneously satisfy anomaly cancellation, SM gauge-coupling unification, and vacuum stability. The GS structure then correlates the transformation properties of the gravitational and gauge sectors, so that the string-induced nonperturbative contribution carries the corresponding gravitational (or GS) charge. Consequently, its contribution to the axion potential is constrained by $U(1)_X$ gauge invariance and accompanied by the $U(1)_X$ charge-field insertions required for gauge invariance.
As shown in Eq.(\ref{sup_01}), in the $\kappa_R\rightarrow0$ limit, the theory admits a supersymmetric Minkowski vacuum, and the surviving axion $A_X$ acquires a string-induced gravitational potential. 

Using the canonical normalization Eq.(\ref{act2}) with $\theta_X=a_\theta/8\pi^2 f_\theta$, $V(\theta_X)$ reduces to $V(a_\theta)$. As discussed in Sec.\ref{NGm}, after the St{\"u}ckelberg mechanism, in which the NG mode $G$ is eaten by the $U(1)_X$ gauge boson, with impose the unitary gauge $G=0$. Thus, $a_\theta$ in the potential can be expressed in term of the surviving low-energy axion $A_X$.
The normalization of the $U(1)_X$ gauge coupling is related to the GS coefficient through $\tilde{\kappa}_X=\delta^X_X/\delta^{\rm GS}_X$ in Eq.(\ref{gt02}): at the flavored-GUT matching scale, using Eqs.(\ref{gac0}) and (\ref{d0}) we identify the normalized modulus defined by the effective action Eq.(\ref{act1}) as
\begin{eqnarray}
\sigma(\Lambda_{\rm fGUT})&=&\frac{1}{\tilde{g}^2_X(\Lambda_{\rm fGUT})}\nonumber\\
&=&\frac{\tilde{\kappa}_X}{g^2_X}=\tilde{\kappa}_X\,\sigma^{\rm comp}\,,
\label{d0aa}
\end{eqnarray}
where $\sigma^{\rm comp}$ denotes the modulus in string compactification.
This relation specifies the boundary matching between the compactification and the flavored-GUT descriptions. 
Including $\kappa_R$-dependent terms (see footnote-25), the string-induced axion potential can be written in terms of Eqs.(\ref{u1t001}), (\ref{u1t002}), (\ref{d0ac}), and (\ref{inst01}), together with Eq.(\ref{d0aa}), as
\begin{eqnarray}
&&V_{\rm grav}(A_X)=4\frac{M^4_P}{|\eta(\tau)|^{12}}\Big\{|C_0||A_0|\Delta^{N_\alpha}_\chi\,S_\alpha\,e^{-S_\alpha}\,e^{-T_\alpha}\cos\Big(\frac{A_X}{f_\alpha}N_\alpha+\varphi_{\alpha c}\Big)\nonumber\\
&&-|C_0||B_0|\Delta^{N_\beta}_\chi\,S_\beta\,e^{-S_\beta}\,e^{-T_\beta}\cos\Big(\frac{A_X}{f_\alpha}N_\beta+\varphi_{\beta c}\Big)\nonumber\\
&&-|A_0||B_0|\Delta^{N_\beta+N_\beta}_\chi\,e^{-S_\alpha-S_\beta}\,e^{-T_\alpha-T_\beta}\big(\frac{2}{3}S_\alpha S_\beta+S_\alpha+S_\beta\big)\cos\Big(\frac{A_X}{f_\alpha}(N_\alpha-N_\beta)+\varphi_{\alpha\beta}\Big)\Big\}\,,
  \label{sup_01b}
\end{eqnarray}
where 
\begin{eqnarray}
f_\alpha=f_A/\Big(1+\frac{4}{\delta^R_X\delta^{\rm GS}_X}\Big(\frac{f_A}{f_\theta}\Big)^2\Big)\,,
  \label{sup_01c}
\end{eqnarray}
in which $f_\alpha$ is approximated to $f_\alpha\approx f_A$ for $1\gg (4/|\delta^R_X\delta^{\rm GS}_X|)(f_A/f_\theta)^2$. 
The nonperturbative stringy effects that generate the above periodic potential for the surviving low-energy axion $A_X$ can spoil the QCD axion solution to the strong CP problem due to the nonvanishing phases $\varphi_{ij}$. An important point is that, within the flavored-GUT framework, the modulus value matched at the flavored-GUT scale, Eq.(\ref{d0aa}), feeds directly into the gravitational contribution.

We now include the full effects of supersymmetry breaking. In the presence of a non-vanishing SUSY-breaking background, $\langle W\rangle\sim m_{3/2}M^2_P$ (see Eq.(\ref{mp32})), the supergravity scalar potential contains additional contributions from the $|W|^2$ terms and are proportional to the same exponential terms of Eq.(\ref{sup_01b}). Therefore, after including the full supergravity corrections, the string-induced gravitational contribution to the axion potential can be simply expressed as
\begin{eqnarray}
V_{\rm grav}(A_X)=\sum^3_{i=1}(\Lambda^{\rm eff}_i)^4\cos\Big(\frac{A_X}{f^{\rm eff}_i}+\psi_i\Big)\,,
  \label{gr_01}
\end{eqnarray}
where the physical phases $\psi_i$ receive contributions from both the phases $\varphi_{\alpha}, \varphi_{\beta}, \varphi_{c}$ of $A_0, B_0, C_0$ and the SUSY-breaking interference terms between supersymmetric terms and supersymmetry breaking terms, after accounting for field redefinitions.
Each amplitude $(\Lambda^{\rm eff}_i)^4$ becomes
\begin{eqnarray}
&&(\Lambda^{\rm eff}_1)^4=4|C_0||A_0|\Delta^{N_\alpha}_\chi\,S_\alpha\,e^{-S_\alpha}\,e^{-T_\alpha}\frac{M^4_P}{|\eta(\tau)|^{12}}\Big(1+{\cal O}\big(\frac{m^2_{3/2}}{M^2_P}\big)\Big)\,,\nonumber\\
&&(\Lambda^{\rm eff}_2)^4=-4|C_0||B_0|\Delta^{N_\beta}_\chi\,S_\beta\,e^{-S_\beta}\,e^{-T_\beta}\frac{M^4_P}{|\eta(\tau)|^{12}}\Big(1+{\cal O}\big(\frac{m^2_{3/2}}{M^2_P}\big)\Big)\,,\nonumber\\
&&(\Lambda^{\rm eff}_3)^4=-4|A_0||B_0|\Delta^{N_\alpha+N_\beta}_\chi\,e^{-S_\alpha-S_\beta}\,e^{-T_\alpha-T_\beta}\frac{M^4_P}{|\eta(\tau)|^{12}}\Big(1+{\cal O}\big(\frac{m^2_{3/2}}{M^2_P}\big)\Big)\,.
  \label{gr_01a}
\end{eqnarray}
And the effective decay constants are given by
\begin{eqnarray}
f^{\rm eff}_1=\frac{f_\alpha}{N_\alpha}\,,\qquad f^{\rm eff}_2=\frac{f_\alpha}{N_\beta}\,,\qquad f^{\rm eff}_3=\frac{f_\alpha}{N_\alpha-N_\beta}\,,
  \label{gr_01ab}
\end{eqnarray}
with $f_\alpha$ given in Eq.(\ref{sup_01c}).
Unlike the QCD axion potential Eq.(\ref{qcd_01}), which arises from QCD instantons and is associated with chiral rotation of SM quarks characterized by the color anomaly coefficient $\delta^G_X$, the string-induced gravitational potential is governed by the quantized nonperturbative indices $N_i$ and the nonperturbative actions $S_i$ and $T_i$ ($i=\alpha,\beta$). Although larger values of $S_i\propto N_i/\tilde{g}^2_X(\Lambda_{\rm fGUT})$ and/or $T_i\propto N_i\kappa_R$ tend to suppress the string-induced gravitational contribution, $V_{\rm grav}(A_X)\propto \Delta^{N_i}_\chi S_i\,e^{-S_i}\,e^{-T_i}\rightarrow0$, the values of $S_i$ and $T_i$ are not freely adjustable, but are constrained by the vacuum-stability condition in Eq.(\ref{bac00b}) and the gauge-coupling normalization at the flavored-GUT scale in Eq.(\ref{d0aa}).

Minimizing the total axion potential $V(A_X)=V_{\rm QCD}(A_X)+V_{\rm grav}(A_X)$ with respect to $A_X$, the vacuum satisfies $d V(A_X)/dA_X=0$:
\begin{eqnarray}
\frac{\Lambda^4_{\rm QCD}}{F_a}\sin\Big(\vartheta_{\rm eff}+\frac{A_X}{F_a}\Big)+\sum^3_{i=1}\frac{(\Lambda^{\rm eff}_i)^4}{f^{\rm eff}_i}\sin\Big(\frac{A_X}{f^{\rm eff}_i}+\psi_i\Big)=0\,,
  \label{gr_003}
\end{eqnarray}
which gives a shifted vacuum and a nonzero $\vartheta_{\rm phys}$. Assuming $(\Lambda^{\rm eff}_i)^4\ll\Lambda^4_{\rm QCD}$, {\it i.e.}, negligible gravity, let $A_X=\langle A_X\rangle+\delta A$ the physical angle defined by $\vartheta_{\rm phys}=\delta A/F_a$ is given from Eq.(\ref{gr_003}) as
\begin{eqnarray}
\vartheta_{\rm phys}=\sum^3_{i=1}\Big(\frac{\Lambda^{\rm eff}_i}{\Lambda_{\rm QCD}}\Big)^4\frac{F_a}{f^{\rm eff}_i}\sin\Big(\frac{\langle A_X\rangle}{f^{\rm eff}_i}+\psi_i\Big)\equiv\sum^3_{i=1}x_i\,\sin\Big(\frac{\langle A_X\rangle}{f^{\rm eff}_i}+\psi_i\Big)<10^{-10}\,,
  \label{gr_004}
\end{eqnarray}
which is bounded by the experimental limits. Here, $x_i\equiv\big(\frac{\Lambda^{\rm eff}_i}{\Lambda_{\rm QCD}}\big)^4\frac{F_a}{f^{\rm eff}_i}$, are the axion-quality parameters, evaluated at the matched flavored-GUT-scale modulus value $\sigma_0(\Lambda_{\rm fGUT})$ in Eq.(\ref{d0aa}).

Once the anomaly coefficients $\delta^R_X$, $\delta^G_X$, and $\delta^X_X$ are fixed by the flavor structure, see Eqs.(\ref{SAo}) and (\ref{grA}), the gauge coupling $g_X$, the unified gauge coupling $\alpha^{-1}_{\rm fGUT}=4\pi\langle S\rangle$, and the flavor-dynamics scale $M_X$ in Eq.(\ref{gbm01m}) are determined through RG running and the SM gauge-coupling unification, as discussed above Eq.(\ref{fppe}). The string-induced axion potential in Eq.(\ref{gr_01}), evaluated at the matched modulus value $\sigma_0(\Lambda_{\rm fGUT})$, with $\tilde{\kappa}_X=\delta^X_X/\delta^{\rm GS}_X$ in Eq.(\ref{gt02}) and $\delta^X_X=-64572$ from Table-\ref{reps_q} and -\ref{reps_l}, is then characterized by the same five parameters as in Eq.(\ref{fpp}).
As discussed in the previous subsection, these parameters are subject to the vacuum-stability condition in Eq.(\ref{bac00b}), with the modulus mass $m_\sigma$ evaluated at the original compactification minimum $\sigma_0^{\rm comp}$ and determined by the same parameter set, as given in Eq.(\ref{sig_01}). The additional condition in Eq.(\ref{Ux_100}) must also be satisfied.

\begin{figure}[t]
\begin{minipage}[h]{8.0cm}
\epsfig{figure=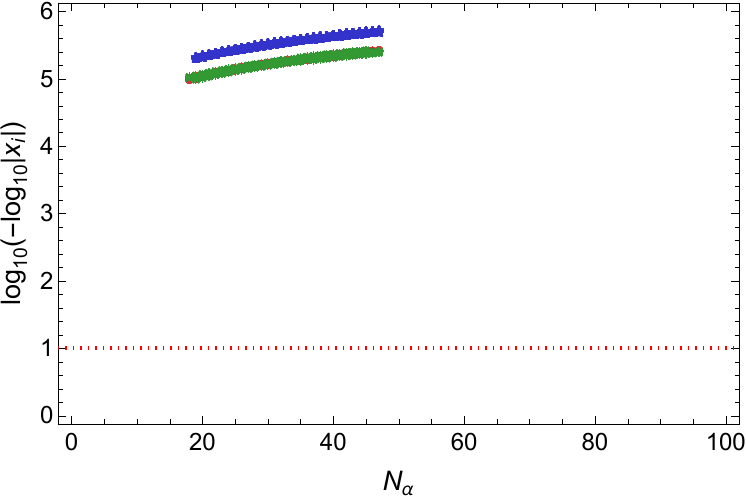,width=8.0cm,angle=0}
\end{minipage}
\begin{minipage}[h]{8.0cm}
\epsfig{figure=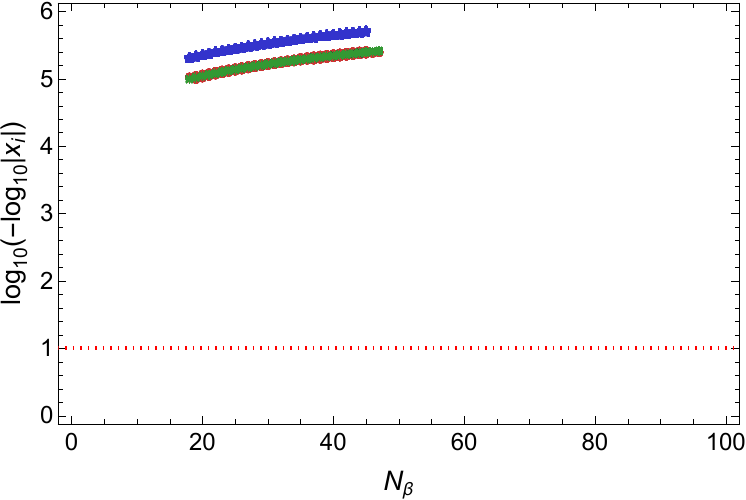,width=8.0cm,angle=0}
\end{minipage}
\caption{\label{FigC1}  The axion-quality constraints $x_i$ in Eq.(\ref{gr_004}) as a function of the quantized nonperturbative indices $N_\alpha$ (left panel) and $N_\beta$ (right panel). The $y$-axis shows the double‑logarithmic quantity $\log_{10}(-\log_{10}|x_i|)$. The red, green, and blue points correspond to $|x_1|$, $|x_2|$, and $|x_3|$, respectively. The horizontal red-dotted line indicates the experimental bound, $|x_i|<10^{-10}$.}
\end{figure}

Evaluated at the matched flavored-GUT-scale modulus value specified by Eq.(\ref{d0aa}), the quantities $x_i$, defined in Eq.(\ref{gr_004}) and entering the axion-quality constraint, are estimated from Fig.\ref{FigC1} to be vanishingly small (effectively zero) for the same parameter set satisfying the vacuum-stability condition in Eq.(\ref{fppe}) and Eq.(\ref{Ux_100}) (see also Eq.(\ref{m32_m2})).
Thus, anomaly cancellation and SM gauge coupling unification, as fundamental consistency conditions, constrain the framework in a parameter region where vacuum-stability is maintained and the string-induced gravitational contribution to the axion potential is effectively vanishing. This suppression is achieved through the normalized $U(1)_X$ gauge coupling at the flavored-GUT scale, as specified by Eq.(\ref{d0aa}), together with the large anomaly coefficient $\delta^X_X$ in Eq.(\ref{SAo}), thereby dynamically realizing a qualified QCD axion within the flavored-GUT framework. Therefore, the axion solution to the strong CP problem provides a compelling motivation for SM gauge-coupling unification within this framework.

\section{Quark, Lepton, and Flavored-QCD axion}
\label{visu}
In the flavored-GUT framework, the  $SL(2,\mathbb{Z})\times U(1)_X\times U(1)_{B-L}$ plays a central role in organizing the fermion sector. The flavor structure is not introduced independently but is constrained by the requirement that anomalies cancel consistently over the fermion spectrum while maintaining gauge coupling unification (see Secs.-\ref{dgSM}, \ref{NGm}, and -\ref{gcu}). As a consequence, the framework predicts the QCD axion mass (see Eqs.(\ref{axide1}) and (\ref{axiMass2})). 
Imposing the anomaly cancellation conditions (including ${\cal A}_C={\cal A}_E=0$, ${\cal A}_L=-{\cal A}_Y/c_Y=0$, ${\cal A}_X=0$, and ${\cal A}_{B-L}={\cal A}_{\rm grav}=0$ discussed in Sec.\ref{dgSM}, together with $\delta^X_{B-L}=0$ in Eq.(\ref{SAo1})), as well as the gauge coupling unification condition Eq.(\ref{th0003})), places strong constraints on the allowed charge assignments and modular weights of quark and lepton fields. 
A representative viable charge assignment is presented in Table-\ref{reps_q} and -\ref{reps_l}. In this framework, Yukawa interactions are not arbitrary but are restricted to unit-magnitude complex coefficients, reflecting the underlying modular structure. Consequently, the flavor dynamics scale $M_X$ (see Eqs.(\ref{FN01}) and (\ref{gbm01m})) is closely connected to physical scales such as the axion mass scale through the PQ mechanism (see Eqs.(\ref{AFN1}) and (\ref{stcp})), and is determined by the SM gauge coupling unification.
By contrast, the $U(1)_{B-L}$ breaking scale associated with the seesaw mechanism can be determined jointly by the anomaly cancellation condition and gauge coupling unification.

\subsection{Modular-invariant Yukawa superpotentials for quark and lepton}
\begin{table}[h]
\centering
\caption{\label{reps_q} Representations of the SM quark and charged-lepton field s under $SL(2,\mathbb{Z})\times U(1)_{X}$ and modular weight $k_I$ with $h=3$. For quarks: $Q_i$ ($i=1,2,3$) are left-handed doublets, $(d^c,s^c,b^c)$ are right-handed down-type quarks, $(u^c,c^c,t^c)$ are right-handed up-type quarks. For leptons: $L_i$ ($i=e,\mu,\tau$) are left-handed doublets, $(e^c,\mu^c,\tau^c)$ are right-handed charged leptons\,\protect\footnote{Note that the modular weight can equivalently be expressed as $k_{\hat{\psi}}=\frac{h}{4}-\frac{1}{2}k_I$, see Eq.(\ref{cr02}).}.}
\label{reps_combined}
\begin{ruledtabular}
\begin{tabular}{cccccccccc}
Field & $Q_1$ & $Q_2$ & $Q_3$ & $d^c$ & $s^c$ & $b^c$ & $u^c$ & $c^c$ & $t^c$ \\
\hline
$k_I$ & $\frac{h}{2}-p-8$ & $\frac{h}{2}-p-4$ & $\frac{h}{2}-p$ & $\frac{h}{2}+p+8$ & $\frac{h}{2}+p+4$ & $\frac{h}{2}+p$ & $\frac{h}{2}+p+8$ & $\frac{h}{2}+p+4$ & $\frac{h}{2}+p$ \\
$U(1)_{X}$ & $-6$ & $-7$ & $-12$ & $-12$ & $18$ & $16$ & $-17$ & $16$ & $12$ \\
\end{tabular}
\begin{tabular}{ccccccc}
Field & $L_e$ & $L_\mu$ & $L_\tau$ & $e^c$ & $\mu^c$ & $\tau^c$ \\
\hline
$k_I$ & $\frac{h}{2}+3p+16$ & $\frac{h}{2}+3p+12$ & $\frac{h}{2}+3p+8$ & $\frac{h}{2}-3p-16$ & $\frac{h}{2}-3p-12$ & $\frac{h}{2}-3p-8$ \\
$U(1)_{X}$ & $12$ & $12$ & $13$ & $-35$ & $-24$ & $-8$ \\
\end{tabular}
\end{ruledtabular}
\end{table}
According to Table-\ref{reps_q}, the quark and charged-lepton Yukawa superpotential read
\begin{eqnarray}
 W_q &=&
 \Big[y_{t}\,t^cQ_{3}+y_{c}\Big(\frac{\tilde{\chi}}{M_X}\Big)^{9}c^cQ_2+y_{u}\Big(\frac{\chi}{M_X}\Big)^{23}u^cQ_1\nonumber\\
  &+&y_{t2}\Big(\frac{\tilde{\chi}}{M_X}\Big)^{5}Y^{(4)}_{{\bf1}}t^cQ_2+y_{t1}\Big(\frac{\tilde{\chi}}{M_X}\Big)^{6}Y^{(8)}_{\bf1}t^cQ_1+y_{c1}\Big(\frac{\tilde{\chi}}{M_X}\Big)^{10}Y^{(4)}_{\bf1}c^cQ_1\Big]H_u\nonumber\\
  &+& \Big[y_{b}\Big(\frac{\tilde{\chi}}{M_X}\Big)^{4}b^cQ_{3}+y_{b2}\Big(\frac{\tilde{\chi}}{M_X}\Big)^{9}Y^{(4)}_{\bf1}b^cQ_{2}+y_{b1}\Big(\frac{\tilde{\chi}}{M_X}\Big)^{10}Y^{(8)}_{\bf1}b^cQ_{1}\nonumber\\
  &+&y_{s}\Big(\frac{\tilde{\chi}}{M_X}\Big)^{11}s^cQ_{2}+y_{s1}\Big(\frac{\tilde{\chi}}{M_X}\Big)^{12}Y^{(4)}_{\bf1}s^cQ_{1}+y_{d}\Big(\frac{\chi}{M_X}\Big)^{18}d^cQ_{1}\Big]H_d+....\,,
 \label{lagrangian_q}
 \end{eqnarray}
\begin{eqnarray}
 W_{\ell} &=&
  \Big[y_{\tau}\Big(\frac{\tilde{\chi}}{M_X}\Big)^{5}\tau^cL_{\tau}+ y_{\mu}\Big(\frac{\chi}{M_X}\Big)^{12}\mu^cL_{\mu}+y_{e}\Big(\frac{\chi}{M_X}\Big)^{23}e^cL_{e}\nonumber\\
  &+&y_{e2}\Big(\frac{\chi}{M_X}\Big)^{23}Y^{(4)}_{\bf1}e^cL_\mu+y_{e3}\Big(\frac{\chi}{M_X}\Big)^{22}Y^{(8)}_{\bf1}e^cL_\tau+y_{\mu3}\Big(\frac{\chi}{M_X}\Big)^{11}Y^{(4)}_{\bf1}\mu^cL_\tau\Big]H_d+...\,.
 \label{lagrangian_l0}
 \end{eqnarray}
where the flavor dynamics scale $M_X$ is identified with the mass scale of $U(1)_X$ gauge boson that is integrated out, see Eq.(\ref{FN01}).
Here all Yukawa coefficients $y_i$ are complex numbers with unit-magnitude, and dots represent higher-order contributions compactly expressed as $\sum^{\infty}_{n=1}(\frac{\chi\tilde{\chi}}{\Lambda^{2}})^n\times\text{\it leading terms}$.
These corrections modify the effective Yukawa coefficients $y_i$ (similarly, $\beta^{(n)}_{i\ell}$, $\gamma^{(n)}_{ij}$ in Eq.(\ref{modu6})), constrained by
\begin{eqnarray}
1-\frac{\Delta^2_\chi}{1-\Delta^2_\chi}\leq|y_i|\leq1+\frac{\Delta^2_\chi}{1-\Delta^2_\chi}\qquad\text{with}~\Delta_\chi\equiv \frac{v_\chi}{\sqrt{2}\,M_X}\,.
  \label{AFN1}
\end{eqnarray}
According to the canonically normalized fields (see Eq.(\ref{cn0})), the Yukawa coefficients transform as
{\begin{eqnarray}
&&y_{c1}\rightarrow (2{\rm Im}\,\tau)^{-2}\,y_{c1}\,,\quad y_{t1}\rightarrow (2{\rm Im}\,\tau)^{-4}\,y_{t1}\,,\quad y_{t2}\rightarrow (2{\rm Im}\,\tau)^{-2}\,y_{t2}\,,\nonumber\\
&&y_{s1}\rightarrow (2{\rm Im}\,\tau)^{-2}\,y_{s1}\,,\quad y_{b1}\rightarrow (2{\rm Im}\,\tau)^{-4}\,y_{b1}\,,\quad y_{b2}\rightarrow (2{\rm Im}\,\tau)^{-2}\,y_{b2}\,,\nonumber\\
&&y_{e3}\rightarrow (2{\rm Im}\,\tau)^{-2}\,y_{e3}\,,\quad y_{\mu3}\rightarrow (2{\rm Im}\,\tau)^{-4}\,y_{\mu3}\,,\quad y_{\mu1}\rightarrow (2{\rm Im}\,\tau)^{-2}\,y_{\mu1}\,,
\label{nYc0}
 \end{eqnarray}}
 while $y_{u,c,t}$, $y_{d,s,b}$, and $y_{e,\mu,\tau}$ remain unchanged. The modular forms of weights 4 and 8 under $SL(2,\mathbb{Z})$ is given by $Y^{(4)}_{\bf 1}=Y^2_1+2Y_2Y_3=E_4$ and $Y^{(8)}_{\bf 1}=(Y^2_1+2Y_2Y_3)^2=E_8=E^2_4$ with Eq.(\ref{mfc}). Modular forms of even weight for $SL(2,\mathbb{Z})$ can be expressed as polynomials in the Eisenstein series $E_4$ and $E_6$\,\cite{Feruglio:2017spp,Petcov:2024vph}, see Eq.(\ref{lagrangian_l}). 
 
Under chiral rotation of the quark fields, the QCD anomaly term reduces to
 \begin{eqnarray}
 {\cal L}_\vartheta=\Big(\vartheta_{\rm eff}+\frac{A_X}{F_a}\Big)\frac{\alpha_3}{16\pi}G^{a\mu\nu}\tilde{G}^a_{\mu\nu}\qquad\text{with}~F_a=\frac{f_A}{\delta^G_X}\,,
 \label{stcp}
 \end{eqnarray}
where $\alpha_3=g^2_3/4\pi$, $F_a$ is the axion decay constant with $f_A$ Eq.(\ref{NGboson}), and $\vartheta_{\rm eff}$ is the effective strong CP phase of Eq.(\ref{qcdph}) with the vanishing modular anomaly condition, ${\cal A}_C=0$ (see Eq.(\ref{cr04})). At low energies $A_X$ will get a VEV, $\langle A_X\rangle=-F_a\vartheta_{\rm eff}$, eliminating the constant $\vartheta_{\rm eff}$ term. The QCD axion then is the excitation of the $A_X$ field, $a_X=A_X-\langle A_X\rangle$. The quark quantum numbers in Table-\ref{reps_q} yield the color anomaly coefficient for $U(1)_X \times [SU(3)_C]^2$ (defined in Eq.(\ref{SAo})) as
 \begin{eqnarray}
  \delta^G_X=-17\,,
 \label{dGi}
 \end{eqnarray}
 determining the domain-wall number $N_{\rm DW}= |\delta_X^G|$. To avoid cosmological domain walls, either $N_{\text{DW}} = 1$ or the PQ transition must occur during/before inflation for $N_{\text{DW}} > 1$.

Below the $U(1)_X$ symmetry breaking scale, the effective interactions of QCD axion with the weak and hypercharge gauge bosons and with the photon are expressed through the chiral rotation of Eq.(\ref{X-tr}).
The electromagnetic anomaly coefficient $E$ of $U(1)_{X}\times[U(1)_{EM}]^2$ is defined by $E=2\sum_{\psi_f} X_{\psi_f}(Q^{\rm em}_{\psi_f})^2$ where $Q^{\rm em}_{\psi_f}$ is the electric charge of the field $\psi_f$. For the $U(1)_X$ charges (see Table-\ref{reps_q} and -\ref{reps_l}), this evaluates to
  \begin{eqnarray}
  E&=&\delta^W_X+\frac{\delta^Y_X}{c_Y}=-\frac{298}{3}\,, 
  \label{eano}
 \end{eqnarray}
where the anomaly coefficients $\delta^W_X=-38$, $\delta^Y_X=-80$ with $c_Y=30/23$ are obtained (see Eq.(\ref{SAo})).
The physical quantities of QCD axion, such as axion mass $m_a$ and axion-photon coupling $g_{a\gamma\gamma}$, depend on the ratio of electromagnetic anomaly coefficient $E$ to the color anomaly coefficient $\delta^G_X$ (see Fig.\ref{Fig4}).
\begin{table}[h]
\caption{\label{reps_l} Representations of the lepton fields under $SL(2,\mathbb{Z})\times U(1)_{X}$ and modular weight $k_I$ with $h=3$, where $N^c_j$ ($j=1,2,3$) are the right-handed neutrinos.}
\begin{ruledtabular}
\begin{tabular}{cccc}
Field &$N^c_{1}$ & $N^c_{2}$ & $N^c_3$\\
\hline
$k_I$&$\frac{h}{2}-3p-244$&$\frac{h}{2}-3p-244$&$\frac{h}{2}-3p+452$\\
$U(1)_{X}$&$0$&$0$&$2$
\end{tabular}
\end{ruledtabular}
\end{table}
According to Table-\ref{reps_q} and \ref{reps_l}, the neutrino Yukawa superpotential reads
\begin{eqnarray}
 W_{\nu} &=&\Big[\beta_{1e}Y^{(228)}_{\bf1}N^c_1L_{e}\Big(\frac{\tilde{\chi}}{M_X}\Big)^{12}+\beta_{1\mu}Y^{(232)}_{\bf1}N^c_1L_{\mu}\Big(\frac{\tilde{\chi}}{M_X}\Big)^{12}+\beta_{1\tau}Y^{(236)}_{\bf1}N^c_1L_{\tau}\Big(\frac{\tilde{\chi}}{M_X}\Big)^{13}\nonumber\\
 &&+\beta_{2e}Y^{(228)}_{\bf1}N^c_2L_{e}\Big(\frac{\tilde{\chi}}{M_X}\Big)^{12}+\beta_{2\mu}Y^{(232)}_{\bf1}N^c_2L_{\mu}\Big(\frac{\tilde{\chi}}{M_X}\Big)^{12}+\beta_{2\tau}Y^{(236)}_{\bf1}N^c_2L_{\tau}\Big(\frac{\tilde{\chi}}{M_X}\Big)^{13}\Big]H_u\nonumber
 \end{eqnarray}
\begin{eqnarray}
  &+&\frac{1}{2}\Big[\gamma_{11} Y^{(6p+488)}_{{\bf1}}N^c_1N^c_1+\gamma_{12} Y^{(6p+488)}_{\bf1}N^c_1N^c_2+\gamma_{22} Y^{(6p+488)}_{\bf1}N^c_2N^c_2\nonumber\\
  &&\quad+\gamma_{13} Y^{(6p-208)}_{{\bf1}}N^c_1N^c_3\Big(\frac{\tilde{\chi}}{M_X}\Big)^{2}+\gamma_{23} Y^{(6p-208)}_{{\bf1}}N^c_2N^c_3\Big(\frac{\tilde{\chi}}{M_X}\Big)^{2}\nonumber\\
  &&\quad+\gamma_{33}Y^{(6p-904)}_{\bf1}N^c_3N^c_3\Big(\frac{\tilde{\chi}}{M_X}\Big)^{4}\Big]\rho
  +...
\,.
 \label{lagrangian_l}
 \end{eqnarray}
Recall that, analogous to the quark sector, all Yukawa coefficients are effectively determined by Eq.(\ref{AFN1}), after accounting for the contributions of all higher-dimensional operators induced by $\chi\tilde{\chi}$. 
According to the canonically normalized fields (see Eq.(\ref{cn0})), the Yukawa coefficients transform as
\begin{eqnarray}
&&\beta_{ie}\rightarrow (2{\rm Im}\,\tau)^{-114}\,\beta_{ie}\,,\,\quad\qquad \beta_{i\mu}\rightarrow (2{\rm Im}\,\tau)^{-116}\,\beta_{i\mu}\,, \qquad\quad \beta_{i\tau}\rightarrow (2{\rm Im}\,\tau)^{-118}\,\beta_{i\tau}\,,\nonumber\\
&&\gamma_{11}\rightarrow (2{\rm Im}\,\tau)^{-3p-244}\,\gamma_{11}\,,\quad \gamma_{12}\rightarrow (2{\rm Im}\,\tau)^{-3p-244}\,\gamma_{12}\,, \quad\gamma_{22}\rightarrow (2{\rm Im}\,\tau)^{-3p-244}\,\gamma_{22}\,,\nonumber\\
&&\gamma_{13}\rightarrow (2{\rm Im}\,\tau)^{-3p+104}\,\gamma_{13}\,,\,\,\quad\gamma_{23}\rightarrow (2{\rm Im}\,\tau)^{-3p+104}\,\gamma_{23}\qquad\gamma_{33}\rightarrow (2{\rm Im}\,\tau)^{-3p+452}\,\gamma_{33}\,.
\label{nYc1}
 \end{eqnarray}
For the neutrino operators, the effects of higher modular-weight operators are systematically absorbed by the relation $E^2_6-E^3_4\simeq-E^3_4$ at $\tau\approx i$ in Eq.(\ref{Dta_01}) into a reduced set of effective higher-order Yukawa coefficients entering the seesaw formula, see Eq.(\ref{Ynu1eff}).
 
\subsection{Mass, Mixing, and Flavored-QCD axion}
\label{vis_ql}
The QCD axion decay constant is predicted in a manner consistent with exact SM gauge coupling unification, as shown in Eq.(\ref{axide1}). By contrast, the seesaw scale ($U(1)_{B-L}$ breaking scale) is not fixed  solely by SM gauge coupling unification, since it depends on the value of the gauge coupling $g_{B-L}(\Lambda_{\rm fGUT})$. Instead, its scale is constrained  through the combined requirements of gauge coupling unification and anomaly cancellation conditions.
The undetermined modular weight $p$ appearing in Eq.(\ref{lagrangian_l}) (see Table-\ref{reps_q} and -\ref{reps_l}) is likewise constrained by gauge coupling unification and anomaly cancellation, and determines  the seesaw scale consistently with the neutrino oscillation data (see Table-\ref{exp_nu}). For instance, among the possible modular weight $p=\{-10,-8,-6,-4,-2,0,2,4,6,8\}$ allowed by anomaly cancellation (consequently the heavy neutrino sector is reduced to Eq.(\ref{MR1})), the gauge coupling unification setup further restricts the viable values to $p=\{-10,-8,-6,-4,-2,0,2\}$. The corresponding VEV of scalar $\rho$ is approximately $5\times10^{6}$ GeV for $p=-10$, and $5\times10^{11}$ GeV for $p=-2$. 

 At energies below the electroweak scale when $H_{u(d)}$ acquire non-zero VEVs all quarks and leptons obtain masses.
The relevant quark and lepton interactions are given from Eqs.(\ref{lagrangian_q},\ref{lagrangian_l0},\ref{lagrangian_l}) by 
 \begin{eqnarray}
  -{\cal L} &\supset&
  \overline{q^{u}_{R}}\,\mathcal{M}_{u}\,q^{u}_{L}+\overline{q^{d}_{R}}\,\mathcal{M}_{d}\,q^{d}_{L}+\frac{g}{\sqrt{2}}W^+_\mu\overline{q^u_L}\gamma^\mu\,q^d_L \nonumber\\
  &+&\overline{\ell_{R}}\,{\cal M}_{\ell}\,\ell_{L}+ \frac{1}{2} \begin{pmatrix} \overline{\nu^c_L} & \overline{N_R} \end{pmatrix} \begin{pmatrix} 0 & m^T_{D}  \\ m_{D} & ~~~ M_R  \end{pmatrix} \begin{pmatrix} \nu_L \\ N^c_R  \end{pmatrix}+\frac{g}{\sqrt{2}}W^-_\mu\overline{\ell_L}\gamma^\mu\,\nu_L+\text{h.c.}\,,
  \label{AxionLag2}
 \end{eqnarray}
where $g$ is the $SU(2)_L$ coupling constant, $q^{u}=(u,c,t)$, $q^{d}=(d,s,b)$, $\ell=(e, \mu, \tau)$, $\nu=(\nu_e,\nu_\mu,\nu_\tau)$, and $N=(N_1,N_2, N_3)$. $M_R$ contains a VEV of $\chi $ in Eq.(\ref{NGboson}). The explicit forms of $\mathcal{M}_{u,d,l}$ will be given
later.
The above Lagrangian of the fermions, including their kinetic terms, should be invariant under $U(1)_X$:
 \begin{eqnarray}
  \psi_f\rightarrow e^{iX_{\psi_f}\frac{\gamma_5}{2}\beta}\psi_f\,,\quad t=\text{invariant}\,,\quad N\rightarrow e^{i\frac{\gamma_5}{2}\beta}N
 \label{X-tr}
 \end{eqnarray}
where $\psi_f=\{u,c,d,s,b,e,\mu,\tau,\nu\}$ and $\beta$ is a transformation constant parameter.

With the VEV of Eq.(\ref{super_02}) the mass matrices ${\cal M}_{u}$, ${\cal M}_{d}$, and  ${\cal M}_{\ell}$ for up- and down-type quarks and charged-leptons are described in terms of $\Delta_\chi$ and  modular forms $Y^{(4)}_{\bf 1}$ and $Y^{(8)}_{\bf 1}$: 
  \begin{eqnarray}
  &{\cal M}_{u}=C^u_R{\left(\begin{array}{ccc}
 y_u\Delta^{23}_\chi &  0 &  0 \\
 y_{c1}(2{\rm Im}\,\tau)^{-2}\Delta^{10}_\chi\, Y^{(4)}_{\bf1}  &  y_c\Delta^{9}_\chi &  0   \\
 y_{t1}(2{\rm Im}\,\tau)^{-4}\Delta_\chi^{6}\,Y^{(8)}_{\bf 1}  &  y_{t2}(2{\rm Im}\,\tau)^{-2}\Delta^{5}_\chi\,Y^{(4)}_{\bf1}  & y_t
 \end{array}\right)}C^u_L\,v_u\,,
 \label{Ch2}
  \end{eqnarray}
 \begin{eqnarray}
 &{\cal M}_{d}=C^d_R{\left(\begin{array}{ccc}
 y_{d}\,\Delta^{18}_\chi & 0 & 0 \\
 y_{s1}(2{\rm Im}\,\tau)^{-2}\Delta^{12}_\chi\,Y^{(4)}_{\bf1} &  y_{s}\,\Delta^{11}_\chi  &  0   \\
 y_{b1}(2{\rm Im}\,\tau)^{-4}\Delta_\chi^{10}\,Y^{(8)}_{\bf 1} &  y_{b2}(2{\rm Im}\,\tau)^{-2}\Delta^{9}_\chi\,Y^{(4)}_{\bf1}  & y_{b}\,\Delta^{4}_\chi
 \end{array}\right)}C^d_L\,v_d\,,  
 \label{Ch1}
 \end{eqnarray}
  \begin{eqnarray}
 {\cal M}_{\ell}&=& C^\ell_R{\left(\begin{array}{ccc}
 y_{e}\,\Delta^{23}_\chi & y_{e2}Y^{(4)}_{\bf 1}\Delta^{23}_\chi(2{\rm Im}\,\tau)^{-2} &   y_{e3}Y^{(8)}_{\bf 1}\Delta^{22}_\chi(2{\rm Im}\,\tau)^{-4}  \\
0 &  y_{\mu}\,\Delta^{12}_\chi & y_{\mu3}Y^{(4)}_{\bf 1}\Delta^{11}_\chi(2{\rm Im}\,\tau)^{-2} \\
0 &   0 &  y_{\tau}\,\Delta^{5}_\chi
 \end{array}\right)}C^\ell_L\,v_d\,,
 \label{ChL1}
 \end{eqnarray}
where $v_{d}\equiv\langle H_{d}\rangle=v\cos\beta/\sqrt{2}$, $v_{u}\equiv\langle H_{u}\rangle =v\sin\beta/\sqrt{2}$ with $v\simeq246$ GeV, and
\begin{eqnarray}
  &&C^u_R={\rm diag}(e^{i29\frac{A_X}{f_A}}, e^{-i4\frac{A_X}{f_A}},1)\,,\qquad\qquad\quad C^u_L={\rm diag}(e^{-i6\frac{A_X}{f_A}}, e^{-i5\frac{A_X}{f_A}},1)\,,\nonumber\\
  &&C^d_R={\rm diag}(e^{i24\frac{A_X}{f_A}}, e^{-i6\frac{A_X}{f_A}}, e^{-i4\frac{A_X}{f_A}})\,,\qquad~\, C^d_L={\rm diag}(e^{-i6\frac{A_X}{f_A}}, e^{-i5\frac{A_X}{f_A}}, 1)\,,\nonumber\\
   &&C^\ell_R={\rm diag}(e^{i27\frac{A_X}{f_A}}, e^{i16\frac{A_X}{f_A}}, 1)\,,\qquad\qquad\quad\, C^\ell_L={\rm diag}(e^{-i4\frac{A_X}{f_A}}, e^{-i4\frac{A_X}{f_A}}, e^{-i5\frac{A_X}{f_A}}) \,.
\label{axi01}
 \end{eqnarray}

Similarly, the heavy Majorana neutrino mass matrix $M_{R}$ and the Dirac neutrino mass matrix $m_D$ in the Lagrangian\,(\ref{AxionLag2}) are expressed in terms of $\Delta_\chi$ and the modular forms $Y^{(6p+488)}_{\bf 1}$, $Y^{(228)}_{\bf 1}$, $Y^{(232)}_{\bf 1}$, and $Y^{(236)}_{\bf 1}$ as,
 \begin{eqnarray}
 M_{R}&=& (2{\rm Im}\,\tau)^{-3p-244}{\left(\begin{array}{cc}
\gamma_{11}\,Y^{(6p+488)}_{{\bf 1}} & \gamma_{12}\,Y^{(6p+488)}_{{\bf 1}}  \\
\gamma_{12}\,Y^{(6p+488)}_{{\bf 1}} & \gamma_{22}\,Y^{(6p+488)}_{{\bf 1}} 
 \end{array}\right)}\langle\rho\rangle\,,
 \label{MR1}
  \end{eqnarray}
  \begin{eqnarray}
m_{D}&=&{\left(\begin{array}{ccc}
 \beta_{1e}\, Y^{(228)}_{{\bf1}} &  \beta_{1\mu}\,Y^{(232)}_{\bf1}  (2{\rm Im}\,\tau)^{-2}  &  \beta_{1\tau}\,Y^{(236)}_{\bf1}  (2{\rm Im}\,\tau)^{-4}\Delta_\chi e^{-i\frac{A_X}{f_A}}  \\
 \beta_{2e}\,Y^{(228)}_{{\bf1}}   &  \beta_{2\mu}\,Y^{(232)}_{{\bf1}} (2{\rm Im}\,\tau)^{-2} &  \beta_{2\tau}\,Y^{(236)}_{\bf1}  (2{\rm Im}\,\tau)^{-4}\Delta_\chi e^{-i\frac{A_X}{f_A}}
 \end{array}\right)}\nonumber\\
 &&\times(2{\rm Im}\,\tau)^{-114}\Delta^{12}_\chi e^{-i12\frac{A_X}{f_A}} v_u\,.
 \label{Ynu1}
 \end{eqnarray}
Here, the modular forms with weights 228, 232, 236, and 476 (for $p=-2$ in Eq.(\ref{MR1})) can be expanded as
\begin{eqnarray}
  &&\beta_{ie}Y^{(228)}_{\bf 1}=\sum^{19}_{n=0}\beta^{(n)}_{ie} E^{57-3n}_4(E^2_6-E^3_4)^n\,,\quad \beta_{i\mu}Y^{(232)}_{\bf 1}=\sum^{19}_{n=0}\beta^{(n)}_{i\mu} E^{58-3n}_4(E^2_6-E^3_4)^n\,,\nonumber\\
  &&\beta_{i\tau}Y^{(236)}_{\bf 1}=\sum^{19}_{n=0}\beta^{(n)}_{i\tau} E^{59-3n}_4(E^2_6-E^3_4)^n\,,\quad \gamma_{ij}Y^{(476)}_{\bf 1}=\sum^{39}_{n=0}\gamma^{(n)}_{ij} E^{119-3n}_4(E^2_6-E^3_4)^n\,,
\label{modu6}
 \end{eqnarray}
where the Yukawa coefficients $\beta^{(n)}_{i\ell}, \gamma^{(n)}_{ij}$ parametrize the expansion of the modular form in the basis of $E_4$ and $E_6$. At $\langle\tau\rangle\approx i$ in Eq.(\ref{Dta_01}), the modular-form dependence of $\beta_{ie}Y^{(228)}_{{\bf1}}$, $\beta_{i\mu}Y^{(232)}_{{\bf1}}$, $\beta_{i\tau}Y^{(236)}_{{\bf1}}$, and $\gamma_{ij}Y^{(476)}_{{\bf1}}$ can be reduced\,\footnote{At the specific point $\tau=i$, the combination $E^2_6-E^3_4$ reduces to $-E^3_4$. Consequently, the higher-order terms $(E^2_6-E^3_4)^n$ for $n\geq1$ are not independent because $(E^2_6-E^3_4)|_{\tau=i}=-[E_4(i)]^3$. This allows the modular forms to be approximated by powers of the basic modular form $(Y^2_1+2Y_2Y_3)$ with exponents 57,58,59, and 119, respectively.} to effective Yukawa coefficients and the Majorana couplings, such that they are approximately represented as 
  \begin{eqnarray}
    \beta^{\rm eff}_{ie}(Y^2_1+2Y_2Y_3)^{57}\,,\quad \beta^{\rm eff}_{i\mu}(Y^2_1+2Y_2Y_3)^{58}\,,\quad \beta^{\rm eff}_{i\tau}(Y^2_1+2Y_2Y_3)^{59}\,,\quad \gamma^{\rm eff}_{ij}(Y^2_1+2Y_2Y_3)^{119}\,,
 \label{Ynu1eff}
 \end{eqnarray}
where $|\beta^{\rm eff}_{i\ell}|$, $|\gamma^{\rm eff}_{ij}|\sim{\cal O}(1-10)$ are determined by fitting to the neutrino oscillation data, see Eq.(\ref{neu01}) and (\ref{neu010}).
The different corrections appearing in Eq.(\ref{Ynu1}), originating from the canonical normalization of the matter fields in Eq.(\ref{cn0}), ensure the observed atmospheric and solar neutrino mass-squared differences, $\Delta m^2_{\rm Atm}$ and $\Delta m^2_{\rm Sol}$, and consequently determine whether the neutrino mass ordering is normal or inverted, see Sec.\ref{num_nu}.

As a representative example, taking $p=-2$ (see Eq.(\ref{MR1})) and $g_{B-L}(\Lambda_{\rm fGUT})\sim g_X(\Lambda_{\rm fGUT})$, the requirement of SM gauge coupling unification reproducing Eq.(\ref{th0301}) leads to
 \begin{eqnarray}
  &&M_X=2.228\times10^{15}\,{\rm GeV}\,,~ F_a=1.625\times10^{14}\,{\rm GeV}\,,~ M_{B-L}=3.897\times10^{11}\,{\rm GeV}\,,\nonumber\\
  &&g_X(\Lambda_{\rm fGUT})=0.292\,,~ g_{B-L}(\Lambda_{\rm fGUT})=0.28\,,~\alpha^{-1}_{\rm fGUT}=27.355\,,~ M_{\rm SUSY}=12\,{\rm TeV}\,,
  \label{pheno01}
 \end{eqnarray}
 together with $\langle\chi\rangle=1.381\times10^{15}$ GeV and $\langle\rho\rangle=4.921\times10^{11}$ GeV.

\subsubsection{Quark and charged-lepton masses, mixing, and QCD axion interactions}
The quark mass matrices ${\cal M}_{u}$ of Eq.(\ref{Ch2}) and ${\cal M}_{d}$ of Eq.(\ref{Ch1}) generate the up- and down-type quark masses:
 $\hat{\mathcal{M}}_{u}=V^{u}_{R}\,{\cal M}_{u}\,V^{u\dag}_{L}
 ={\rm diag}(m_{u},m_{c},m_{t})$ and  $\hat{\mathcal{M}}_{d}=V^{d}_{R}\,{\cal M}_{d}\,V^{d\dag}_{L}
 ={\rm diag}(m_{d},m_{s},m_{b})$ with the approximate relations
 \begin{eqnarray}
&&m_u\simeq|y_u|\Delta^{23}_\chi\,v_u\,,\quad m_c\simeq|y_c|\Delta^{9}_\chi\,v_u\,,\quad m_t\simeq|y_t|\,v_u\,,\nonumber\\
&&m_d\simeq|y_d|\Delta^{18}_\chi\,v_d\,,\quad m_s\simeq|y_s|\Delta^{11}_\chi\,v_d\,,\quad m_b\simeq|y_b|\Delta^{4}_\chi\,v_d\,.
 \label{qkm01}
 \end{eqnarray}
 The physical structure of the up- and down-type quark Lagrangian should match up with the empirical results calculated at $90\%$ C.L. from the Particle Data Group (PDG)\,\cite{PDG}:
 \begin{eqnarray}
 m_u&=&2.16\pm0.07\,{\rm MeV}\,,\quad m_c=1.2730\pm0.0046\,{\rm GeV}\,,\quad m_t=172.56\pm0.31\,{\rm GeV}\,,\nonumber\\
 m_d&=&4.70\pm0.07\,{\rm MeV}\,,\quad m_s=93.5\pm0.8\,{\rm MeV}\,,\qquad\quad~m_b=4.183\pm0.007\,{\rm GeV}\,, 
\label{qumas}
 \end{eqnarray}
where $t$-quark mass is the pole mass, $c$- and $b$-quark masses are the running masses in the $\overline{\rm MS}$ scheme, and the light $u$-, $d$-, $s$-quark masses are the current quark masses in the $\overline{\rm MS}$ scheme at the momentum scale $\mu\approx2$ GeV. Below the scale of spontaneous $SU(2)_L\times U(1)_Y$ gauge symmetry breaking, the running masses of $c$- and $b$-quark receive corrections from QCD and QED loops\,\cite{PDG}. The top quark mass at scales below the pole mass is unphysical since the $t$-quark decouples at its scale, and its mass is determined more directly by experiments\,\cite{PDG}.

Diagonalizing the Hermitian matrices ${\cal M}_f^\dagger {\cal M}_f$ and ${\cal M}_f {\cal M}_f^\dagger$ ($f=u,d$) determines the left- and right-handed mixing matrices $V_L^f$ and $V_R^f$, respectively\,\cite{Ahn:2011yj}. Given the structure of the up- and down-type quark mass matrices in Eqs.(\ref{Ch2}) and (\ref{Ch1}), together with their associated quantum numbers\,\footnote{See the 12- and 22-components and the 31-and 32-components in Eqs.(\ref{Ch2}) and (\ref{Ch1}).}, the left-handed quark mixing matrices $V_L^f$ ($f=u,d$) take the common parametric form 
 \begin{eqnarray}
 V^f_{L}=\tilde{C}_f{\left(\begin{array}{ccc}
 1-\frac{1}{2}\lambda^2_f & -\lambda_f\,e^{i\varphi_f} & \lambda^3_f(A_fe^{i\varphi_f} -B_f)  \\
\lambda_f\,e^{-i\varphi_f}  & 1-\frac{1}{2}\lambda^2_f & -A_f\lambda^2_f   \\
 B_f\lambda^3_f & A_f\lambda^2_f  & 1
 \end{array}\right)}\tilde{K}_f+{\cal O}(\lambda^4_f)\,,
 \label{ckm01}
 \end{eqnarray}
where $\varphi_f=2\alpha^f_3-\alpha^f_2$, $\tilde{C}_f={\rm diag}(e^{i(\alpha^f_2-\alpha^f_1-\alpha^f_3)}, e^{i(\alpha^f_3-\alpha^f_1)}, e^{i(\alpha^f_2-\alpha^f_1)})$, and $\tilde{K}_f={\rm diag}(e^{i(\alpha^f_1-2\alpha^f_2)}, 1, e^{i2\alpha^f_1})$. Here we define
 \begin{eqnarray}
&&A_u\lambda^2_u=\Big|\frac{y_{t2}}{y_t}Y^{(4)}_{\bf1}\Big|\Delta^5_\chi(2\,{\rm Im}\,\tau)^{-2}\,,\qquad A_d\lambda^2_d=\Big|\frac{y_{b2}}{y_b}Y^{(4)}_{\bf1}\Big|\Delta^5_\chi(2\,{\rm Im}\,\tau)^{-2}\,,\nonumber\\
&&B_u\lambda^3_u=\Big|\frac{y_{t1}}{y_t}Y^{(8)}_{\bf1}\Big|\Delta^6_\chi(2\,{\rm Im}\,\tau)^{-4}\,,\qquad B_d\lambda^3_d=\Big|\frac{y_{b1}}{y_b}Y^{(8)}_{\bf1}\Big|\Delta^6_\chi(2\,{\rm Im}\,\tau)^{-4}\,,\nonumber\\
&&\,\quad\lambda_u=\Big|\frac{y_{c1}}{y_c}Y^{(4)}_{\bf1}\Big|\Delta_\chi(2\,{\rm Im}\,\tau)^{-2}\,,\qquad\,\quad\lambda_d=\Big|\frac{y_{s1}}{y_s}Y^{(4)}_{\bf1}\Big|\Delta_\chi(2\,{\rm Im}\,\tau)^{-2}\,.
 \label{ckm001}
 \end{eqnarray}
The phases are approximately given by
 \begin{eqnarray}
&&\alpha^u_1\simeq\frac{1}{2}\arg(y^\ast_{t2}Y^{(4)\ast}_{\bf 1})\,,\qquad\qquad\qquad\qquad \alpha^d_1\simeq\frac{1}{2}\arg(y^\ast_{b2}Y^{(4)\ast}_{\bf 1})\,,\nonumber\\
&&\alpha^u_2\simeq\frac{1}{2}\arg(y^\ast_{t1}Y^{(8)\ast}_{\bf 1})-\frac{1}{2}\alpha^u_1\,,\qquad\qquad\quad \alpha^d_2\simeq\frac{1}{2}\arg(y^\ast_{b2}Y^{(8)\ast}_{\bf 1})-\frac{1}{2}\alpha^d_1\,,\nonumber\\
&&\alpha^u_3\simeq\frac{1}{2}\arg(y^\ast_{c1}y_cY^{(4)\ast}_{\bf 1})+\frac{1}{2}(\alpha^u_1-\alpha^u_2)\,,\quad \alpha^d_3\simeq\frac{1}{2}\arg(y^\ast_{s1}y_sY^{(4)\ast}_{\bf 1})+\frac{1}{2}(\alpha^d_1-\alpha^d_2)\,.
 \label{ckm002}
 \end{eqnarray}
 Using the quark fields redefinitions, the unphysical phases can be absorbed into the quark fields, such that the CKM (Cabibbo-Kobayashi-Maskawa) matrix $V_{\rm CKM}=V_L^u V_L^{d\dagger}$ takes the Wolfenstein parametrization up to ${\cal O}(\lambda^4)$\,\cite{Wolfenstein:1983yz} (with high precision\,\cite{Ahn:2011fg})
 \begin{eqnarray}
 V_{\rm CKM}={\left(\begin{array}{ccc}
 1-\frac{1}{2}\lambda^2 & \lambda & A_q\lambda^3(\rho-i\eta)  \\
-\lambda & 1-\frac{1}{2}\lambda^2 & A_q\lambda^2   \\
 A_q\lambda^3(1-\rho-i\eta) & -A_q\lambda^2  & 1
 \end{array}\right)}+{\cal O}(\lambda^4)\,.
 \label{ckm0}
 \end{eqnarray}
where $\lambda=0.22504^{+0.00061}_{-0.00065}$, $A_q=0.821^{+0.015}_{-0.027}$, $\bar{\rho}=\rho/(1-\lambda^2/2)=0.156^{+0.031}_{-0.014}$, and $\bar{\eta}=\eta/(1-\lambda^2/2)=0.356^{+0.019}_{-0.021}$ with $3\sigma$ errors\,\cite{ckm}.
The quark masses and mixing parameters must be matched to the empirical values provided in Eqs.(\ref{ckmmixing}) and (\ref{qumas}).
 The current best-fit values of the CKM mixing angles in the standard parameterization\,\cite{Chau:1984fp} read in the $3\sigma$ range\,\cite{ckm}
 \begin{eqnarray}
  \theta^q_{23}[^\circ]=2.384^{+0.041}_{-0.076}\,,~\theta^q_{13}[^\circ]=0.214^{+0.011}_{-0.010}\,,~\theta^q_{12}[^\circ]=13.005^{+0.036}_{-0.038}\,,~\delta^q_{CP}[^\circ]=66.2^{+2.1}_{-4.6}\,.
 \label{ckmmixing}
 \end{eqnarray} 
 
 The charged-lepton mass matrix ${\cal M}_\ell$ in Eq.(\ref{ChL1}) generates the charged-lepton mass eigenvalues through $\hat{\mathcal{M}}_{\ell}=V^{\ell}_{R}\,{\cal M}_{\ell}\,V^{\ell\dag}_{L}={\rm diag}(m_{e},m_{\mu},m_{\tau})$, with the approximate expressions
 \begin{eqnarray}
 &m_{e}\simeq|y_{e}|\,\Delta^{23}_{\chi}\,v_d\,,\qquad m_{\mu}\simeq|y_{\mu}|\,\Delta^{12}_{\chi}\,v_d\,,\qquad
 m_{\tau}\simeq|y_{\tau}|\,\Delta^{5}_{\chi}\,v_d\,.
 \label{cLep1}
 \end{eqnarray}
 These should reproduce the experimentally measured PDG values\,\cite{PDG}, 
 \begin{eqnarray}
 &m_{e}=0.511\,{\rm MeV}\,,\qquad m_{\mu}=105.658\,{\rm MeV}\,,\qquad
 m_{\tau}=1776.93\pm0.09\,{\rm MeV}\,.
 \label{clm01}
 \end{eqnarray}
The left-handed charged-lepton mixing matrix $V^\ell_L$ enters the PMNS (Pontecorvo-Maki-Nakagawa-Sakata) lepton mixing matrix defined in Eq.(\ref{PMNS1}). In the present framework, it is approximately given by $V^\ell_L={\bf I}+{\cal O}(\lambda^4)$, as shown in Eq.(\ref{VlI}), implying negligible charged-lepton corrections to lepton mixing.

After diagonalizing the quark mass matrices in Eqs.(\ref{Ch2},\ref{Ch1}), the leading flavored-QCD axion interactions with quarks, up to $\mathcal{O}(\lambda^4)$, are given by\,\footnote{The right-handed quark and charged-lepton mixing matrices, $V^{d(u)}_R$ and $V^\ell_R$, which also contribute to the flavored-QCD axion couplings to up-type quarks, down-type quarks, and charged leptons, are determined by Eq.(\ref{qrma}) for $V^{d(u)}_R$ and Eq.(\ref{cLep2}) for $V^\ell_R$.} 
\begin{eqnarray}
-{\cal L}^{aq}&\simeq&\frac{\partial_\mu a_X}{2f_A}\Big\{-(23+\lambda^2_u)\,\bar{u}\gamma^\mu\gamma_5 u-(9+\lambda^2_u)\,\bar{c}\gamma^\mu\gamma_5c+(18+\lambda^2_d)\,\bar{d}\gamma^\mu\gamma_5 d\nonumber\\
&&\qquad\quad-(11+\lambda^2_d)\,\bar{s}\gamma^\mu\gamma_5s-4\,\bar{b}\gamma^\mu\gamma_5b\Big\}\nonumber\\
&+&\frac{\partial_\mu a_X}{2f_A}\Big\{(\lambda_u-\frac{1}{2}\lambda^3_u)\,\bar{u}\gamma^\mu(1-\gamma_5) c-\lambda^3_u(5A_ue^{i(\alpha^u_3-\alpha^u_2)}-6B_ue^{-\alpha^u_3})\bar{u}\gamma^\mu(1-\gamma_5) t\nonumber\\
&&\qquad\quad+\lambda^2_uA_u\,e^{i(\alpha^u_3-\alpha^u_2)}\bar{c}\gamma^\mu(1-\gamma_5) t+{\rm h.c.}\Big\}\nonumber\\
&+&\frac{\partial_\mu a_X}{2f_A}\Big\{(\lambda_d-\frac{1}{2}\lambda^3_d+30\theta^d_3)\,\bar{d}\gamma^\mu s-(\lambda_d-\frac{1}{2}\lambda^3_d-30\theta^d_3)\,\bar{d}\gamma^\mu\gamma_5 s\nonumber\\
&&\qquad\quad+\lambda^3_d(6B_de^{-i\alpha^d_3}-5A_de^{i(\alpha^d_3-\alpha^d_2)})\bar{d}\gamma^\mu(1-\gamma_5) b\nonumber\\
&&\qquad\quad+5A_d\lambda^2_de^{i(\alpha^d_3-\alpha^d_2)}\bar{s}\gamma^\mu(1-\gamma_5) b+{\rm h.c.}\Big\}\nonumber\\
&+&\sum_{q=d,s,b,u,c,t}\big(m_q\bar{q}q-\bar{q}i\! \! \not\! \partial\, q\big)\,.
\label{fl_in}
\end{eqnarray}
Here $V^{u,d}_{L}$ in Eq.(\ref{ckm01}) and $V^{u,d}_{R}$ in Appendix \ref{mmR} have been used. These interactions are the result of a direct interaction of SM gauge-singlet scalars $\chi, \tilde{\chi}$ coupling to $U(1)_X$-charged quarks.
Similarly, the flavored-QCD axion interactions with charged leptons, up to ${\cal O}(\lambda^4)$, are given by
 \begin{eqnarray}
  -{\cal L}^{a\ell} \simeq&-&\frac{\partial_\mu a_X}{2f_A}\Big\{23\,\bar{e}\gamma^\mu\gamma_5 e+12\,\bar{\mu}\gamma^\mu\gamma_5\mu-5\,\bar{\tau}\gamma^\mu\gamma_5\tau\Big\}\nonumber\\
  &+&\frac{\partial_\mu a_X}{2f_A}\Big\{16\theta^{r}_{1}e^{i(\alpha^\ell_3-\alpha^\ell_2)}\,\bar{\mu}\gamma^\mu(1+\gamma_5) \tau+{\rm h.c.}\Big\}
+\sum_{\ell=e,\mu,\tau}\big(m_\ell\bar{\ell}\ell-\bar{\ell}i\! \! \not\! \partial\, \ell\big)\,.
\label{fla_1}
 \end{eqnarray}
 where the charged-lepton mixing $V^\ell_R$ in Eq.(\ref{cLep2}) has been contributed. 
The flavored-QCD axion $a_X$ is produced by flavor-changing neutral Yukawa interactions in Eqs.(\ref{fl_in}) and (\ref{fla_1}), which leads to induced rare flavor-changing processes such as $s\rightarrow d+a_X$\,\cite{Wilczek:1982rv,Bolton:1988af, Artamonov:2008qb,raredecay,Berezhiani:1989fp}, $b\rightarrow s+a_X$, $b\rightarrow d+a_X$\,\cite{CLEO:2001acz}, and $\tau\rightarrow\mu +a_X$\,\cite{BaBar:2009hkt,ARGUS:1995bjh}. 
The most stringent constraint on the axion decay constant arises from the flavor-changing process $K^+\rightarrow\pi^++a_X$\,\cite{Wilczek:1982rv, Bolton:1988af, Artamonov:2008qb, raredecay}, rather than $B^\pm \to K^\pm + a_X$ and/or $\tau \to \mu + a_X$, which typically imply only $f_A \gtrsim 10^{5-6}$ GeV\, \cite{Bjorkeroth:2018dzu, Calibbi:2016hwq, delaVega:2021ugs}.
The flavor-violating $s-d-$axion interaction relevant for $K^+\rightarrow\pi^++a_X$ is given in Eq.(\ref{fl_in}).
 The corresponding decay width is
 \begin{eqnarray}
   \Gamma(K^+\rightarrow\pi^++a_X)=\frac{m^3_K}{16\pi}\Big(1-\frac{m^2_{\pi}}{m^2_{K}}\Big)^3\Big|\frac{1}{2\,F_{a}\delta^G_X}\Big(\lambda_d-\frac{\lambda^3_d}{2}+30\,\theta^d_3\Big)\Big|^2\,.
  \label{Gkp}
 \end{eqnarray}
Here $m_{K^{\pm}}=493.677\pm0.013$ MeV and $m_{\pi^{\pm}}=139.57061\pm0.00024$ MeV haven been taken from Ref.\cite{PDG}.
Using $F_a\delta^G_X=2.762\times10^{15}$ GeV from Eq.(\ref{pheno01}) (see also Eq.(\ref{axide1})), together with the numerical values of $\lambda_d=0.2439$ and $\theta^d_3=0.0128$ obtained in Eq.(\ref{qdata}), the branching ratio is predicted to be
 \begin{eqnarray}
   {\rm Br}(K^+\rightarrow\pi^++a_X)\simeq5.36\times10^{-19}\,,
  \label{brktp}
 \end{eqnarray}
where the central value of ${\rm Br}(K^+\rightarrow\pi^+\nu\bar{\nu})=(10.6^{+4.0}_{-3.4}|_{\rm stat}\pm0.9_{\rm syst})\times10^{-11}$ at $68\%$CL\,\cite{NA62:2021zjw} has been used, which is much lower than the present experimental upper bound ${\rm Br}(K^+\rightarrow\pi^+a_X)<(3-6)\times10^{-11} (1\times 10^{-11})$ for $m_{a}=0-110$ (160-260) MeV at $90\%$ CL.
 
The QCD axion mass $m_a$ in terms of the pion mass and pion decay constant reads\,\cite{Ahn:2014gva, Ahn:2016hbn}
 \begin{eqnarray}
 m^{2}_{a}F^{2}_{a}=m^{2}_{\pi^0}f^{2}_{\pi}F(z,w)\,,
\label{axiMass2}
 \end{eqnarray}
where $f_\pi\simeq92.1$ MeV\,\cite{PDG} and
 $F(z,w)=z/(1+z)(1+z+w)$ with $\omega=0.315\,z$.
Here the Weinberg value lies in $z\equiv m^{\overline{\rm MS}}_u(2\,{\rm GeV})/m^{\overline{\rm MS}}_d(2\,{\rm GeV})=0.47^{+0.06}_{-0.07}$\,\cite{PDG}.
After integrating out the heavy $\pi^{0}$ and $\eta$ at low energies, there is an effective low energy Lagrangian with an axion-photon coupling $g_{a\gamma\gamma}$:
${\cal L}_{a\gamma\gamma}= -g_{a\gamma\gamma}\,a\,\vec{E}\cdot\vec{B}$
where $\vec{E}$ and $\vec{B}$ are the electromagnetic field components. The axion-photon coupling is expressed in terms of the QCD axion mass, pion mass, pion decay constant, $z$ and $w$,
 \begin{eqnarray}
 g_{a\gamma\gamma}=\frac{\alpha_{\rm em}}{2\pi}\frac{m_a}{f_{\pi}m_{\pi^0}}\frac{1}{\sqrt{F(z,w)}}\left(\frac{E}{\delta^G_X}-\frac{2}{3}\,\frac{4+z+w}{1+z+w}\right)\,,
 \label{gagg}
 \end{eqnarray}
where $E/\delta^G_X=298/51$ (see Eqs(\ref{dGi}) and (\ref{eano})).
\begin{figure}[t]
\hspace*{-0.5cm}
\begin{minipage}[h]{8.3cm}
\includegraphics[width=8.3cm]{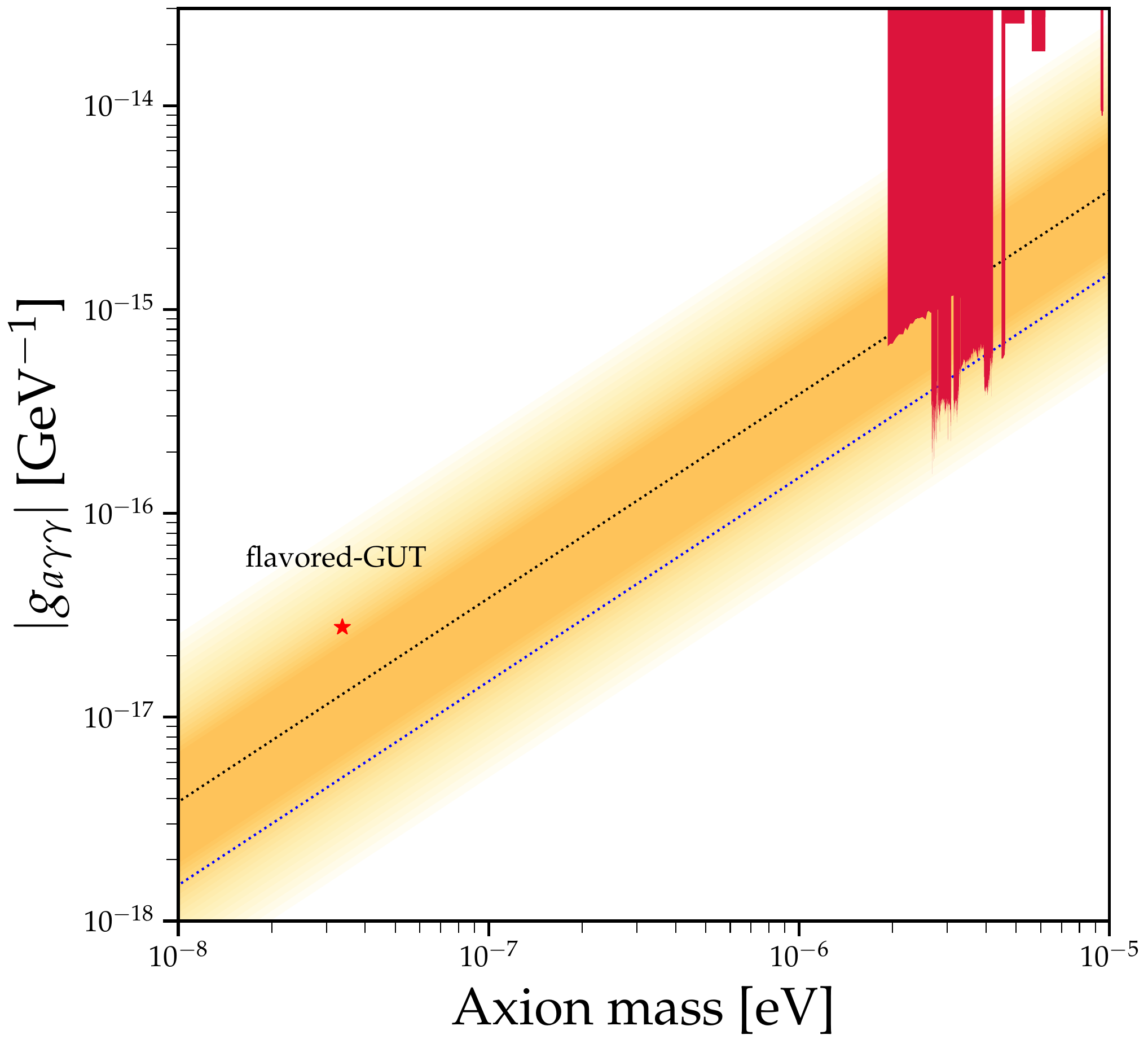}
\end{minipage}
\begin{minipage}[h]{8.3cm}
\includegraphics[width=8.3cm]{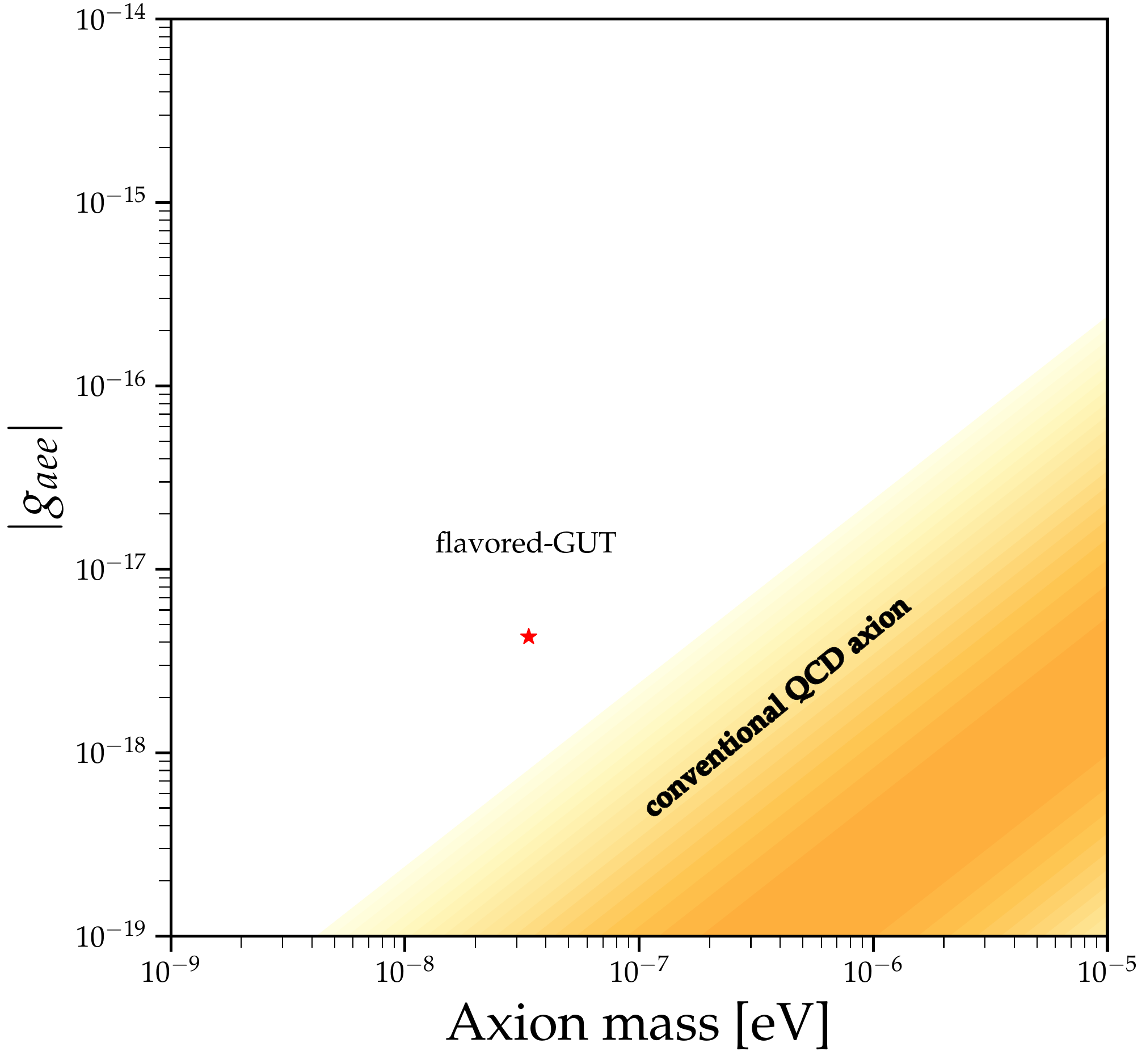}
\end{minipage}
\caption{\label{Fig4} Prediction (red star) of flavored-GUT for axion-photon coupling $|g_{a\gamma\gamma}|=2.74\times10^{-17}\,{\rm GeV}^{-1}$ (left) and axion-electron coupling $|g_{aee}|=4.26\times10^{-18}$ (right) as a function of the flavored-QCD axion mass $m_{a}=3.35\times10^{-8}\,{\rm eV}$. Black- and blue-dotted line indicate KSVZ\,\cite{KSVZ} and DFSZ\,\cite{DFSZ} model. Orange shaded region and vertical red lines indicate the conventional QCD axion predictions and the exclusion region of various axion search experiments, respectively, see Ref.\cite{PDG}.}
\end{figure}
The flavored-QCD axion mass, axion-photon coupling, and axion-electron coupling given by $g_{aee} = 23 m_e / f_A$, are predicted, as depicted in Fig.\ref{Fig4}, as
 \begin{eqnarray}
 m_{a}=3.35\times10^{-8}\,{\rm eV}\,,\qquad |g_{a\gamma\gamma}|=2.74\times10^{-17}\,{\rm GeV}^{-1}\,,\qquad |g_{aee}|=4.26\times10^{-18}\,,
\label{axiMass2}
 \end{eqnarray} 
while there are constraints on $g_{a\gamma\gamma}$ and $g_{aee}$ by the recent analysis of the horizontal branch stars in galactic globular clusters $ |g_{a\gamma\gamma}|<6.6\times10^{-11}\,{\rm GeV}^{-1}\,(95\%\,{\rm CL})$\,\cite{Ayala:2014pea}, by red giant branch (RGB) stars $|g_{aee}| < 4.3 \times 10^{-13} \quad (95\%\,\text{CL})$\,\cite{Viaux:2013lha}, 
and by white dwarf (WD) cooling $|g_{aee}| \lesssim 2.8 \times 10^{-13}$ \cite{Bertolami:2014wua}, though theoretical uncertainties persist.

{\bf Numerical simulation for quark mass and  mixing:} 
To reproduce the experimental quark masses and CKM mixing parameter given in Eqs.(\ref{ckmmixing}) and (\ref{qumas}), we perform a numerical analysis using the linear algebra tools from Ref.\cite{Antusch:2005gp}.
The quark Yukawa matrices in Eqs.(\ref{Ch2}) and (\ref{Ch1}) are defined at the $U(1)_X$ symmetry-breaking scale, where their parameters generally receive quantum corrections. In principle, these matrices should be evolved down to the top quark mass scale ($m_t$) via RG running equations and diagonalized. However, for the hierarchical  fermion mass textures considered here, the one-loop renormalization group running effects on the physical observables are expected to be sufficiently small. Therefore, for simplicity, we assume that the Yukawa matrices at the scale of $U(1)_X$-breaking scale are approximately identical to those at the scale $m_t$, since the one-loop RG running effect on observables for hierarchical mass spectra is expected to be negligible.

The low-energy Yukawa couplings required for experimental values are obtained from the physical masses and mixing angles compiled by the PDG\,\cite{PDG} and CKMfitter\,\cite{ckm}. Using $\langle\tau\rangle\approx i$ in Eq.(\ref{Dta_01}) together with
   \begin{eqnarray}
\Delta_\chi=0.62\,,\qquad \tan\beta=6.7\,,
\label{delchi}
 \end{eqnarray}
with effective Yukawa coefficients in the range $0.38\lesssim|y_i|\lesssim1.62$ from Eq.(\ref{AFN1}), we obtain, for the quantum numbers listed in Table-\ref{reps_q}, the following reference inputs
  \begin{eqnarray}
 & y_u=0.755, \,y_c=0.536,\, y_{c1}=0.441,\, y_t=1.003,\, y_{t1}=0.629,\, y_{t2}=0.677,\nonumber\\
 &\arg(y_u)=5.289\,, \arg(y_c)=3.872\,, \arg(y_{c1})=2.670\,, \arg(y_{t1})=3.863\,, \arg(y_{t2})=3.843\,\nonumber\\
 &y_d=1.014,\, y_s=0.679,\, y_{s1}=0.734,\, y_b=1.102,\, y_{b1}=0.854,\, y_{b2}=0.717,\\
 &\arg(y_d)=5.256, \arg(y_s)=2.989, \arg(y_{s1})=2.918, \arg(y_{b1})=4.547, \arg(y_{b2})=6.166.\nonumber
\label{qdata}
 \end{eqnarray}
This benchmark parameter set successfully reproduces the experimental constraints in Eqs.(\ref{ckmmixing}) and (\ref{qumas}), yielding the following physical observables
 $\theta^q_{23}=2.323^{\circ}, \theta^q_{13}=0.218^{\circ}, \theta^q_{12}=13.011^{\circ}$, $\delta^q_{CP}=66.162^{\circ}$;
 $m_d=4.634$ MeV,  $m_s=93.319$ MeV,  $m_b=4.181$ GeV,  $m_u=2.144$ MeV,  $m_c=1.270$ GeV,  $m_t=172.604$ GeV.

\subsubsection{Neutrino mass and mixing}
\label{num_nu}
Taking the seesaw scale $\langle\rho\rangle$ in Eq.(\ref{pheno01}), and after integrating out the right-handed heavy Majorana neutrinos, the effective light-neutrino mass matrix ${\cal M}_{\nu}$ is given at leading order by
 \begin{eqnarray}
 {\cal M}_{\nu}\simeq -m^T_DM^{-1}_Rm_D=U^{\ast}_{\nu}\,{\rm diag.}(m_{\nu_1}, m_{\nu_2}, m_{\nu_3})\,U^{\dag}_{\nu}\,,
   \label{neut2}
 \end{eqnarray}
where $U_{\nu}$ is the unitary matrix diagonalizing ${\cal M}_{\nu}$, and $m_{\nu_i}$ ($i=1,2,3$) are the light neutrino masses. 
Equivalently,
 \begin{eqnarray}
  U^T_{\nu}\,{\cal M}_{\nu}\,U_{\nu}={\rm diag.}(m_{\nu_1}, m_{\nu_2}, m_{\nu_3})\,.
 \label{nu_mas}
 \end{eqnarray}
The experimentally observed hierarchy $|\Delta m^{2}_{\rm Atm}|= |m^{2}_{\nu_3}-(m^{2}_{\nu_1}+m^{2}_{\nu_2})/2|\gg\Delta m^{2}_{\rm Sol}\equiv m^{2}_{\nu_2}-m^{2}_{\nu_1}>0$, together with the requirement of a Mikheyev-Smirnov-Wolfenstein resonance\,\cite{Wolfenstein:1977ue} for solar neutrinos, allows two possible neutrino mass orderings: normal mass ordering (NO) $m^2_{\nu_1}<m^2_{\nu_2}<m^2_{\nu_3}$ and inverted mass ordering (IO) $m^2_{\nu_3}<m^2_{\nu_1}<m^2_{\nu_2}$. 

From Eq.(\ref{AxionLag2}) the PMNS mixing matrix is given by
  \begin{eqnarray}
 U_{\rm PMNS}=V^\ell_L\,U_\nu\simeq U_\nu\,,
   \label{pmns0}
 \end{eqnarray}
 where the left-handed charged-lepton mixing matrix $V^\ell_L$ is approximately equal to the unit matrix up to ${\cal O}(\lambda^4)$, as shown in Eq.(\ref{VlI}).
 The matrix $U_{\rm PMNS}$ is expressed in terms of three mixing angles, $\theta_{12}, \theta_{13}, \theta_{23}$, and a Dirac type \cp ~violating phase $\delta_{CP}$ and two additional \cp~ violating phases $\varphi_{1,2}$ if light neutrinos are Majorana particle as\,\cite{PDG}
 \begin{eqnarray}
  U_{\rm PMNS}=
  {\left(\begin{array}{ccc}
   c_{13}c_{12} & c_{13}s_{12} & s_{13}e^{-i\delta_{CP}} \\
   -c_{23}s_{12}-s_{23}c_{12}s_{13}e^{i\delta_{CP}} & c_{23}c_{12}-s_{23}s_{12}s_{13}e^{i\delta_{CP}} & s_{23}c_{13}  \\
   s_{23}s_{12}-c_{23}c_{12}s_{13}e^{i\delta_{CP}} & -s_{23}c_{12}-c_{23}s_{12}s_{13}e^{i\delta_{CP}} & c_{23}c_{13}
   \end{array}\right)}Q_{\nu}\,,
 \label{PMNS1}
 \end{eqnarray}
where $s_{ij}\equiv \sin\theta_{ij}$, $c_{ij}\equiv \cos\theta_{ij}$ and $Q_{\nu}={\rm diag.}(e^{-i\varphi_1/2}, e^{-i\varphi_2/2},1)$.
Thus, Eqs.(\ref{PMNS1}) and (\ref{nu_mas}) contain nine physical observables: $\theta_{23}$, $\theta_{13}$, $\theta_{12}$, $\delta_{CP}$, $\varphi_1$, $\varphi_2$, $m_{\nu_1}$, $m_{\nu_2}$, and $m_{\nu_3}$. 
\begin{table}[h]
\caption{\label{exp_nu} The global fit of three-flavor oscillation parameters at the best-fit and $3\sigma$ level with Super-Kamiokande atmospheric data\,\cite{Esteban:2020cvm}. NO = normal neutrino mass ordering; IO = inverted mass ordering. And $\Delta m^{2}_{\rm Sol}\equiv m^{2}_{\nu_2}-m^{2}_{\nu_1}$, $\Delta m^{2}_{\rm Atm}\equiv m^{2}_{\nu_3}-m^{2}_{\nu_1}$ for NO, and  $\Delta m^{2}_{\rm Atm}\equiv m^{2}_{\nu_2}-m^{2}_{\nu_3}$ for IO.}
\begin{ruledtabular}
\begin{tabular}{cccccccccccc} &$\theta_{13}[^{\circ}]$&$\delta_{CP}[^{\circ}]$&$\theta_{12}[^{\circ}]$&$\theta_{23}[^{\circ}]$&$\Delta m^{2}_{\rm Sol}[10^{-5}{\rm eV}^{2}]$&$\Delta m^{2}_{\rm Atm}[10^{-3}{\rm eV}^{2}]$\\
\hline
$\begin{array}{ll}
\hbox{NO}\\
\hbox{IO}
\end{array}$&$\begin{array}{ll}
8.58^{+0.33}_{-0.35} \\
8.57^{+0.37}_{-0.34}
\end{array}$&$\begin{array}{ll}
232^{+118}_{-88} \\
276^{+68}_{-82}
\end{array}$&$33.41^{+2.33}_{-2.10}$&$\begin{array}{ll}
42.2^{+8.8}_{-2.5} \\
49.0^{+2.5}_{-9.1}
\end{array}$&$7.41^{+0.62}_{-0.59}$
 &$\begin{array}{ll}
2.507^{+0.083}_{-0.080} \\
2.486^{+0.084}_{-0.080}
\end{array}$ \\
\end{tabular}
\end{ruledtabular}
\end{table}
Recent global fits\,\cite{Esteban:2018azc, deSalas:2017kay, Capozzi:2018ubv} of neutrino oscillations have enabled a more precise determination of the mixing angles and mass squared differences, with large uncertainties remaining for $\theta_{23}$ and $\delta_{CP}$ at 3$\sigma$. The most recent analysis\,\cite{Esteban:2020cvm} lists global fit values and $3\sigma$ intervals for these parameters in Table-\ref{exp_nu}.
Furthermore, recent constraints on the rate of $0\nu\beta\beta$ decay have added to these findings. Specifically, the most tight upper bounds for the effective Majorana mass (${\cal M}_{\nu})_{ee}$, which is the modulus of the $ee$-entry of the effective neutrino mass matrix, are given by
  \begin{eqnarray}
 ({\cal M}_{\nu})_{ee}< 0.036-0.156\,{\rm eV}\,~ (^{136}\text{Xe-based experiment\,\cite{KamLAND-Zen:2022tow}})
 \label{nubb}
 \end{eqnarray}
at $90\%$ CL.

{\bf Numerical simulation for lepton mass and mixing:}
Similar to the quark sector, to simulate and match experimental results for charged-leptons and neutrinos, Eqs.(\ref{ckmmixing}) and (\ref{qumas}), we use linear algebra tools from Ref.\cite{Antusch:2005gp}.

Using the reference values Eq.(\ref{delchi}) and $\langle\tau\rangle\approx i$ in Eq.(\ref{Dta_01}), the charged lepton masses and the left-handed charged-lepton mixing matrix are reproduced from Eq.(\ref{ChL1}) by the parameter choice
\begin{eqnarray}
&y_e=1.185685\,,\quad y_{e2}=1.521752\,,\quad y_{e3}=1.396277\,,\nonumber\\
&y_{\mu}=1.276033\,,\quad y_{\mu3}=1.077281\,,\quad y_{\tau}=0.755033\,,\nonumber\\
&\arg(y_e, y_{e2}, y_{e3},y_\mu,y_{\mu3}, y_\tau)=[0, 2\pi]\,,
 \label{chl01}
 \end{eqnarray} 
 yielding excellent agreement with the observed charged-lepton masses in Eq.(\ref{clm01}).
Note that here the phases could be arbitrary because the left‑handed charged‑lepton mixing matrix is the unit matrix up to ${\cal O}(\lambda^4)$ (see Eq.(\ref{VlI}) and  the texture of the charged‑lepton mass matrix in Eq.(\ref{ChL1})).

\begin{figure}[t]
\begin{minipage}[h]{8.0cm}
\epsfig{figure=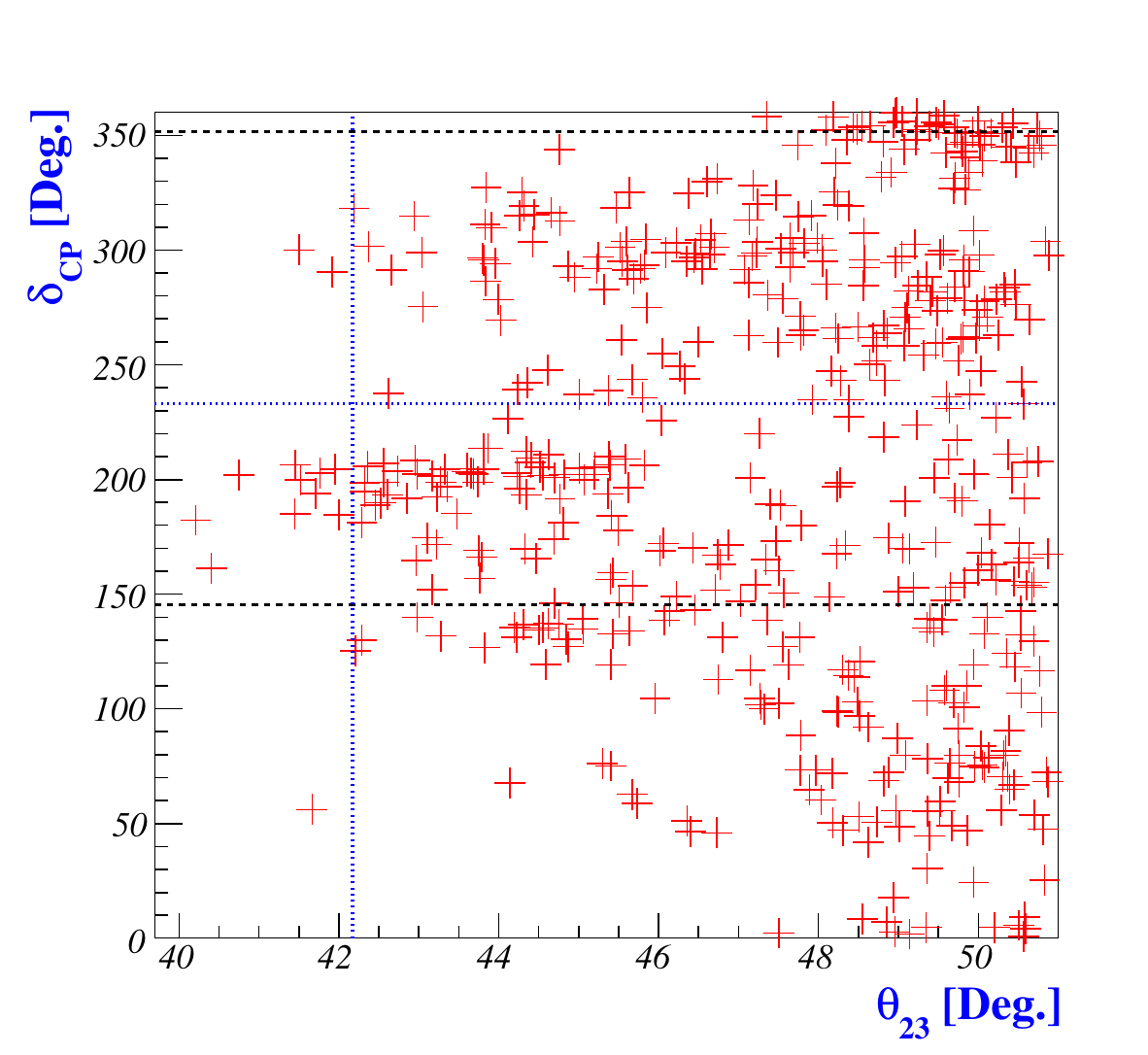,width=8.0cm,angle=0}
\end{minipage}
\begin{minipage}[h]{8.0cm}
\epsfig{figure=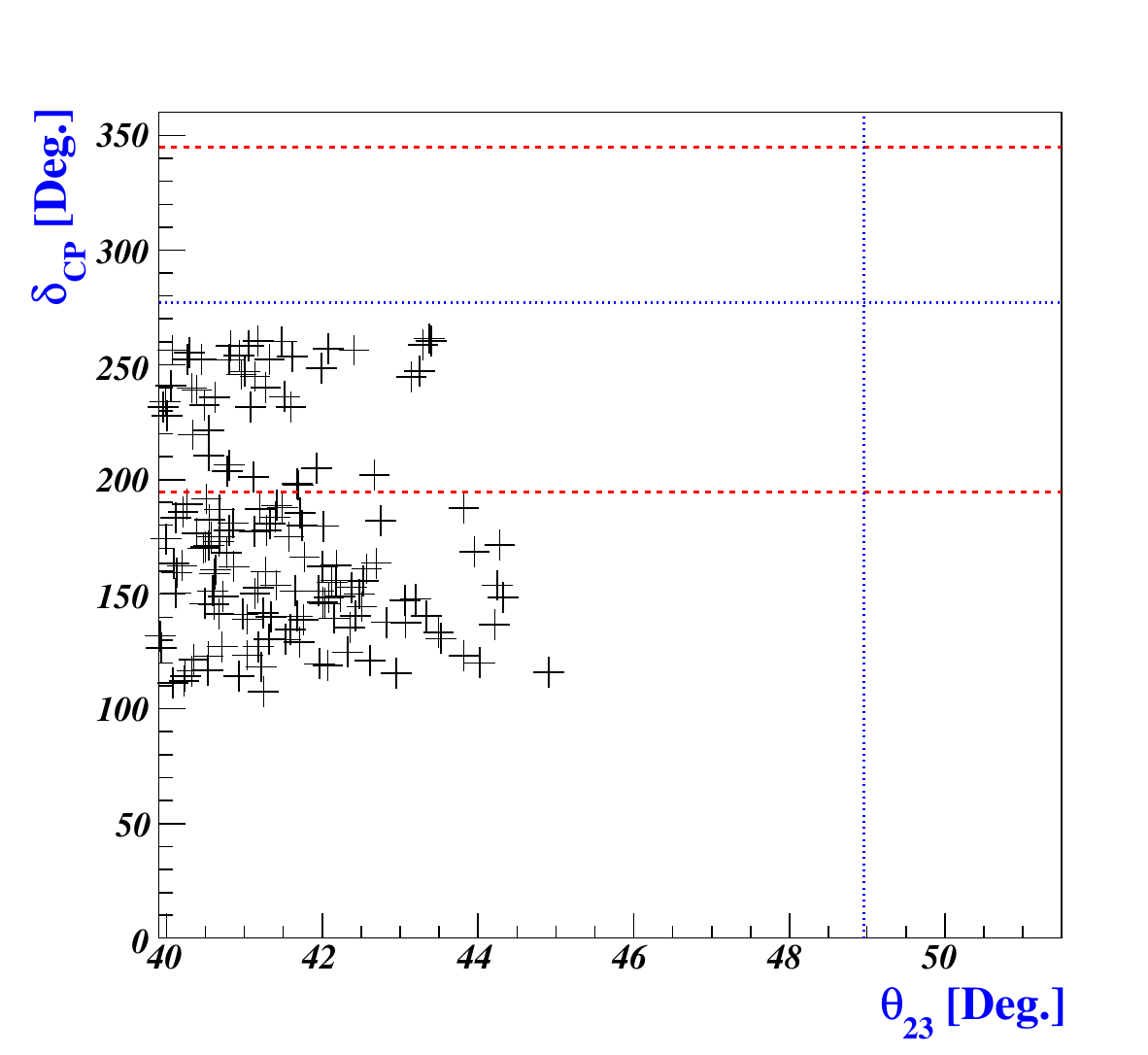,width=8.0cm,angle=0}
\end{minipage}
\caption{\label{Fig5}  Plots of the leptonic Dirac CP phase $\delta_{\rm CP}$ as a function of the atmospheric mixing angle $\theta_{23}$ for normal ordering (NO, left panel) and inverted ordering (IO, right panel). The vertical and horizontal blue dotted lines indicate the $1\sigma$ best-fit value of $\theta_{23}$ and $\delta_{\rm CP}$, respectively,
while the horizontal dashed lines denote the $3\sigma$ allowed ranges of $\delta_{\rm CP}$ listed in Table-\ref{exp_nu}.}
\end{figure}
\begin{figure}[t]
\begin{minipage}[h]{8.0cm}
\epsfig{figure=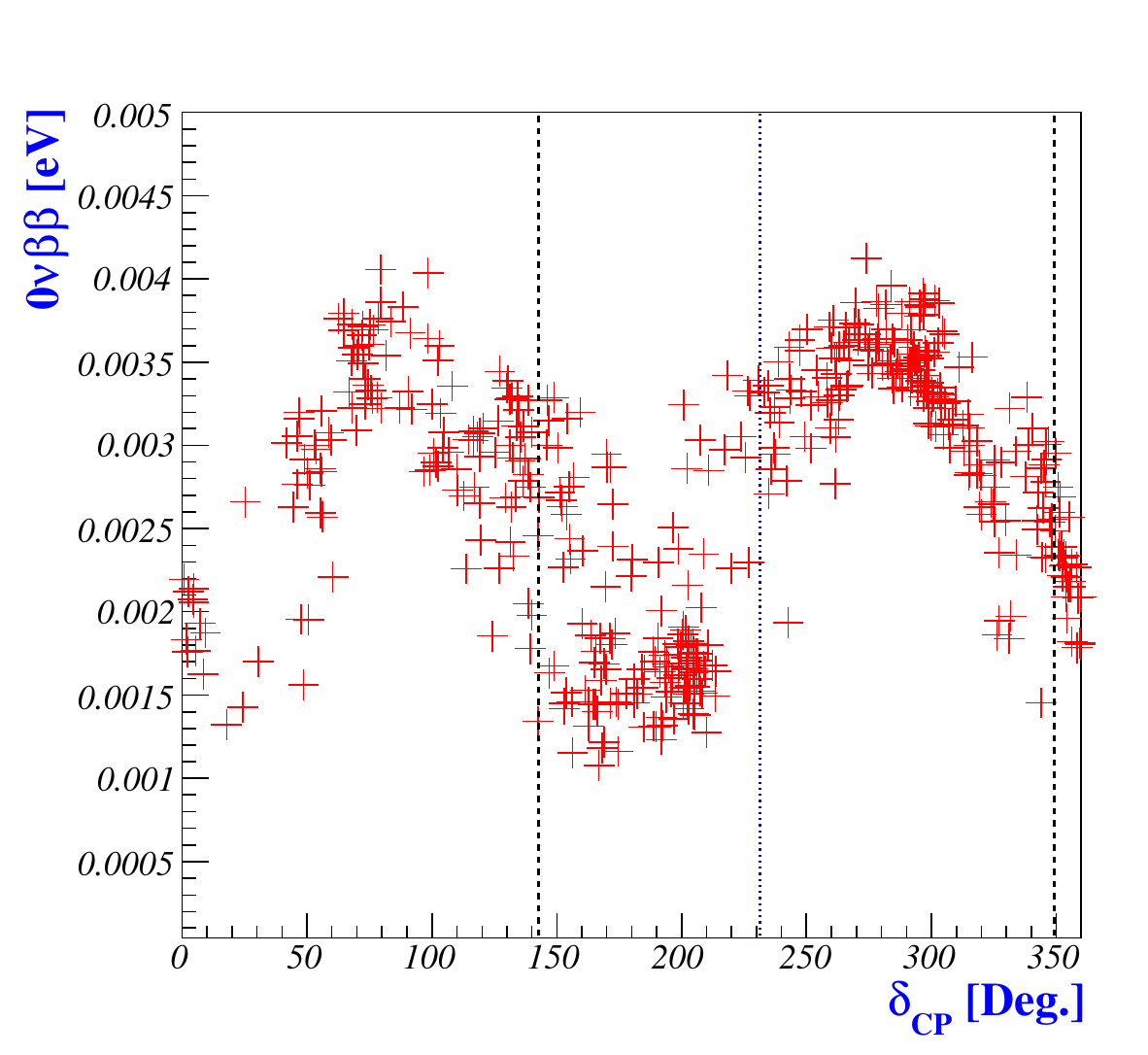,width=8.0cm,angle=0}
\end{minipage}
\begin{minipage}[h]{8.0cm}
\epsfig{figure=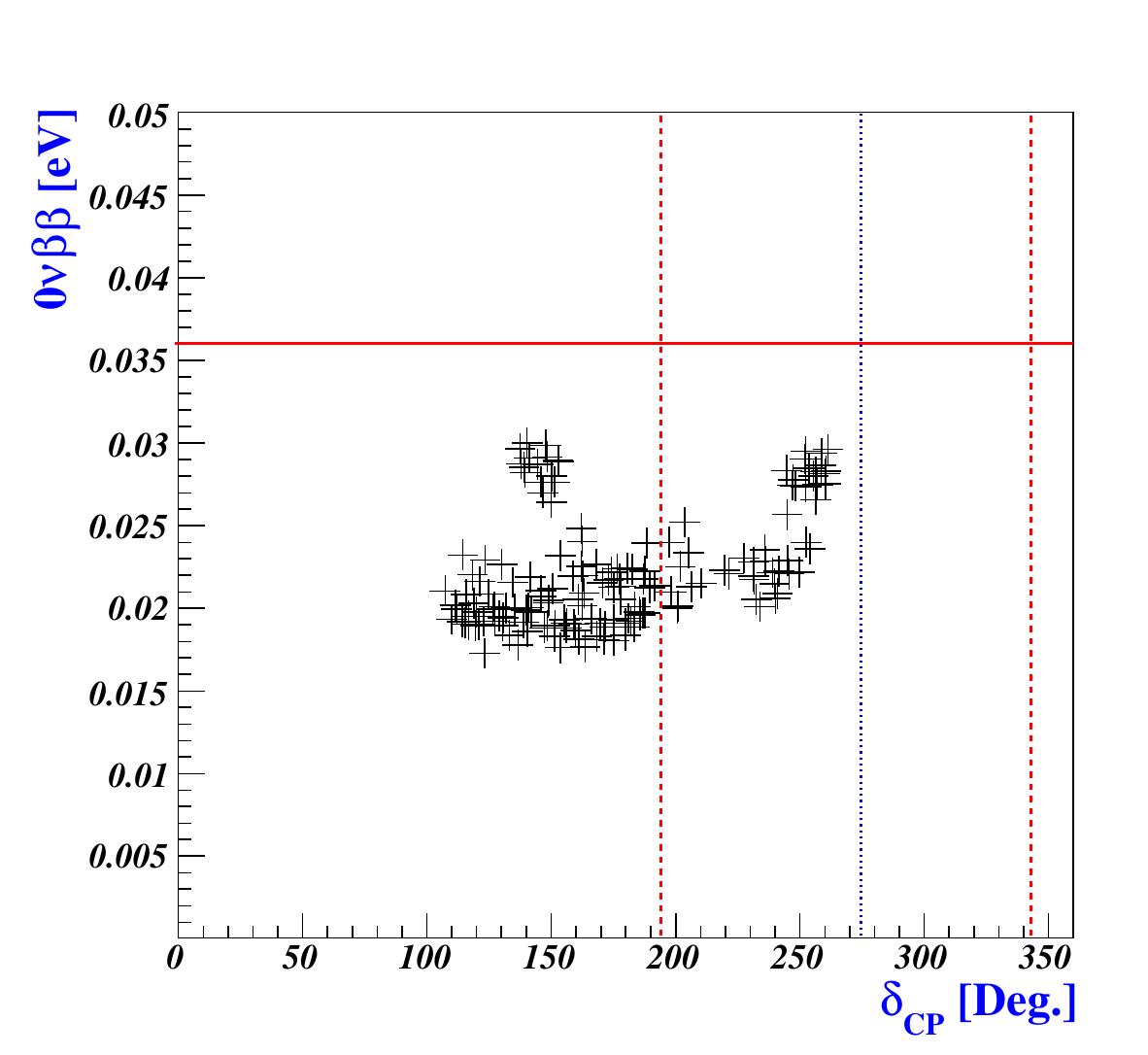,width=8.0cm,angle=0}
\end{minipage}
\caption{\label{Fig6} Plots of the $0\nu\beta\beta$-decay rate as a function of the leptonic Dirac CP phase $\delta_{\rm CP}$ for NO (left panel) and IO (right panel). The vertical dashed lines and the dotted line indicate the $3\sigma$ bounds and the $1\sigma$ best-fit value of $\delta_{\rm CP}$, respectively, taken from Table-\ref{exp_nu}. The horizontal red-line indicates the upper bound from the KamLAND-Zen result\,\cite{KamLAND-Zen:2022tow} (Eq.(\ref{nubb})).}
\end{figure}

The seesaw mechanism in Eq.(\ref{neut2}) operates at the $U(1)_{B-L}$ symmetry breakdown scale, while its implications are measured by experiments below the electroweak scale. Therefore, RG corrections to neutrino masses and mixing parameters can be crucial, especially for quasi-degenerate neutrino spectra\,\cite{Antusch:2005gp}. 
However, our numerical calculation shows that the neutrino mass spectra at the $U(1)_{B-L}$ breakdown scale is hierarchical (see Fig.\ref{Fig6}). Therefore, RG effects on the low-energy neutrino observables are expected to be negligible and are ignored in the following analysis.
To show the viable parameter space, we scan the experimentally allowed $3\sigma$ ranges of $\{\theta_{13}$, $\theta_{23}$, $\theta_{12}$, $\Delta m^2_{\rm Sol}$, $\Delta m^2_{\rm Atm}\}$ at $3\sigma$ listed in Table-\ref{exp_nu}. Using the reference values from Eq.(\ref{delchi}) and $\langle\tau\rangle\approx i$ in Eq.(\ref{Dta_01}), together with $\langle\rho\rangle=4.921\times10^{11}$ GeV from Eq.(\ref{pheno01}), we determine the allowed parameter regions of Eqs.(\ref{MR1}) and (\ref{Ynu1}) at the $U(1)_{B-L}$ breaking scale.
Using the effective Yukawa coefficients in Eq.(\ref{Ynu1eff}), we find parameter regions compatible with both the NO and IO. Adopting the normalization condition $\xi^2_D/\xi_M=1$ for the seesaw sector, the effective Majorana neutrino coefficients of Eq.(\ref{MR1}) are parameterized as
 \begin{eqnarray}
&\gamma^{\rm eff}_{11}=\xi_M[0.70, 1.10],\qquad\gamma^{\rm eff}_{12}=\xi_M[0.36, 0.95],\qquad\gamma^{\rm eff}_{22}=\xi_M[0.90, 1.70],\nonumber\\
&\arg(\gamma^{\rm eff}_{11})=[0.49, 4.3],\qquad \arg(\gamma^{\rm eff}_{12})=[0.31, 4.73],\qquad \arg(\gamma^{\rm eff}_{22})=[0.15, 4.04],
 \label{neu010}
 \end{eqnarray} 
for NO;
 \begin{eqnarray}
&\gamma^{\rm eff}_{11}=\xi_M[1.26, 1.51],\qquad\gamma^{\rm eff}_{12}=\xi_M[1.41, 1.52],\qquad\gamma^{\rm eff}_{22}=\xi_M[1.31, 1.47],\nonumber\\
&\arg(\gamma^{\rm eff}_{11})=[0.04, 0.60],\qquad \arg(\gamma^{\rm eff}_{12})=[0.94, 1.75],\qquad \arg(\gamma^{\rm eff}_{22})=[5.51, 2\pi],
 \label{neu0101}
 \end{eqnarray} 
 for IO,
where $\xi_M$ is the overall normalization factor of the effective coefficients $\gamma^{\rm eff}_{ij}$ in Eq.(\ref{Ynu1eff}). Similarly, the effective Dirac neutrino coefficients of Eq.(\ref{Ynu1}) are given by
\begin{eqnarray}
&\beta^{\rm eff}_{1e}=\xi_D[0.086, 0.195],\qquad\beta^{\rm eff}_{1\mu}=\xi_D[1.151, 1.674],\qquad \beta^{\rm eff}_{1\tau}=\xi_D[2.394, 3.396],\nonumber\\
&\beta^{\rm eff}_{2e}=\xi_D[0.064, 0.297],\qquad\beta^{\rm eff}_{2\mu}=\xi_D[0.871, 1.882], \qquad\beta^{\rm eff}_{2\tau}=\xi_D[1.950, 2.736],\nonumber\\
&\arg(\beta^{\rm eff}_{1e})=[0, 2\pi],\qquad \arg(\beta^{\rm eff}_{1\mu})=[1.56, 6.19],\qquad \arg(\beta^{\rm eff}_{1\tau})=[0, 2\pi],\nonumber\\
&\arg(\beta^{\rm eff}_{2e})=[3.0, 2\pi],\qquad \arg(\beta^{\rm eff}_{2\mu})=[0, 2\pi],\qquad \arg(\beta^{\rm eff}_{2\tau})=[0, 3.85],
 \label{neu01}
 \end{eqnarray} 
 for NO;
 \begin{eqnarray}
&\beta^{\rm eff}_{1e}=\xi_D[1.633, 1.780],\qquad\beta^{\rm eff}_{1\mu}=\xi_D[1.463, 1.574],\qquad \beta^{\rm eff}_{1\tau}=\xi_D[2.811, 3.249],\nonumber\\
&\beta^{\rm eff}_{2e}=\xi_D[0.712, 0.804],\qquad\beta^{\rm eff}_{2\mu}=\xi_D[0.815, 1.055], \qquad\beta^{\rm eff}_{2\tau}=\xi_D[3.149, 3.398],\nonumber\\
&\arg(\beta^{\rm eff}_{1e})=[2.70, \pi],\qquad\qquad\qquad \arg(\beta^{\rm eff}_{1\mu})=[2.15, 2.93]\cup[5.12,5.63],\nonumber\\ 
&\arg(\beta^{\rm eff}_{1\tau})=[3.78, 4.68]\cup[5.4,2\pi],\qquad\qquad\qquad \arg(\beta^{\rm eff}_{2e})=[0.97, 1.72],\nonumber\\
&\arg(\beta^{\rm eff}_{2\mu})=[1.78, 2.97],\qquad\qquad\qquad \arg(\beta^{\rm eff}_{2\tau})=[0.75, 1.40]\cup[4.80,2\pi],
 \label{neu0101}
 \end{eqnarray} 
for IO, where $\xi_D$ is the overall normalization factor of the effective coefficients $\beta^{\rm eff}_{i\ell}$ in Eq.(\ref{Ynu1eff}).

\begin{figure}[t]
\begin{minipage}[h]{8.0cm}
\epsfig{figure=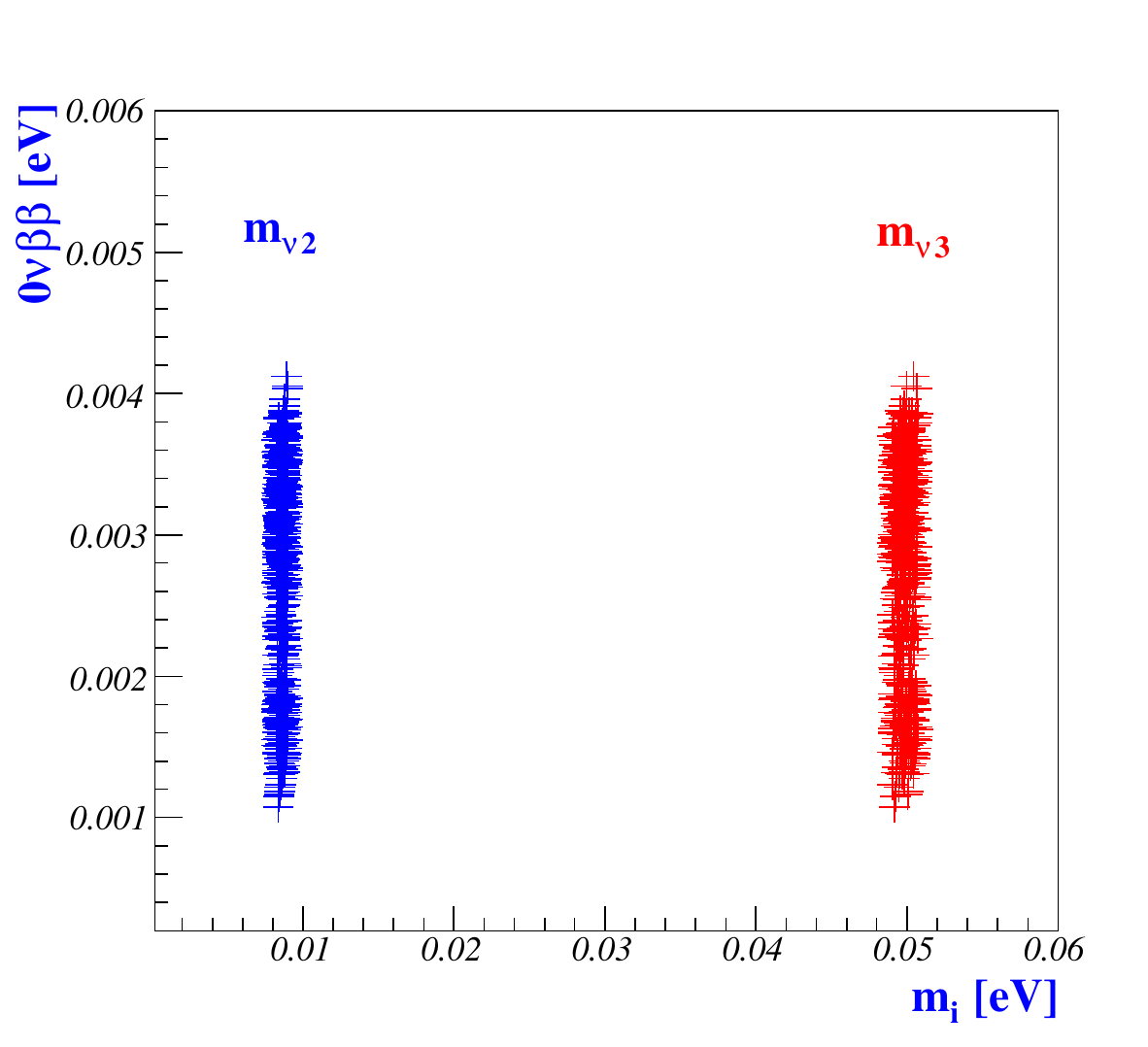,width=8.0cm,angle=0}
\end{minipage}
\begin{minipage}[h]{8.0cm}
\epsfig{figure=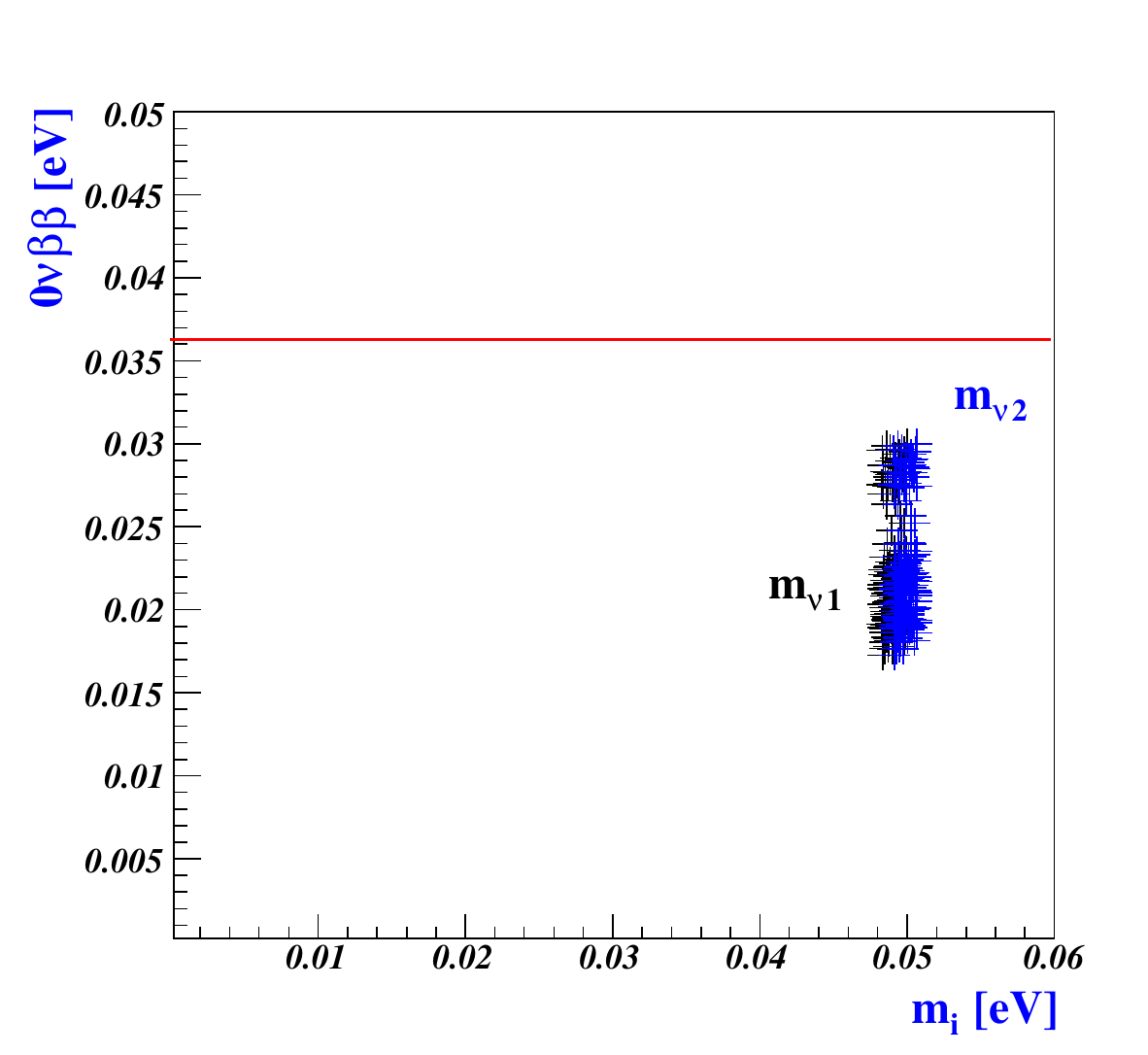,width=8.0cm,angle=0}
\end{minipage}
\caption{\label{Fig7} Plots of the $0\nu\beta\beta$-decay rate as a function of the neutrino masses $m_{\nu_i}$ for NO (left panel) with $m_{\nu_1}=0$, and for IO (right panel) with $m_{\nu_3}=0$. Horizontal red-line indicate the upper bound of KamLAND-Zen result\,\cite{KamLAND-Zen:2022tow} of Eq.(\ref{nubb}).}
\end{figure}

Neutrino oscillation experiments currently aim to make precise measurements of the Dirac CP-violating phase $\delta_{CP}$ and atmospheric mixing angle $\theta_{23}$. 
For the parameter regions explored in our model, distinct favored regions emerge in the $\theta_{23}$-$\delta_{CP}$ plane, as shown in left (NO) and right (IO) panels of Fig.\ref{Fig5}. Ongoing experiments like DUNE\,\cite{DUNE:2018tke}, together with proposed next-generation experiments such as Hyper-K\,\cite{Hyper-Kamiokande:2018ofw}, are expected to significantly improve the precision of these measurements. Their results will therefore provide important tests of the parameter space and predictive structure of the present framework.
As shown in the left (right) panel of Fig.\ref{Fig6} for the NO (IO) case, the effective $0\nu\beta\beta$-decay rate is plotted as a function of the leptonic Dirac CP phase $\delta_{CP}$. The predicted values lie entirely below the experimental upper bound given in Eq.(\ref{nubb}), indicated by the horizontal dotted line in the right panel.
 Moreover, ongoing and future experiments on $0\nu\beta\beta$-decay like NEXT\,\cite{NEXT:2020amj}, SNO$+$\,\cite{SNO:2022trz}, KamLAND-Zen\,\cite{KamLAND-Zen:2022tow}, Theia\,\cite{Theia:2019non}, SuperNEMO\,\cite{Arnold:2004xq} are expected to probe the NO and IO predictions of the present model.
Cosmological and astrophysical measurements provide powerful constraints on the sum of neutrino masses. The upper bound on the sum of the three active neutrino masses can be summarized as $\sum m_{\nu}=m_{\nu_1}+m_{\nu_2}+m_{\nu_3}<0.120$ eV at $95\%$ CL for TT, TE, EE+lowE+lensing+BAO\,\cite{Planck:2018vyg}. 
As shown in the left (NO) and right (IO) panels of Fig.\ref{Fig7}, which plot the effective $0\nu\beta\beta$-decay rate versus neutrino masses, the sum of neutrino masses lies in the range of $0.0574$ to $0.0596$ eV for NO, while for IO it lies in the range of $0.0974$ to $0.101$ eV. Both ranges are consistent with the current cosmological bound.
\section{conclusion}
\label{conc}
We have proposed a framework for flavored-GUT within string-derived supergravity based on $G_{\rm SM} \times SL(2,\mathbb{Z}) \times U(1)_X\times U(1)_{B-L}$, where the additional $U(1)$s are gauged and gravity is intrinsically incorporated. A central result of this work is that anomaly cancellation and SM gauge coupling unification act as fundamental consistency conditions that strongly constrain the flavor structure, rather than treating flavor as an independent input.
The string-theoretic origin of the $SL(2,\mathbb{Z})$ modular symmetry fundamentally constrains the superpotential, K{\"a}hler potential, and gauge and gravitational kinetic functions, while the cancellation of modular-, gauge-, and gravity-mixed anomalies imposes stringent restrictions on the chiral matter content, modular weights, and charge assignments. 
To simultaneously address flavor and the strong CP problem, we introduced an anomalous flavored $U(1)_X$ symmetry. Note that a particular anomaly-free limit reproduces the conventional $U(1)_{B-L}$ symmetry as a special case of the flavored $U(1)_X$ construction. 
The spontaneous breaking of the gauged $U(1)_X$ generates a pseudo-NG mode, while the associated accidental flavored-PQ symmetry is explicitly broken only higher-dimensional operators that preserve the local $U(1)_X$ gauge symmetry and are suppressed by the flavor dynamics scale, which we identify with the $U(1)_X$ gauge boson mass $M_X$. 
After the $U(1)_X$ gauge boson acquires a mass and decouples at this scale, a residual global anomalous symmetry survives in the low-energy effective theory, giving rise to a flavored QCD axion. 
The successful realization of SM gauge coupling unification through RG evolution, together with the reproduction of realistic quark and lepton mass hierarchies and mixing patterns while restricting all Yukawa coefficients to unit-magnitude complex numbers, points to new interactions at a very high energy scale where gauge and flavor dynamics are closely correlated. Within the flavored-GUT framework, these same consistency conditions correlate the flavor dynamics scale with the QCD axion sector, thereby significantly reducing the arbitrariness of conventional flavor constructions.

We have shown that the modular symmetry, realized as a discrete-gauge symmetry in the 4D effective theory, must remain anomaly free. The modular anomalies induced by K{\"a}hler transformations are matched by those arising from the chiral rotations of gauginos. Moreover, despite the nontrivial transformation properties of SM fermions under $SL(2,\mathbb{Z})$, the strong CP phase remains invariant, even in the global supersymmetry limit $M_P\rightarrow\infty$, provided that the symmetry-breaking scalar fields carry vanishing modular weights and therefore acquire modular-invariant VEVs.  
The vanishing modular anomalies $SL(2,\mathbb{Z}) \times [SU(3)_C]^2$ and $SL(2,\mathbb{Z}) \times [U(1)_{\rm EM}]^2$ guaranteed by modular- and SM gauge-invariance with non-negative weight modular forms, together with the cancellation of the mixed modular, gauge, and gravitational anomalies, imposes highly nontrivial constraints on the modular weights and $U(1)_X$ charge assignments of quarks and leptons. Remarkably, these anomaly cancellation conditions determine the flavor structure in a manner compatible with exact SM gauge coupling unification. We note that a particular anomaly-free limit reproduces the conventional $U(1)_{B-L}$ symmetry as a special case of the flavored $U(1)_X$ construction.

We have shown that exact SM gauge coupling unification emerges naturally within the flavored-GUT framework. Unlike conventional GUT scenarios, the unification is controlled by anomaly coefficients, Green-Schwarz contributions, kinetic-mixing effects, and flavored $U(1)$ gauge sectors associated with the flavor structure. Consequently, the experimentally measured low energy values of $\sin^2\theta_W(M_Z)$ and $\alpha_3(M_Z)$ determine not only the flavored-GUT scale but also the flavor scale $M_X$, the axion decay constant $F_a$, the $U(1)_{B-L}$-breaking scale associated with the seesaw mechanism, and the supersymmetry-breaking scale of order ${\cal O}(10)$ TeV.

A particularly significant result is that the QCD axion sector is no longer independent of flavor and gauge unification. The same consistency conditions that constrain the flavor structure also fix the flavor dynamics scale $M_X$. Since the ratio $F_a/M_X$ is determined by the flavor structure, the axion decay constant $F_a$ is consequently fixed as well. As a consequence, the QCD axion mass is predicted rather than treated as a free parameter. Using measured SM gauge couplings, we obtain $m_a=3.35\times10^{-8}$ eV, together with definite predictions for the axion-photon coupling $|g_{a\gamma\gamma}|=2.74\times10^{-17}\,{\rm GeV}^{-1}$, the axion-electron coupling $|g_{aee}|=4.26\times10^{-18}$, and the rare decay branching ratio ${\rm Br}(K^+\rightarrow\pi^+a_X)\simeq5.36\times10^{-19}$. This establishes a direct and testable connection between flavor physics, gauge coupling unification, and axion phenomenology.

We have shown that the scalar fields, including the $U(1)_X$-charged scalar fields and the $U(1)_X$-charged modulus, are stabilized through their interactions, together with loop corrections and small perturbations.
In particular, the backreaction between the $U(1)_X$-charged scalar fields and the modulus gives rise to a vacuum-stability condition that constrains the allowed parameter space of the flavored-GUT framework. 
The modulus $\tau$ settles near the modular fixed points, $\tau\approx i$ and $\tau\approx\pm1/2+i\sqrt{3}/2$, with $\langle\tau\rangle\approx i$ selected in our phenomenological analysis of quark and lepton masses and mixings. Interestingly, at $\tau=i$, the combination $E^2_6-E^3_4$ reduces to $-E^3_4$. Consequently, the higher modular-weight contributions to the neutrino operators can be absorbed into a reduced set of effective couplings entering the seesaw formula.

We have shown that $U(1)_X$ gauge invariance of the nonperturbative superpotential requires the nonperturbative indices $N_i$ ($i=\alpha,\beta$) to be quantized. These indices play a crucial role in both $U(1)_X$ modulus stabilization and the axion quality problem. 
The vacuum-stability requirement constrains the allowed values of  $N_i$ to the range $N_i\in[18,47]$, thereby determining the corresponding nonperturbative instanton actions. Furthermore, the allowed values of the quantized nonperturbative indices constrain the range of $g_X$, yielding values consistent with SM gauge-coupling unification.
We have shown that 
 anomaly cancellation and SM gauge coupling unification, as fundamental consistency conditions, constrain the framework in a parameter region where vacuum-stability is maintained and the string-induced gravitational contribution to the axion potential is effectively vanishing. This suppression is realized through the normalized $U(1)_X$ gauge coupling at the flavored-GUT scale, thereby dynamically realizing a qualified QCD axion within the flavored-GUT framework. Therefore, the axion solution to the strong CP problem provides a compelling motivation for SM gauge-coupling unification within this framework. 

We have also found that supersymmetry breaking is governed predominantly by the uplifting contribution associated with the negative vacuum energy of the supersymmetric AdS vacuum. The resulting gravitino mass can be of order $\sim{\cal O}(10)$ TeV, consistent with the SUSY-breaking scale favored by RG running analyses of SM gauge-coupling unification. 

Overall, the flavored-GUT framework establishes a unified connection among flavor physics, anomaly cancellation, neutrino mass generation, SM gauge-coupling unification, gravitational dynamics, and axion physics, with a high-energy scale emerging from the dynamics of flavor symmetry breaking. Its central implication is that these phenomena need not originate from independent sectors but can instead correlated through a common set of consistency conditions rooted in string-derived supergravity. In particular, the GS mechanism links the gauge and gravitational sectors, while $U(1)_X$ gauge invariance quantizes the allowed nonperturbative indices, thereby connecting vacuum stability to the suppression of the string-induced gravitational contribution to the axion potential. Together with the modular structure governing fermion masses and mixings, these features provide a framework in which the flavor-breaking scale, and the QCD axion properties, the seesaw scale, and the supersymmetry-breaking scale are correlated by the underlying theoretical constraints rather than treated as independent inputs.

\acknowledgments{We would like to thank Prof. Tianjun Li (Henan Normal university) and Eung Jin Chun (KIAS) for the thoughtful and in-depth discussions, and Chris Ahn (Port moody secondary school) for pointing out some typos. This work is supported by  NSFC 12135006.
}

\appendix
\section{The canonically normalized action for axions}
To canonically normalize Eq.(\ref{act1}), setting $\theta_X=a_\theta/(8\pi^2\,f_\theta)$ with $f_\theta=\sqrt{2K_{U_X\bar{U}_X}}/8\pi^2$ and using $\varphi_X$ in Eq.(\ref{kf1}), it becomes
\begin{eqnarray}
 &&\frac{1}{2}(\partial^\mu a_\theta)^2-\tilde{\kappa}_i\frac{a_\theta}{f_\theta\,16\pi^2}{\rm Tr}(F^{\mu\nu}_i\tilde{F}_{i\mu\nu})+\tilde{\kappa}_R\frac{a_\theta}{f_\theta\,64\pi^2}R^{\mu\nu\rho\sigma}\tilde{R}_{\mu\nu\rho\sigma}+\frac{1}{2}(\partial_\mu A_X)^2\nonumber\\
 &&+\tilde{\kappa}_i\frac{A_X}{f_X}\frac{\delta^i_X}{32\pi^2}{\rm Tr}(F^{\mu\nu}_i\tilde{F}_{i\mu\nu})-\tilde{\kappa}_R\frac{A_X}{f_X}\frac{\delta^R_X}{128\pi^2}R^{\mu\nu\rho\sigma}\tilde{R}_{\mu\nu\rho\sigma}-A^\mu_X J^X_\mu-A^\mu_\theta J^\theta_\mu\nonumber\\
 &&+\frac{1}{2\tilde{g}^2_X}m^2_A A^\mu_X A_{X\mu}-\frac{1}{4g^2_i}F^{\mu\nu}_iF_{i\mu\nu}-\frac{\tilde{g}^2_X}{2}(-\xi^{\rm FI}_X+X|\varphi_X|^2)^2+|\partial^\mu\varphi_{B-L}|^2\nonumber\\
 &&+\frac{1}{2\tilde{g}^2_{B-L}}m^2_{B-L} A^\mu_{B-L} A_{(B-L)\mu}-A^\mu_{B-L}J^{B-L}_\mu\,.
 \label{act2}
\end{eqnarray}

\section{Modular functions}
\label{mfe}
The Dedekind eta function is defined as
\begin{eqnarray}
 \eta(\tau)=q^{1/24}\prod^{\infty}_{n=1}(1-q^n)\quad\text{with}~q\equiv e^{i2\pi\tau}~\text{and}~{\rm Im}(\tau)>0\,.
\end{eqnarray}
It satisfies the following modular transformation properties
\begin{eqnarray}
 \eta(-1/\tau)=\sqrt{-i\tau}\,\eta(\tau)\,,\qquad\eta(\tau+1)=e^{i\pi/12}\,\eta(\tau)\,.
\end{eqnarray}

The $q$-expansions of the three linearly independent modular functions $Y_i(\tau)$ are given by
\begin{eqnarray}
 &&Y_1(\tau)=1+12q+36q^2+12q^3+...\nonumber\\
 &&Y_2(\tau)=-6q^{1/3}(17q+8q^2+...)\nonumber\\
 &&Y_3(\tau)=-18q^{2/3}(1+2q+5q^2+...)\,.
 \label{mfc}
\end{eqnarray}

\section{Quark and charged-lepton mixing matrices}
\label{mmR}

The right-handed quark mixing matrices $V_R^{u}$ and $V_R^{d}$ are approximately unit matrices up to ${\cal O}(\lambda^4)$, namely,  $V^{d(u)}_R={\bf I}+{\cal O}(\lambda^4)$, with small mixing angles given respectively by 
 \begin{eqnarray}
  &&\theta^u_{1}\simeq\frac{|y_cy^\ast_{t2}Y^{(4)\ast}_{\bf 1}|}{|y_t|^2}\Delta^{14}_\chi(2{\rm Im}\, \tau)^{-2}\,,\qquad \theta^d_{1}\simeq\frac{|y_sy^\ast_{b2}Y^{(4)\ast}_{\bf 1}|}{|y_b|^2}\Delta^{12}_\chi(2{\rm Im}\, \tau)^{-2}\,, \nonumber\\
  &&\theta^u_{2}\simeq\frac{|y_uy^\ast_{t1}Y^{(8)\ast}_{\bf 1}|}{|y_t|^2}\Delta^{29}_\chi(2{\rm Im}\, \tau)^{-4}\,, \qquad \theta^d_{2}\simeq\frac{|y_dy^\ast_{b1}Y^{(8)\ast}_{\bf 1}|}{|y_b|^2}\Delta^{20}_\chi(2{\rm Im}\, \tau)^{-4}\,, \nonumber\\
  &&\theta^u_{3}\simeq\frac{|y_uy^\ast_{c1}Y^{(4)\ast}_{\bf 1}|}{|y_c|^2}\Delta^{15}_\chi(2{\rm Im}\, \tau)^{-2}\,,\qquad \theta^d_{3}\simeq\frac{|y_dy^\ast_{s1}Y^{(4)\ast}_{\bf 1}|}{|y_s|^2}\Delta^{8}_\chi(2{\rm Im}\, \tau)^{-2}\,,
 \label{qrma}
 \end{eqnarray}
where, in particular, $\theta^d_{3}\sim{\cal O}(\lambda^{3-4})$.

The right-handed charged-lepton mixing matrix $V_R^\ell$ is approximately given by
 \begin{eqnarray}
   &V^\ell_R=\tilde{C}_\ell{\left(\begin{array}{ccc}
 1 & 0 &  0 \\
0 &   1  &  -\theta^r_1   \\
0 & \theta^r_1  & 1
 \end{array}\right)}\tilde{K}_\ell+{\cal O}(\lambda^4)\,,
 \label{cLep2}
 \end{eqnarray}
 where $\tilde{C}_\ell={\rm diag}(e^{i(\alpha^\ell_2-\alpha^\ell_1-\alpha^\ell_3)}, e^{i(\alpha^\ell_3-\alpha^\ell_1)}, e^{i(\alpha^\ell_2-\alpha^\ell_1)} )$, and $\tilde{K}_\ell={\rm diag}(e^{i(\alpha^\ell_1-2\alpha^\ell_2)}, 1, e^{i2\alpha^\ell_1} )$. The phases are approximately given by $\alpha^\ell_1=\frac{1}{2}\arg(y_{\mu3}Y^{(4)}_{\bf 1})$, $\alpha^\ell_2\simeq\frac{1}{2}\arg(y_{e3}Y^{(8)}_{\bf 1})-\frac{1}{2}\alpha^\ell_1$, $\alpha^\ell_3\simeq\frac{1}{2}\arg(y^\ast_{\mu}y_{e2}Y^{(4)}_{\bf 1})+\frac{1}{2}\alpha^\ell_1-\frac{1}{2}\alpha^\ell_2$, and the corresponding mixing angles are approximately given by
  \begin{eqnarray}
   &&\theta^r_1\simeq\big|\frac{y_{\mu3}}{y_\tau}Y^{(4)}_{\bf 1}\big|(2\,{\rm Im}\,\tau)^{-2}\Delta^6_\chi\,,\nonumber\\
   &&\theta^r_2\simeq\big|\frac{y_{e3}}{y_\tau}Y^{(8)}_{\bf 1}\big|(2\,{\rm Im}\,\tau)^{-4}\Delta^{17}_\chi\,,\nonumber\\
   &&\theta^r_3\simeq\big|\frac{y_{e2}}{y_\mu}Y^{(4)}_{\bf 1}\big|(2\,{\rm Im}\,\tau)^{-2}\Delta^{11}_\chi\,.
 \label{Vl}
 \end{eqnarray} 
For $\Delta_\chi=0.62$ and ${\rm Im}\,\tau\simeq1$, one typically expects $\theta^r_1\sim{\cal O}(\lambda^3)$, $\theta^r_2\sim{\cal O}(\lambda^7)$, and $\theta^r_3\sim{\cal O}(\lambda^4)$.
On the other hand, the left-handed charged-lepton mixing matrix $V_L^\ell$ is approximately equal to the unit matrix up to ${\cal O}(\lambda^4)$, with
  \begin{eqnarray}
   &&\theta^L_1\simeq\frac{|y_{\mu}y_{\mu3}Y^{(4)}_{\bf 1}|}{|y_\tau|^2}(2\,{\rm Im}\,\tau)^{-2}\Delta^{13}_\chi\,,\nonumber\\
   &&\theta^L_2\simeq\frac{|y^\ast_{e}y_{e3}Y^{(8)}_{\bf 1}|}{|y_\tau|^2}(2\,{\rm Im}\,\tau)^{-4}\Delta^{35}_\chi\,,\nonumber\\
   &&\theta^L_3\simeq\frac{|y^\ast_ey_{e2}Y^{(4)}_{\bf 1}|}{|y_\mu|^2}(2\,{\rm Im}\,\tau)^{-2}\Delta^{22}_\chi\,.
 \label{VlI}
 \end{eqnarray} 
\newpage

\end{document}